\documentclass[manuscript]{aastex}

\usepackage{color}

\newcommand{\OI}{\mbox{[\ion{O}{1}]}}
\newcommand{\CII}{\mbox{[\ion{C}{2}]}}
\newcommand{\SiII}{\mbox{[\ion{Si}{2}]}}

\newcommand{\FeII}{\mbox{[\ion{Fe}{2}]}}

\newcommand{\degree}{\mbox{$^{\circ}$}}
\newcommand{\am}{\mbox{\arcmin}}
\newcommand{\as}{\mbox{\arcsec}}

\shorttitle{L1448-MM}
\shortauthors{Lee. et al.}

\begin{document}

\title{L1448-MM observations by the {\it Herschel} Key program, ``Dust, Ice, and Gas
In Time'' (DIGIT)}

\author{Jinhee Lee, Jeong-Eun Lee}
\affil{Department of Astronomy and Space Science, Kyung Hee University,
   Yongin-shi, Kyungki-do 449-701, Korea}
\email{jeongeun.lee@khu.ac.kr}



\author{Seokho Lee}
\affil{Astronomy Program, Department of Physics and Astronomy, Seoul National
University,
Seoul 151-742, Korea}

\author{Joel. D. Green, Neal J. Evans II}
\affil{Department of Astronomy, University of Texas at Austin,
2515 Speedway, Stop C1400, Austin, TX 78712-1205, USA}

\author{Minho Choi}
\affil{Korea Astronomy and Space Science Institute,
776 Daedeokdaero, Yuseong, Daejeon 305-348, Korea}

\author{Lars Kristensen}
\affil{Harvard-Smithsonian Center for Astrophysics, MS78, Cambridge, MA02138, USA}

\author{Odysseas Dionatos, Jes K. J{\o}rgensen}
\affil{Centre for Star and Planet Formation, Natural History Museum of Denmark,
University of Copenhagen, \O ster Voldgade 5 - 7, 1350, Copenhagen, Denmark}

\and

\author{the DIGIT team}

\begin{abstract}

We present {\it Herschel}/PACS observations of L1448-MM, a Class 0 protostar with a
prominent outflow.
Numerous emission lines are detected at 55 $<$ $\lambda$ $<$ 210 $\mu$m
including CO, OH, H$_2$O, and \OI.
We investigate the spatial distribution of each transition to find that
 lines from low energy levels tend to distribute along the outflow direction
 while lines from high energy levels peak at the central spatial pixel.
 Spatial maps reveal that OH emission lines are
 formed in a relatively small area,
 while \OI\ emission is extended.
According to the rotational diagram analysis, the CO emission can be fitted by two (warm and hot) temperature components.
For H$_2$O, the ortho-to-para ratio is close to 3.
The non-LTE LVG calculations suggest that CO and H$_2$O lines could
instead be formed
 in a high kinetic temperature (T $>$ 1000 K) environment, indicative of a shock origin. 
For OH, IR-pumping processes play an important role in the level population.
The molecular emission in L1448-MM is better explained with a C-shock model,
 but the atomic emission of PACS \OI\ and  {\it Spitzer}/IRS \SiII\ emission is not consistent with C-shocks,
suggesting multiple shocks in this region.
Water is the major line coolant of L1448-MM in the PACS wavelength range,
 and the best-fit LVG models predict that H$_2$O and CO emit (50--80) $\%$ of their line luminosity in the PACS wavelength range.
\end{abstract}


\section{Introduction}
 In the earliest evolutionary stage, protostars are surrounded by optically
thick envelopes,
 still infalling remnants of the molecular cloud core.
 These deeply embedded young stellar objects (YSOs)
can be seen at long wavelengths, from infrared to radio,
as higher frequency photons from the protostars are absorbed
by their envelope material and reemitted at longer wavelengths.
The active accretion process in the embedded Class 0/I objects
launches jets and well collimated outflows that induce shocks and thus
heat the surrounding envelope material, which cools through atomic and 
molecular emission lines. 
The fast-moving jets carve out portions of the envelope as they exit the
protostar, creating dense walls around an evacuated cavity, along which FUV photons
penetrate.  

The gas heated by shocks and high energy UV photons produces line
emission at IR wavelengths.
CO and H$_2$O, which are the most abundant species after H$_2$ in the envelopes of YSOs,
have copious rotational transitions in the FIR regime.
In addition to CO and H$_2$O lines, the FIR {\OI} and OH lines are also frequently 
observed in the shocked and entrained material as presented by \citet{Gia01} and 
\citet{Nis02}.
Based on observations of 17 Class 0 sources with the Long Wavelength Spectrometer (LWS)
on board the {\it Infrared Satellite Observatory (ISO)}, \citet{Gia01} showed that the total
FIR line cooling generated by \OI, OH, H$_2$O, and CO lines can be a direct
tracer of the power deposited in the outflow, which is directly related to the mass accretion 
rate to the central object. Therefore, FIR spectroscopic observations in the wavelength range
of $\sim$50 to 200 $\mu$m is very important to study the cooling budget among 
{\OI}, OH, H$_2$O, and CO, and to understand the related heating mechanisms.

Recently, observations of embedded YSOs 
have been carried out with the Photodetector Array Camera and
Spectrometer (PACS; Poglitsch et al. 2010) aboard {\it Herschel} to reveal 
very rich FIR line forests \citep{Kem10a, Kem10b, Her12, Goi12, Kar13, Gre13}. 
Because of its high sensitivity and relatively high spectral resolution,
relative to the {\it ISO} observations,
PACS can detect weaker lines and resolve different line components.
Those PACS FIR line observations have been
combined with detailed models to find that both C- and J-shocks,
 as well as UV radiation,
are necessary to explain the relative strength of lines \citep{Vis12, Goi12}.

One of the most studied embedded sources with {\it ISO} and {\it Herschel} is L1448-MM 
\citep{Nis99, Nis00, Gia01, Kri11, San12, Nis13}.
L1448-MM, a deeply embedded Class 0 YSO with $L\rm_{bol}$ = 8.4 $L_{\sun}$
\citep{Gre13} with an outflow, was detected first in 2-cm radio observations
\citep{Cur90}.
VLBI parallax measurements of the water maser from L1448-MM yield
a distance, $d = 232\pm18$ pc (Hirota et al. 2011).
The continuum was also detected at millimeter wavelengths \citep{Bac91}.
The H$_2$ $v = 1-0$ vibrational line \citep{Bal93}
and water masers were also detected \citep{Che95, Cla96, Hir11}.
\citet{Bac90} detected high velocity (up to 70 km s$^{-1}$) bullets in
CO J = 2 -- 1 and J = 1 -- 0 observations. 
The bullets in the red-shifted gas are aligned in the SE direction and have a
symmetrical counterpart to the NW, with respect to the L1448-MM position.
\citet{Dut97} also found more evidence for 
bullets by observing the SiO $v=$0, J = 2 -- 1
transition.

The {\it ISO} observations detected molecular emission lines (CO,
OH, and H$_2$O) as well as the
atomic fine structure line (\OI ) in L1448-MM \citep{Nis99}.
\citet{Nis99} suggested that the molecular lines are excited in a region with T$\sim$700-1400 K. 
They also pointed out that  H$_2$O is the main coolant in the region with a high abundance, which
may be associated with non-dissociative shock.
The low spatial resolution (the beam size of LWS is 75$\arcsec$) of ISO left
open the possibility that emission from different regions could have been
mixed together. 

High sensitivity observations with {\it Spitzer} revealed two
point-like infrared objects in L1448-MM \citep{Jor06}.
The newly identified point-like infrared source (L1448-MM(B)) is located 
$8\arcsec$ to the south
of the previously known YSO (L1448-MM(A)).
(These two sources were encompassed by a {\it ISO} beam.) 
\citet{Hir10} found a low velocity ($v_{\rm{flow}}< 15$ km s$^{-1}$) outflow
associated with L1448-MM(B) in their Submillimeter Array (SMA) CO J=3--2 map.

Recently, \citet{Kri11} resolved multiple kinematic components, which consist of 
the high velocity ($|v| >$ 50 km s$^{-1}$) bullets referred to as EHV (Extremely High Velocity)
components and a broad emission component centered at $v\rm_{LSR}$, in water and CO line profiles
observed with {\it Herschel}/HIFI in L1448-MM. \citet{San12} and \citet{Nis13} made 
detailed studies of the excitation conditions in different kinematic components of one of outflow knots and 
on-source, respectively, based on the {\it Herschel}/HIFI water observations.
Therefore, the significantly improved sensitivity and spatial/spectral resolutions of {\it Herschel},
compared to {\it ISO}, is crucial to study L1448-MM, which is very complex spatially and
kinematically. 

Here we present a more detailed study of L1448-MM, which was observed 
by the {\it Herschel} Key Program, DIGIT ``Dust, Ice, and Gas In Time'' (PI: N. Evans), 
in the view of the cooling budget in the gas heated up to a few 100 -- 1000 K.
We model observed molecular line fluxes with a non-LTE radiative tranfer code, 
RADEX, to understand the energy budgets and the excitation conditions in the region, 
and we also compare  observed line fluxes 
with shock models to study the characteristics of associated shocks.

 In this paper, we combine our PACS observations with previously 
 obtained data and information to study L1448-MM in more detail.
Observations and data reduction are described in section 2.
In section 3 and 4, we present observational results and the analyses.
Then we discuss and summarize the results in section 5 and 6.

\section{Observations and Data Reduction}

\subsection{The PACS observation}

PACS is a 5$\times$5 array of $9\farcs4 \times 9\farcs4$ spatial pixels
(hereafter referred
to as spaxels) covering the spectral range from 50-210 $\mu$m with
$\lambda$/$\delta\lambda$ $\sim$ 1000-3000, divided into four segments,
covering
$\lambda \sim$ 50-75, 70-105, 100-145, and 140-210 $\mu$m.
L1448-MM was observed on 2 Feb 2011
($\lambda \sim$ 50-75 and 100-145 $\mu$m; AOR: 1342213683) and 22 Feb 2011
($\lambda \sim$ 70-100 and 140-210 $\mu$m; AOR: 1342214675) 
in the range scan mode of PACS with a single footprint.

The telescope and sky background emission was subtracted using two nod
positions 6\arcmin\ from the source in opposite directions.  Each segment was
reduced using the ``calibration block ''
pipeline from HIPE v8.1 \citep{ott10}, and flux calibrated to an extraction
from HIPE v6.1.  The former extraction produces
spectra of higher S/N, while the latter provides better absolute flux
calibration.   This
process and the reasoning behind using two different HIPE versions are
described in detail in \citet{Gre13}.
\citet{Gre13} applied consistent methods to all sources to correct for the extended emission in 
continuum and line. However, the methods are not optimized for targets with multiple sources.
 In the case of L1448-MM,
we observed clear evidence of {\it multiple} emitting sources within the PACS field-of-view.
As a result, we used modified methods to extract exact continuum/line fluxes, so our fluxes are 
slightly different from those in \citet{Gre13}.
However,  the main concept for our flux measurements is similar to \citet{Gre13}; 
only the HIPE v6.1 reduction was used for the SED extraction, but the combination of the HIPE v6.1 
and HIPE v8.1 reductions was adopted for line fluxes. 

Unlike isolated sources, it is hard to correct for the point spread function in L1448-MM because 
it has multiple sources. As a result, we added the whole fluxes over 25 spaxels to 
present the total continuum flux. 
(The decomposition of the fluxes by the multiple sources is explained in \S 3.2.)  
For the line fluxes, we calculated the equivalent width (EW) of each emission line from the HIPE v8.1
reduction, then multiplied the EW by the local continuum determined from the HIPE v6.1
reduction to calculate total line fluxes over the whole 25 spaxels, fitting a
first-order polynomial baseline to local continuum.
Line widths and continuum levels were fitted with the Spectroscopic Modeling
Analysis and Reduction Tool (SMART; \citet{Hig04}),
which was originally developed for data analysis of the Infrared Spectrograph
(IRS; \citet{Hou04}) on {\it Spitzer}.

 For the total line fluxes over the whole L1448-MM region covered by
PACS, we used the EWs of CO, H$_2$O, and OH lines measured
in the spectra extracted from the two spaxels, where the lines are the strongest 
and two point sources, L1448-MM(A) and L1448-MM(B) are located. 
We compared the EWs measured from these two spaxels with the EWs 
measured from a sum over all 25 spaxels to determine if there was any sign of
spatially extended CO, H$_2$O, or OH lines.  In this analysis we
found that the EWs were not different in the two cases, and the line fluxes increased 
as the point spread function (PSF) would predict.
However, for \OI\ lines, the EW measured from all spaxels
 is much greater than that measured from the two spaxels,
even after accounting for broadening due to the PSF.
\OI\ lines are detectable over 4 spaxels (see Fig.~\ref{map_oi}), and the EWs from the 4 spaxels are
consistent with the EWs from the whole 25 spaxels although the 4-spaxel 
EWs have lower measurement errors.
As a result, we adopted the 4-spaxel EWs for \OI\ lines to
calculate total \OI\ line fluxes. 
The procedure of measuring line fluxes as well as the error calculation is described 
in the Appendix of this work.
According to the measured EWs, all observed lines are spectrally
unresolved in the PACS observations.
As mentioned above, L1448-MM is composed of multiple sources;
therefore, we also measured fluxes of each source separately.
The simple method we used to decompose the multiple sources into separate
fluxes is described in section 3.2.

\subsection{The IRS mapping of the H$_2$ pure rotation lines}

The area around L1448-MM was mapped with {\it Spitzer}/IRS in February 2008 \citep{Neu09, Gia11}.
H$_2$ rotational lines (S(0) - S(7)) in the range of 5 $\sim$ 38 $\mu$m were
detected.  We convolved these H$_2$ maps with the PACS spaxels to compare with our PACS
maps.
Refer to \citet{Neu09} and \citet{Gia11} for the details of these observations.

\subsection{The SRAO CO J = 2 -- 1 observation}

Additionally, we present observations of the CO J $=$ 2 -- 1 transition at
230.537970 GHz, mapped with the 6-m telescope
at Seoul Radio Astronomy Observatory (hereafter, SRAO) in 2010 May.
The beam FWHM is $48\arcsec$ at 230 GHz and  the velocity resolution is 0.127 km s$^{-1}$ after 
binning by two channels.
The main beam efficiency and pointing accuracy are 0.57 and $\sim$ 3$\arcsec$,
respectively.

\subsection{The archival data}
Finally we include {\it Spitzer} observations of L1448-MM.
The IRAC and MIPS images used here have been downloaded from the {\it Spitzer} data archive.
However, we adopt the point source fluxes at the IRAC and MIPS bands from \citet{Jor06}.
The IRS SH and LH data obtained by Nisini in 2006 have been also downloaded from the {\it Spitzer} 
data archive and reduced with the same method used in \citet{Fur06}.

\section{Results}

\subsection{Spitzer images of L1448-MM}

\begin{figure}
\includegraphics[width=0.8\textwidth]{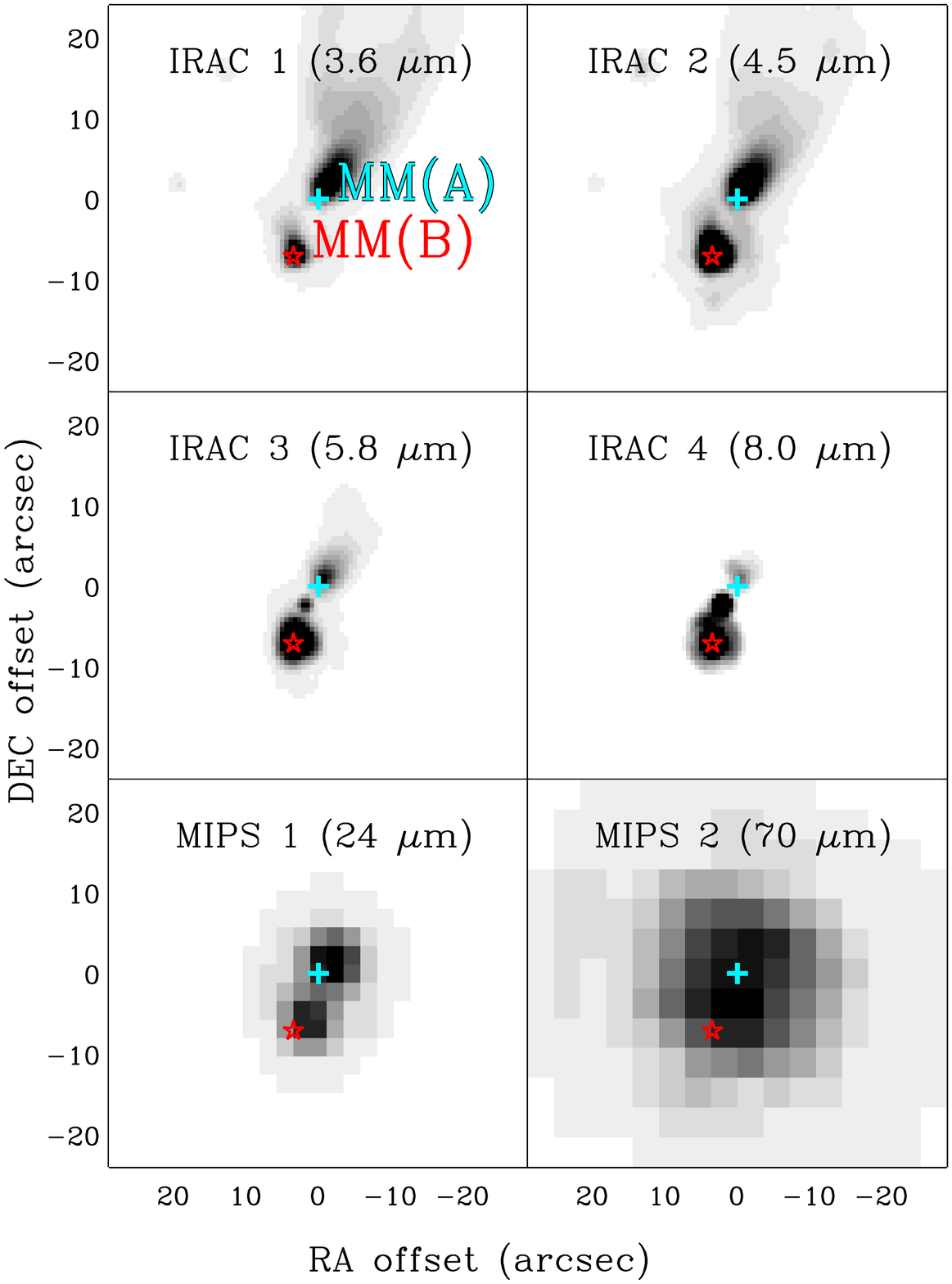}
\caption{{\it Spitzer} images of L1448-MM. Images are $\sim$ 50 $\arcsec$ ($\sim$
11000 AU) on each side.
The cross shows the location of the millimeter emission peak of L1448-MM(A) and
the star represents the location of L1448-MM(B). \citep{Jor07}}
The center of the images corresponds to the coordinates of L1448-MM(A), 
($\alpha$, $\delta$)=(3$^{h}$25$^{m}$38.87$^{s}$, +30\degree 44\am 5.4\as).
The coordinates of L1448-MM(B) are (3$^{h}$25$^{m}$39.14$^{s}$, +30\degree 43\am 58.3\as).

\label{spitzerimages}
\end{figure}

In the IRAC images, the trail of sources aligns in the SE-NW direction
(Fig.~\ref{spitzerimages}), which is consistent with the large-scale outflows.
There is an additional point-like source between L1448-MM(A) and MM(B) in the
IRAC band 3 and 4 images,
which has been reported as a tertiary 1.3 mm continuum source \citep{Mau10}.
 Owing to decreased resolution in MIPS band 2,
L1448-MM(A) and MM(B) are not resolved in that image, although
they seem distinct from each other in images at shorter wavelength.
Hereafter, we designate L1448-MM(A) and L1448-MM(B) as MM(A) and MM(B), respectively.

\subsection{The IRS and PACS spectra}

\begin{figure}
 \plotone{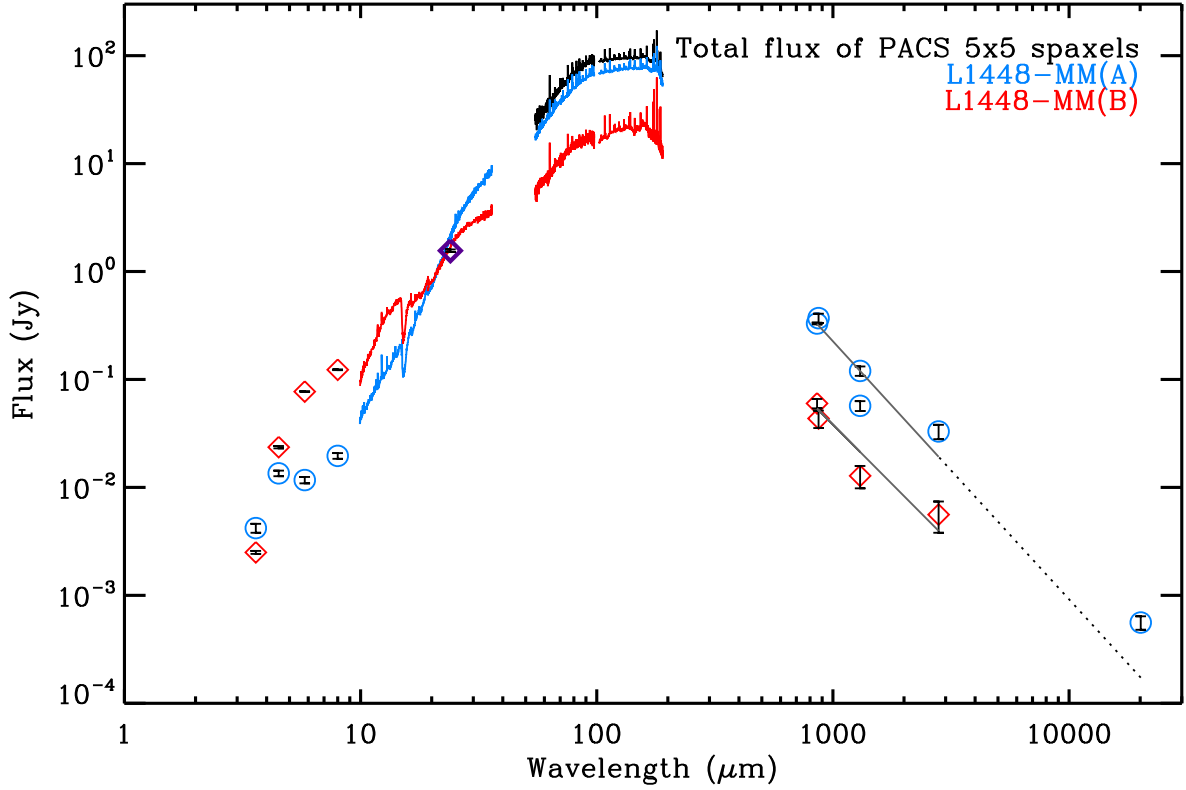}
\caption{The SEDs of MM(A) and MM(B).
The color lines represent spectroscopic observations:
The blue lines show spectra from MM(A) in the range of IRS and PACS.
The red lines represent spectra from MM(B) in the range of IRS and PACS.
The black line displays the spectrum extracted the whole 5 $\times$ 5 array of
PACS.
The longer than 100 $\mu$m of the PACS spectra has been scaled down by the
factor of 0.753 to match the continuum levels around 100 $\mu$m (Green et al.
in prep.).
Points show millimeter and IR photometry observations:
blue circles and red diamonds display MM(A) and MM(B) fluxes, respectively.
Purple diamond represents flux at 24 $\mu$m, where MM(A) and MM(B) are not
separated.
The solid lines present the results of fitting the fluxes in the range of 700
$\mu$m $<$ $\lambda$ $<$ 3000 $\mu$m.
The fitted line for MM(A) is extended to $\lambda$ $\sim$ 20000 $\mu$m as
a dotted line.
}
\label{SED}
\end{figure}

Here we present the SEDs of two point sources, MM(A) and MM(B) collected
for wavelengths ranging from a few microns to sub-mm.
In Fig.~\ref{SED}, the IR data points come from \citet{Jor06} and fluxes at
millimeter wavelengths are taken from
 \citet{Cur90}, \citet{Jor07}, \citet{Mau10}, and \citet{Hir10}.
The IRS spectra of MM(A) and MM(B) are also plotted in Fig.~\ref{SED}.

\begin{figure}
\epsscale{0.8}
\plotone{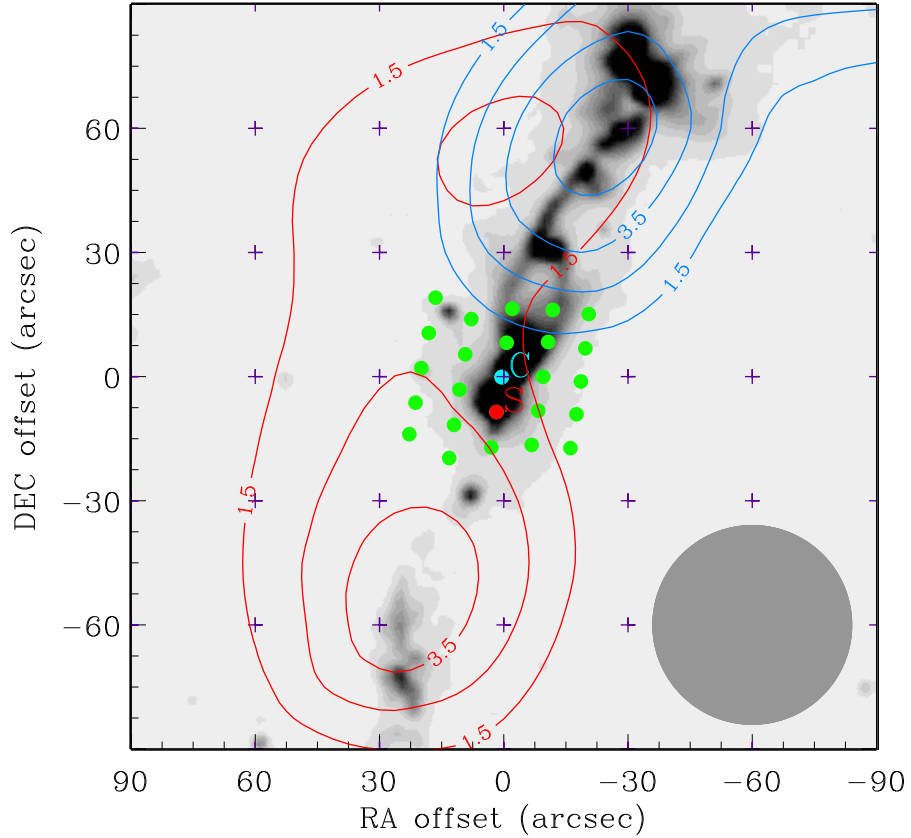}
\caption{PACS footprint (green) against the IRAC 2 image and the SRAO CO J $=$
2--1 outflow map. The spaxels located at the positions of MM(A) and MM(B) are 
designated as spaxels C (blue) and S (red), respectivlely.
The purple crosses are the positions of SRAO observation.
A much higher resolution map of CO J = 2 -- 1 can be found in \citet{Bac90},
 but the general feature is the same.
The grey circle shows the beam FWHM of SRAO.
The blue contours are integrated from $-10.0$ to $-1.2$
 km s$^{-1}$, while the red contours are
integrated from 9.9 to 20 km s$^{-1}$.
The contours start at 1.5 K km s $^{-1}$ and increase by 0.5 and 1.0 K km
s$^{-1}$
for blue and red components, respectively.
}
\label{HerSRAO}
\end{figure}

In the PACS range (Fig.~\ref{SED}), the black line represents the spectrum
extracted from all 25 spaxels (see \S2.1)
while the blue and red lines show the spectra of MM(A) and MM(B), 
separately.
All three spectra (black, blue, and red) have been extracted from the HIPE v6.1
reduction, which provides superior flux calibration.
In the PACS footprint, MM(A) and MM(B) are located near the central spaxel
and the spaxel in the right south of the central spaxel, which are designated as 
spaxel C and S, respectively, as seen in Fig.~\ref{HerSRAO}. 
The coordinates of spaxel C are the same as the coordinates of MM(A), but 
the coordinates of spaxel S, (3$^{h}$25$^{m}$39.0$^{s}$, +30\degree 43\am 56.8\as),
are slightly different from those of MM(B).

To obtain the spectra of MM(A) and MM(B),
we 1) extracted spectra from the spaxels, C and S, respectively,
 2) scaled to match continuum levels around 100 $\mu$m,
and 3) corrected for the effect of the Point Spread Function (PSF) of PACS to each source.

Since the spectra of each segment 
($\lambda \sim$50-75, 70-105, 100-145, 140-210 $\mu$m) are not matched smoothly, 
we scaled up or down SEDs of each segment to match continuum levels around 
100 $\mu$m.
The scale factors applied to the spectrum extracted from spaxel C are 1.052,
0.918, and 0.854
 in the range of $\lambda$ $<$ 72, 101 $\sim$ 142, and $>$ 142 $\mu$m,
respectively.
The scale factors for the spectrum extracted from spaxel S are 0.903, 0.667,
and 0.727, in the same wavelength range.

In order to correct for the PSF, that is, to decompose fluxes of MM(A)
and MM(B),
 we assume that only two point-like sources contribute the 
emission in the two spaxels
 and calculate the contribution of each source to the normalized fluxes of two
spaxels
 using the point spread function\footnote{$\rm
http://pacs.ster.kuleuven.ac.be/PACSPSF\_monochromatic\_ver2.0.tar.gz$}.
 The contribution of a point source to the spaxel where it is located is 80
$\%$ at 50 $\mu$m and 32 $\%$ at 200 $\mu$m,
 and to the adjacent spaxel is $\sim$ 3.5 $\%$ at 50 $\mu$m and  $\sim$10 $\%$
at 200 $\mu$m.
As a result, at a given wavelength, the fluxes of the two point sources are
decomposed by solving two simple linear simultaneous equations as below.

 \begin{equation}
 F_{\lambda ,\rm C}  = P_{\lambda ,\rm AC} * X_{\lambda ,\rm A} + P_{\lambda ,\rm BC} *
X_{\lambda ,\rm B}
 \end{equation}
 \begin{equation}
 F_{\lambda ,\rm S} = P_{\lambda ,\rm AS} * X_{\lambda ,\rm A} + P_{\lambda ,\rm BS} *
X_{\lambda ,\rm B}
\end{equation}
 $F_{\lambda ,\rm C}$ and $F_{\lambda ,\rm S}$ are fluxes measured from the spectra
extracted at spaxel C and S,
 while $X_{\lambda ,\rm A}$ and $X_{\lambda ,\rm B}$ are the decomposed fluxes of
MM(A) and MM(B), respectively.
 $P_{\lambda ,\rm AC}$ and $P_{\lambda ,\rm BC}$ are the contribution to spaxel C  by
MM(A) and MM(B),
  while $P_{\lambda ,\rm AS}$ and $P_{\lambda ,\rm BS}$ are the contribution to spaxel
S by MM(A) and MM(B), respectively.
$P_{\lambda ,\rm AC}$, $P_{\lambda ,\rm BC}$, $P_{\lambda ,\rm AS}$, and $P_{\lambda ,\rm BS}$ 
were calculated with the exact coordinates of MM(A) and MM(B).

The slope of the SEDs (rising into the FIR)  demonstrates that cold
material is present showing
that both MM(A) and MM(B) are deeply embedded.
The continuum luminosity integrated in the PACS wavelength range
is about 3.5 and 0.9 $L_{\sun}$ for MM(A) and MM(B), respectively, as listed in Table~\ref{lumtable}.
The bolometric luminosities of MM(A) and MM(B), which are calculated with
their decomposed SEDs, are $\sim$5.5 and 1.7 $L_{\sun}$ (see \S 5).

In the IRS range, the flux of MM(B) is flatter than that of MM(A)
longward of 15 $\mu$m.
In the IRS spectrum of MM(B), the flux drops shortward of 15 $\mu$m.
The excess flux in the IRAC bands compared to what is extrapolated from the
IRS SED might be attributed to some shocked gas and scattered light around MM(B).

\begin{figure}
\plotone{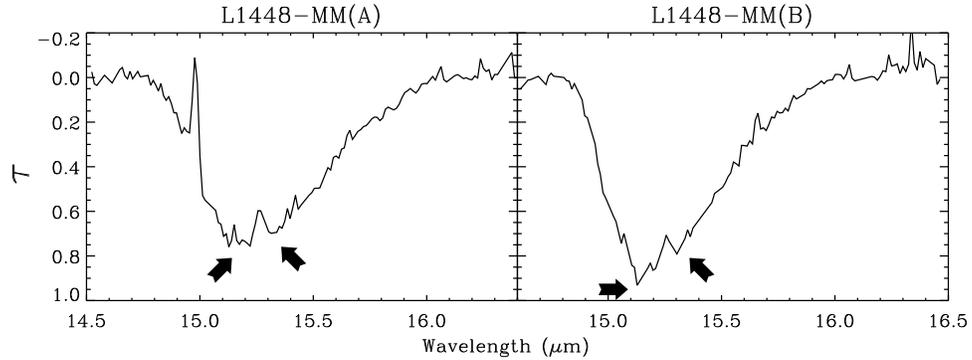}
\caption{The optical depth of the CO$_2$ 15.2 $\mu$m ice absorption feature
toward MM(A) and MM(B).
The emission line around 15 $\mu$m in MM(A) is possibly the CO$_2$ gas
line at 14.98 $\mu$m \citep{Dar98}.  
The arrows indicate the double-peaked feature, which is produced by the pure
CO$_2$ ice component.
}
\label{co2ice}
\end{figure}

A deep CO$_2$ 15.2 $\mu$m ice absorption feature has been detected in both
MM(A) and MM(B) (Fig.~\ref{co2ice}),
indicative of the existence of dense envelopes in both sources.
Both CO$_2$ absorption profiles show the broad red wing, which is attributed to
the CO$_2$ and H$_2$O ice mixture.
Although the S/N ratio is somewhat low, we see clearly a double-peaked feature, 
indicative of the pure CO$_2$ ice component, which exists only toward 
protostars \citep{Pon08}.
In the millimeter wavelength range (700 $<$ $\lambda$ $<$ 3000 $\mu$m),
the spectral index ($-d$log$S_{\nu}$/$d$log$\nu$, where $S_{\nu}$ is the flux density, and $\nu$
is the frequency)
is about 2.4 for MM(A) and 2.2 for MM(B).
These large ($> 2$) positive values imply that
the millimeter continuum emission comes from dust, not from ionized gas.
These SEDs and the CO$_2$ ice absorption feature, including the pure CO$_2$
ice feature, indicate that both sources are deeply embedded YSOs.
The available millimeter data, however,
do not constrain the dust properties further.

\begin{figure}
\includegraphics[width=0.9\textwidth]{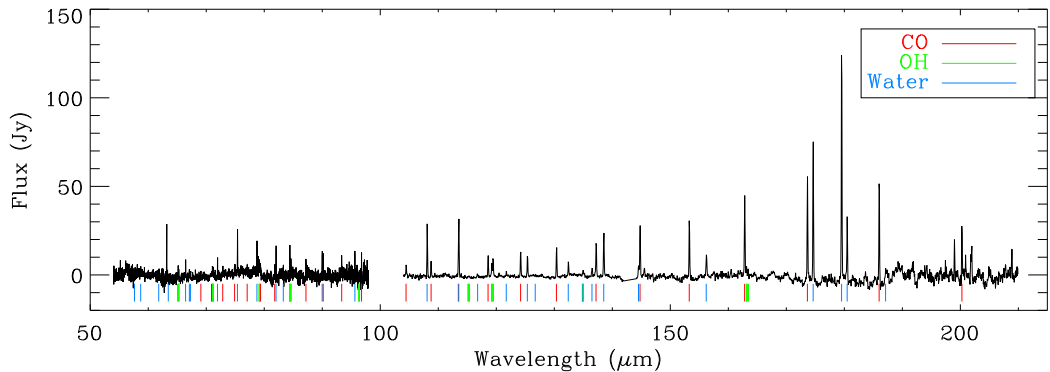}
\caption{{\it Herschel}/PACS continuum-subtracted spectra extracted from all 25
spaxels.  The emission lines are marked:
CO (red dashes), and OH (green dashes), H$_2$O (blue dashes).}
\label{pacsSED}
\end{figure}

\begin{figure}
\includegraphics[width=0.9\textwidth]{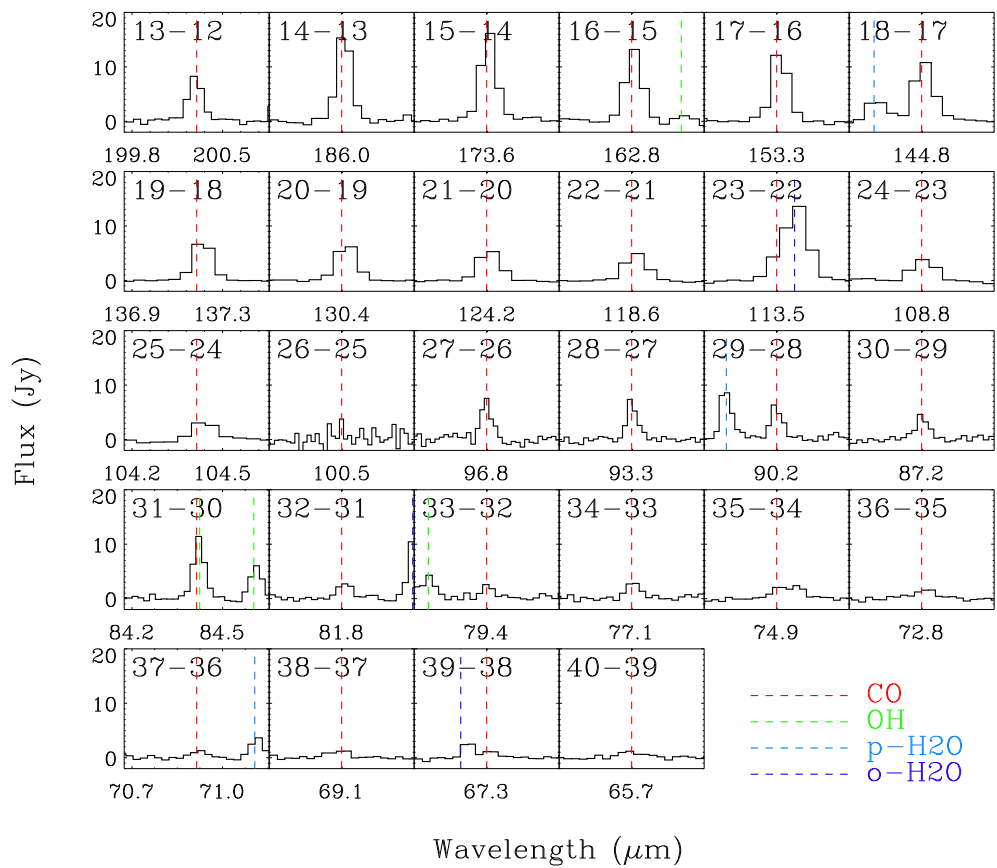}
\caption{CO emission lines (red vertical dashed lines) detected in L1448-MM.
These spectral line are determined from the central spaxel spectrum,
in order to present all detected line transitions.
Emission lines of other species located close to the CO lines are also marked
with different colors
(green for OH, blue for p-H$_2$O, and dark blue for o-H$_2$O).
The rotational transition of each line is noted on the upper left side of each
box.  The J$=$13--12 line
has an artificially lower flux due to a factor of $\sim$ 2 decrease in
sensitivity at the edge of the
PACS array, and this line is not used in subsequent models.}
\label{COlines}
\end{figure}

\begin{figure}
\includegraphics[width=0.9\textwidth]{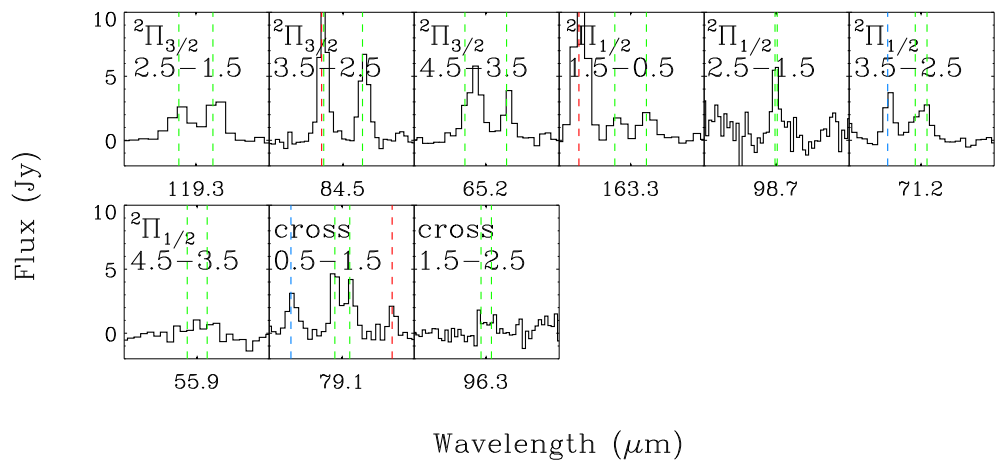}
\caption{The same as Fig.~\ref{COlines} but for OH lines.}
\label{OHlines}
\end{figure}

\begin{figure}
\includegraphics[width=0.9\textwidth]{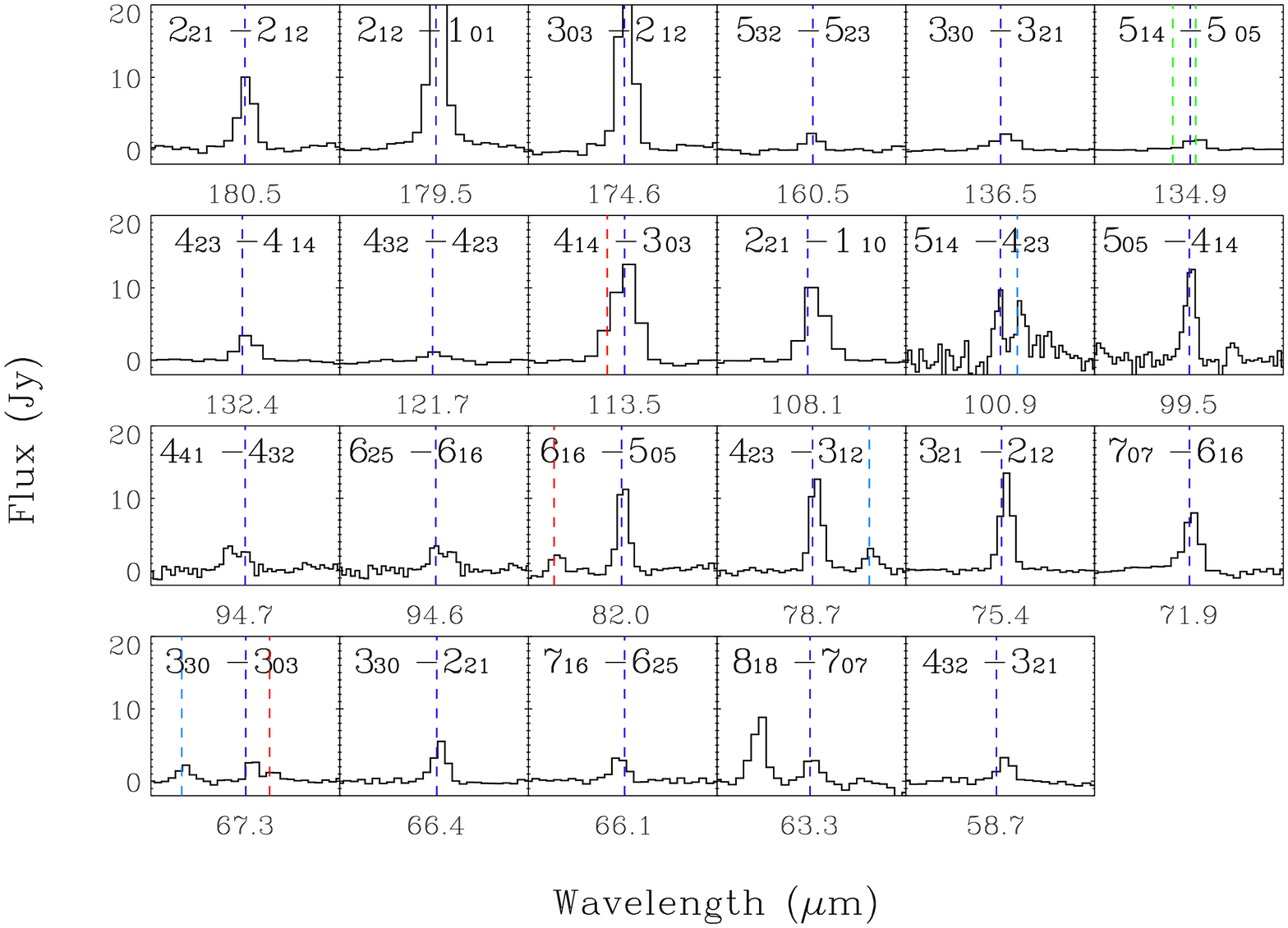}
\caption{The same as Fig.~\ref{COlines} $\&$ ~\ref{OHlines} but for o-H$_2$O
lines.}
\label{oH2Olines}
\end{figure}

\begin{figure}
\includegraphics[width=0.9\textwidth]{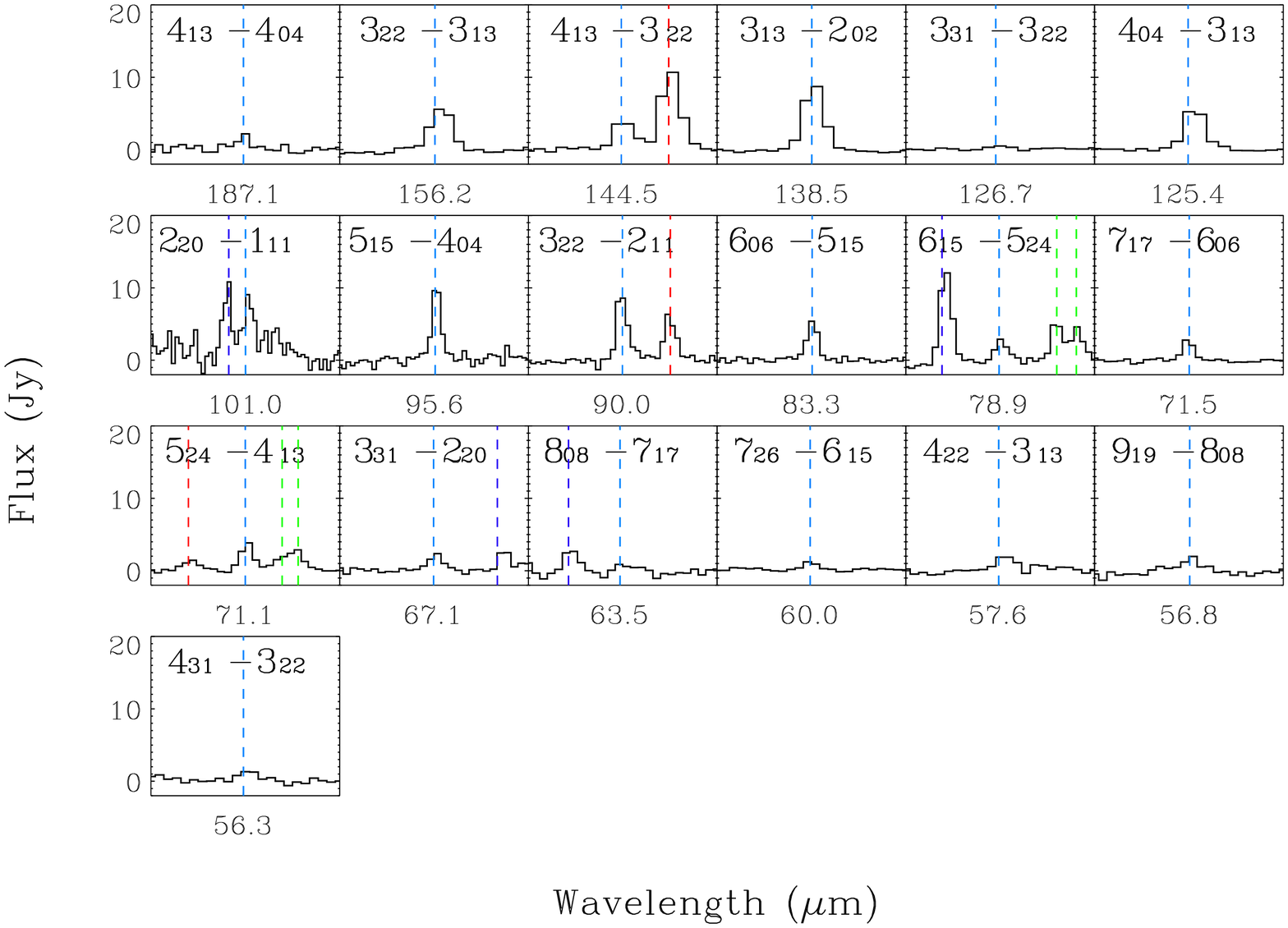}
\caption{The same as Fig.~\ref{COlines}, ~\ref{OHlines} $\&$ ~\ref{oH2Olines}
but for p-H$_2$O lines.}
\label{pH2Olines}
\end{figure}

The PACS spectra show numerous molecular lines of CO, H$_2$O,
and OH in addition to atomic \OI\ lines (Fig.~\ref{pacsSED}).
The detected lines are listed in Table~\ref{fluxtable} including the upper level energy,
$E_{\rm u}={E_{\rm upper}\over k}$ ($k$ is Boltzmann constant), 
and the Einstein coefficient, $A_{\rm ul}$.
The detected lines of each species are presented in Fig.~\ref{COlines} to
~\ref{pH2Olines}, where we present lines extracted only from the central spaxel.
As the central spaxel contains most of the continuum,
 many lines were detected only at the central spaxel.
 \citet{Dio09} calculated a visual extinction ($A_{\rm V}$) of 11 and 32 mag toward MM(A) and
 MM(B), respectively, using the 9.7 $\mu m$ silicate absorption feature of the IRS spectra.
 These $A_{\rm V}$ values attenuate line fluxes by less than 10 \% at $\lambda \ge 60$ $\mu m$.
 Therefore, we did not correct for the reddening in line fluxes.

CO transitions were detected from J = 13--12 up to J = 40--39.
The highest OH transition is $^{2}\Pi_{1/2}$ J = 9/2 -- 7/2.
Both lines of \OI\ at 63 $\mu$m and 145 $\mu$m and 
22 ortho- and 19 para-H$_2$O lines were detected.
No H$_2^{18}$O or $^{13}$CO lines were detected.
The upper limits of H$_2^{18}$O (2$_{12}$-1$_{01}$) and $^{13}$CO (15--14) are
1.70 $\times$ 10$^{-17}$ and 1.73 $\times$ 10$^{-17}$ W m$^{-2}$, respectively.
Assuming $[^{16}\rm O]/[^{18}\rm O] = 550$
and $[^{12}\rm C]/[^{13}\rm C]=90$, the upper limits of optical depth of $^{12}$CO (15--14)
and H$_2^{16}$O (2$_{12}$-1$_{01}$) are about 3 and 8, respectively, in the assumption
of equal excitation temperatures for the isotopologues.

\subsection{The PACS and IRS maps; distribution of continuum and line emission}

Figure~\ref{HerSRAO} compares the infrared image in the {\it Spitzer} IRAC band 2 with
the SRAO
CO J $=$ 2--1 outflow map. 
The footprint of PACS is marked on the image.
The IRAC 2 image shows a jet-like structure along a NW and SE path.
The SRAO CO J$=$2--1 outflow falls along this axis.
The jet-like structure is more prominent to the NW in the IRAC 2 image.
 \citet{Nis00} noted that the adjacent area of the 
southern lobe has a higher visual extinction than
 that of northern lobe.

\begin{figure}
\plottwo{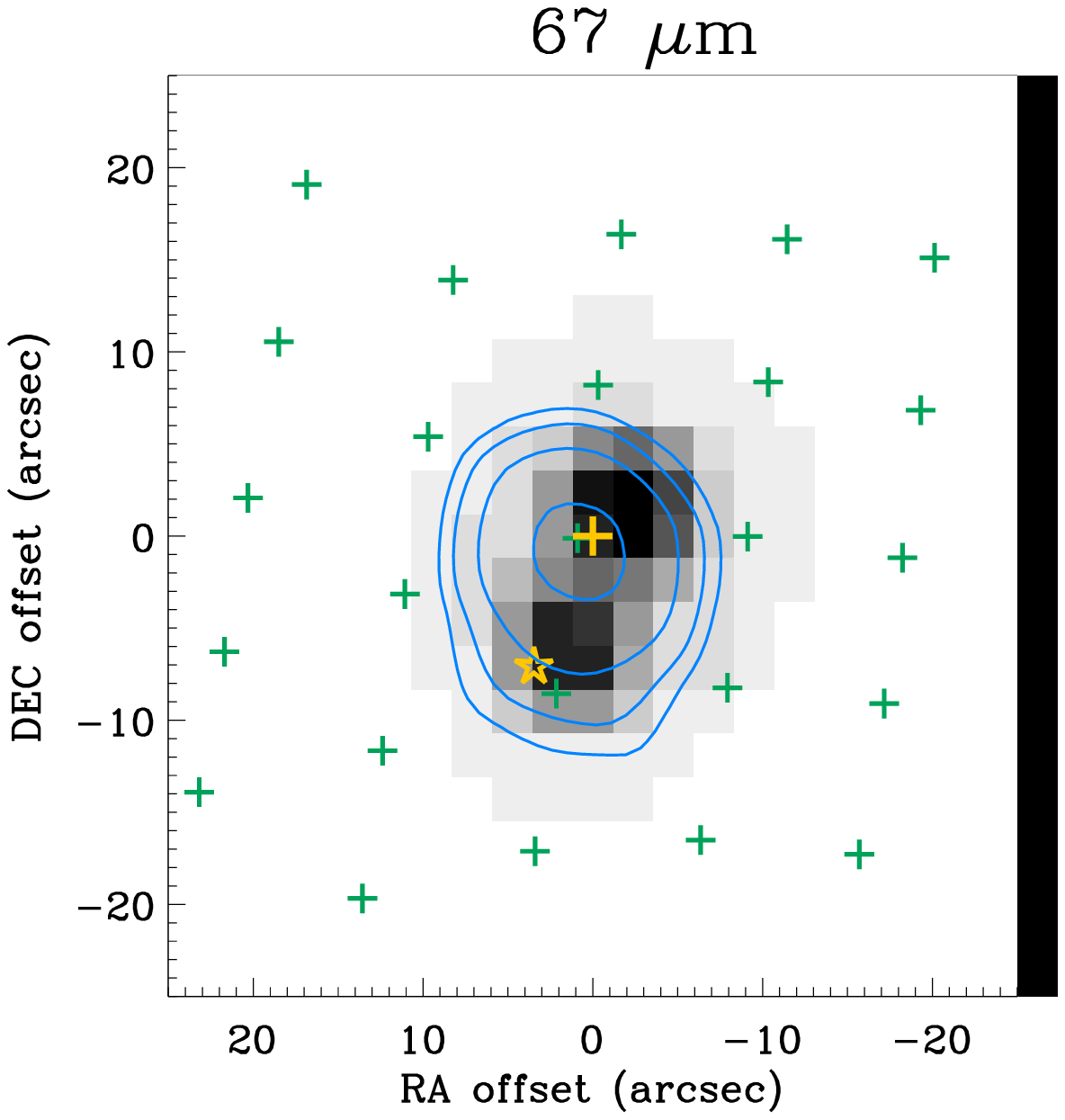}{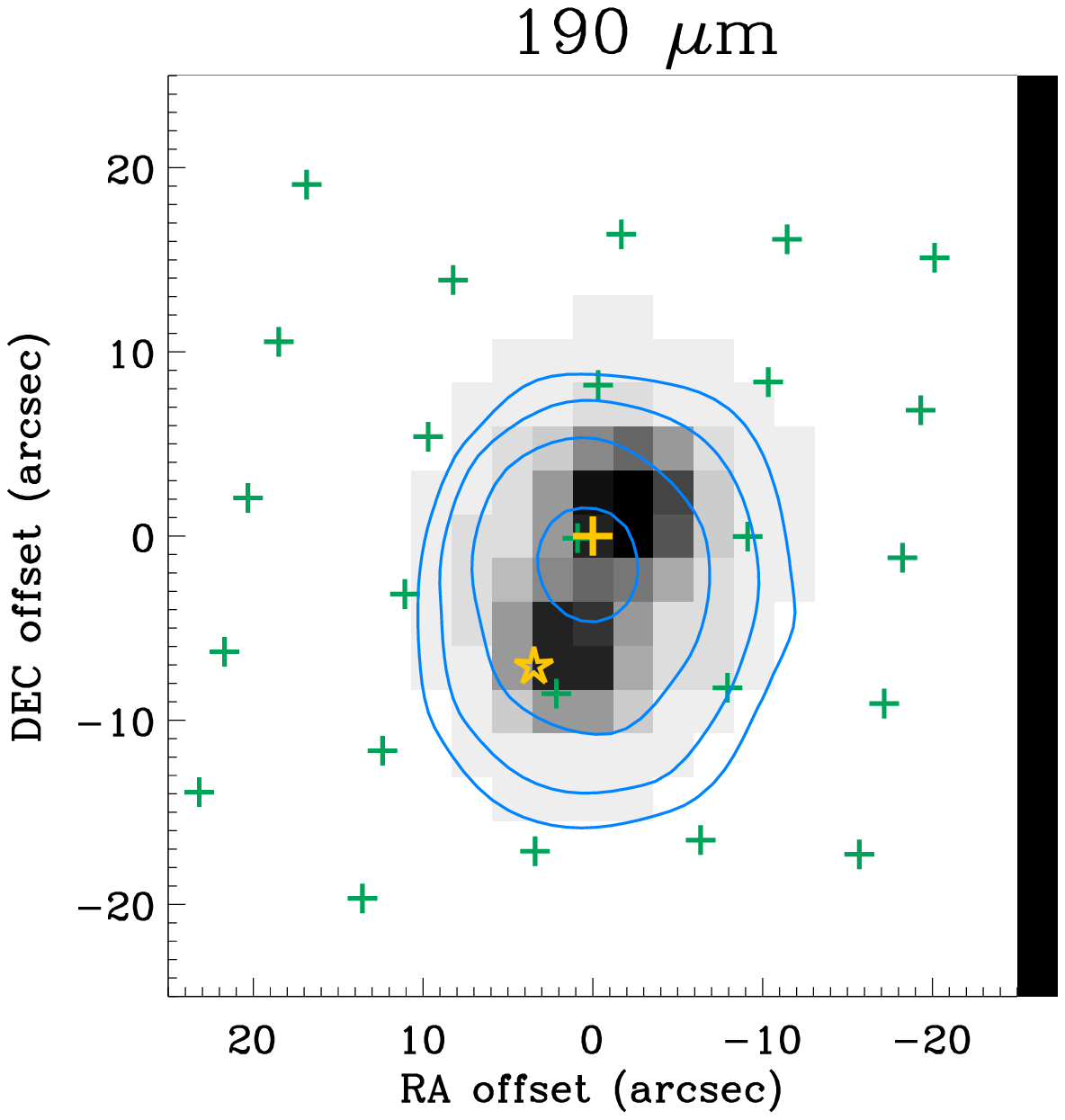}
\caption{The contour maps of continuum emission at 67 $\mu$m and 190 $\mu$m on
top of the MIPS 1 (grey image).
In each map, contour levels are 20, 30, 50, and 90 $\%$ of the peak flux.
The crosses indicate the locations of the 5 $\times$ 5 spaxels.
The sub-mm positions of MM(A) and MM(B) \citep{Jor06}
are marked as yellow cross and star, respectively.
}
\label{map_conti1}
\end{figure}

Figure~\ref{map_conti1} shows the
distribution of continuum emission at 67 $\mu$m and 190 $\mu$m,
 extracted from the PACS data cube, compared to the MIPS 1 image.
The continuum peaks close to MM(A) and slightly extends to MM(B),
which is probably due to the contribution of MM(B) in the dust continuum.
These two sources were decomposed into the continuum emission as described in
section 3.2.

\begin{figure}
\plottwo{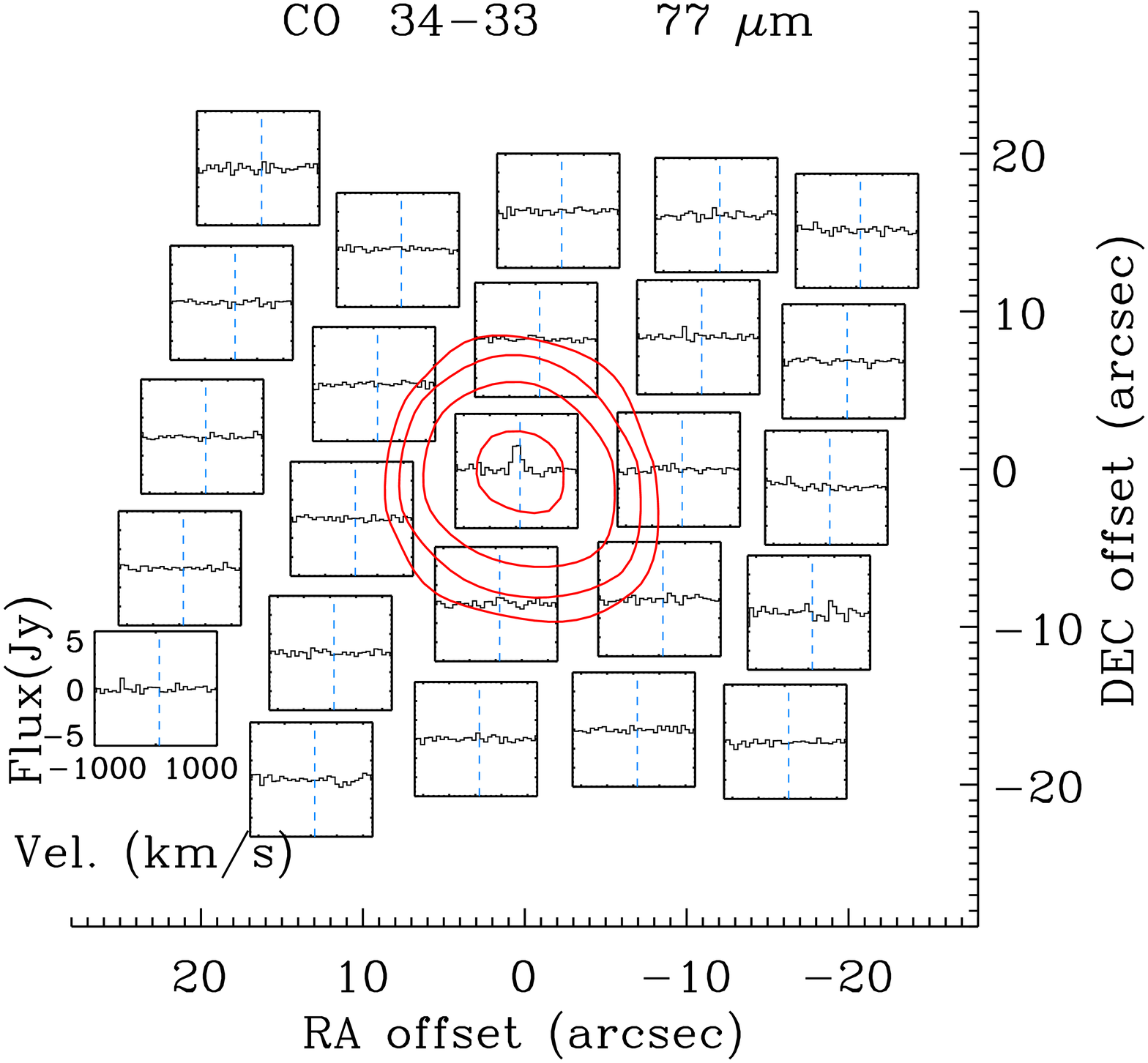}{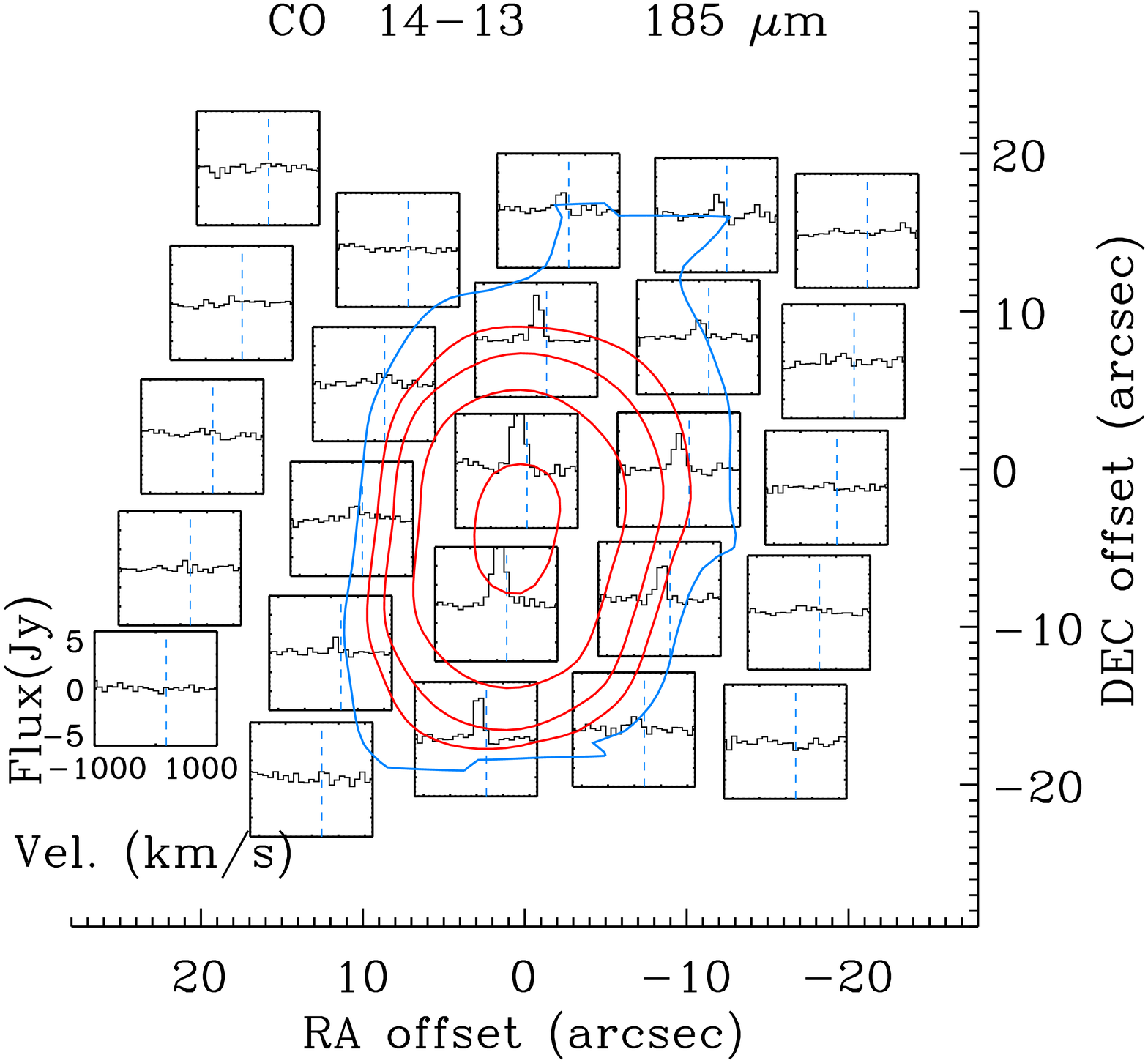}
\caption{Left: The contour map of the CO J $=$ 34 -- 33 line flux on top of its
spectrum map.
The upper level energy is 3279 K.
Right: The same map but for CO J $=$ 14 -- 13.
The upper level energy is 580 K.
The red contour levels are 20, 30, 50, and 90 $\%$ of the peak flux,
 but one more blue(right) contour is for 10 $\%$ of the peak flux to present
the extended emission better.
}
\label{map_co}
\end{figure}

\begin{figure}
\plottwo{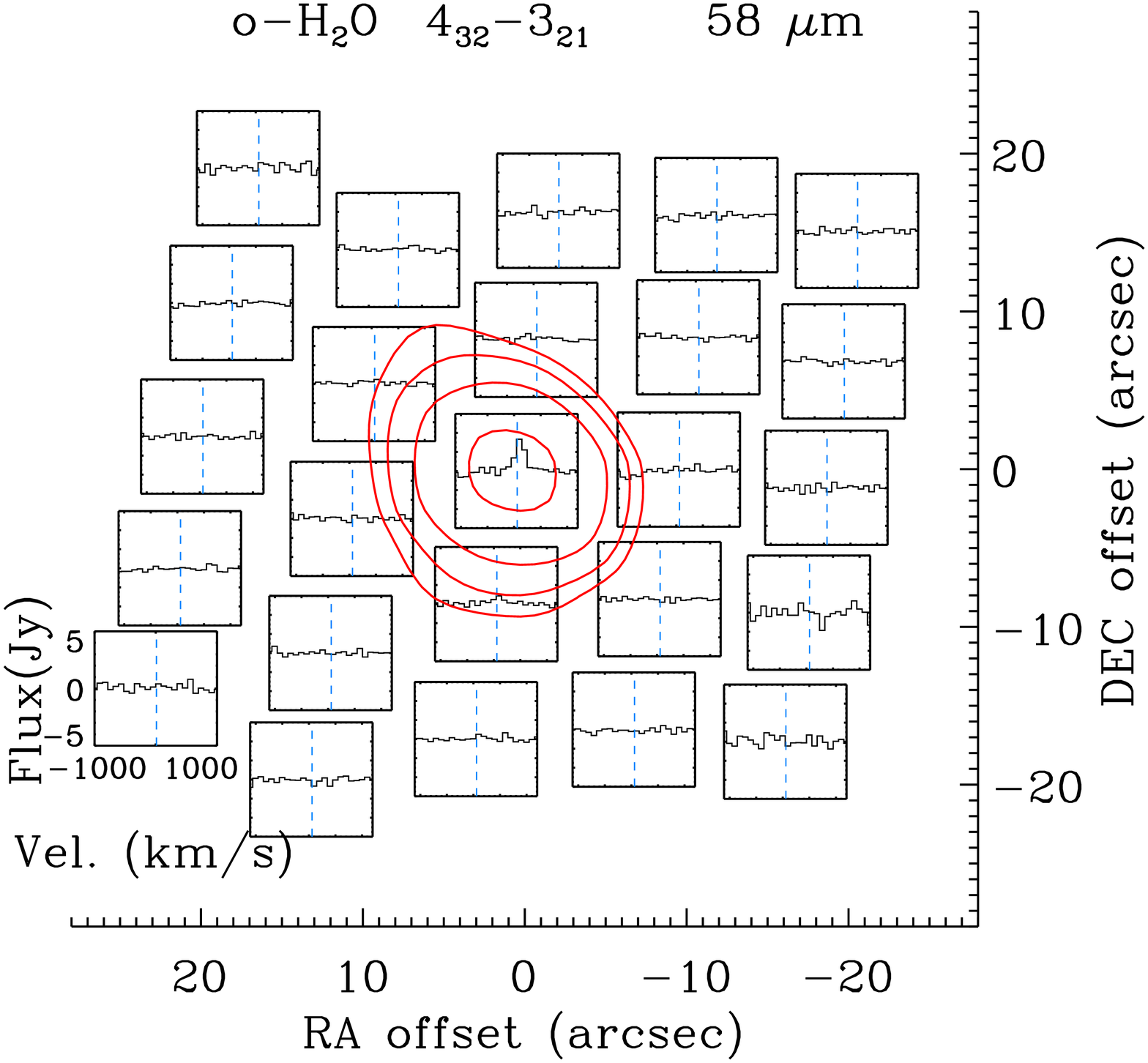}{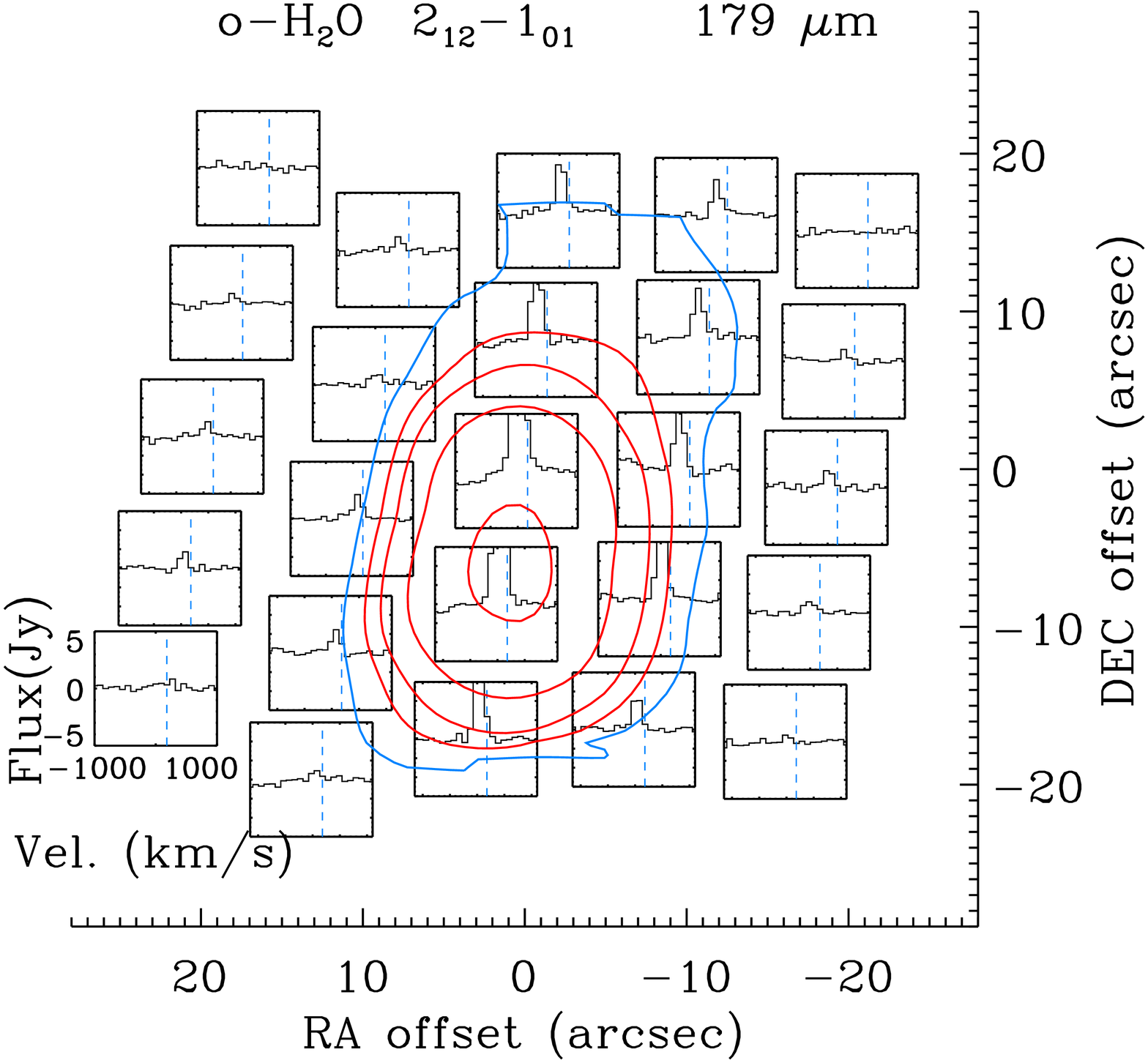}
\caption{Left: The same maps as Fig.~\ref{map_co} but for o-H$_2$O
J$_{K_{-1},K_1}$ =  4$_{32}$ -- 3$_{21}$.
The upper level energy is 550 K.
Right: The same map but for o-H$_2$O J$_{K_{-1},K_1}$ =  2$_{12}$ -- 1$_{01}$.
The upper level energy is 114 K.
}
\label{map_oh2o}
\end{figure}

\begin{figure}
\plottwo{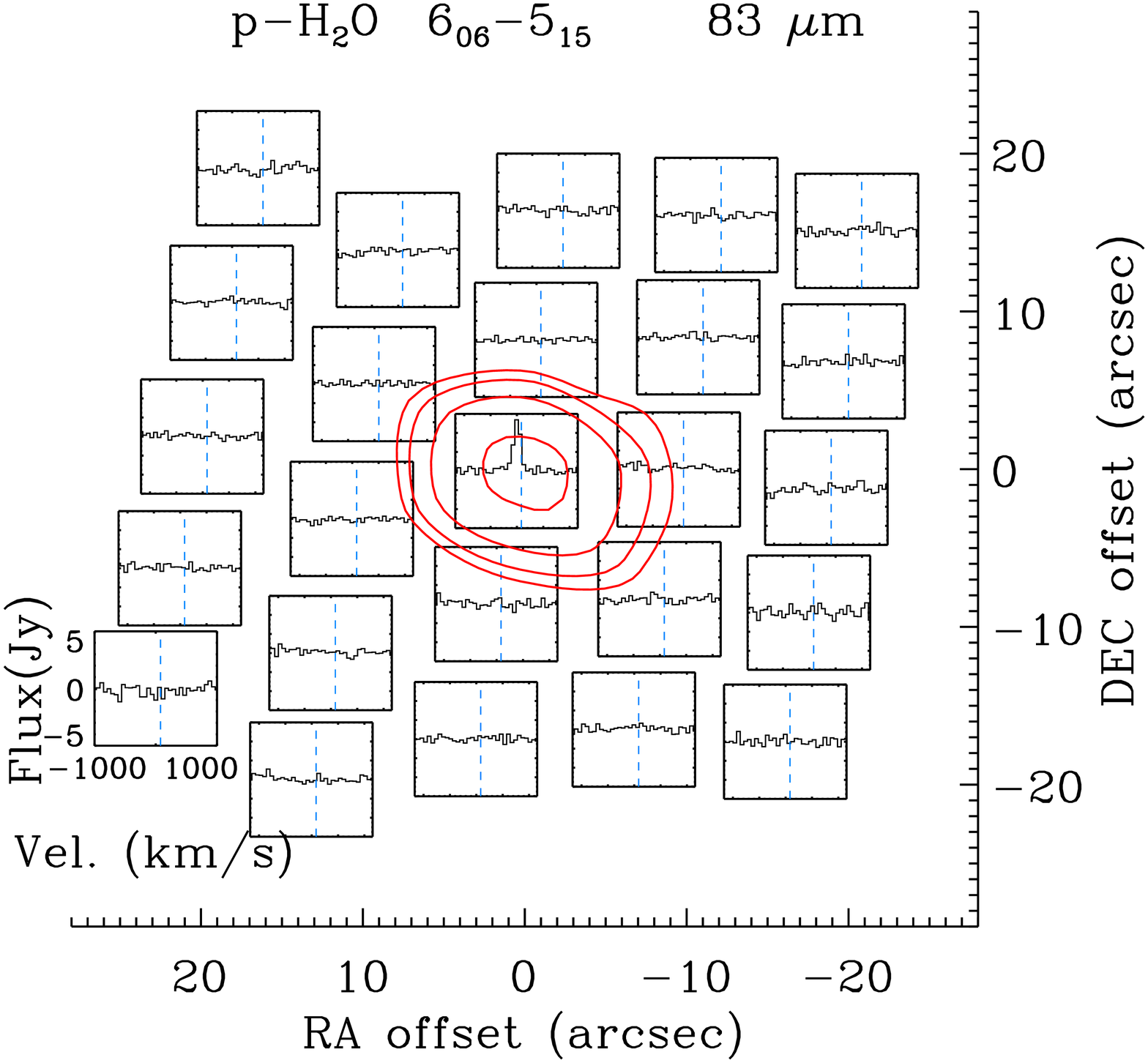}{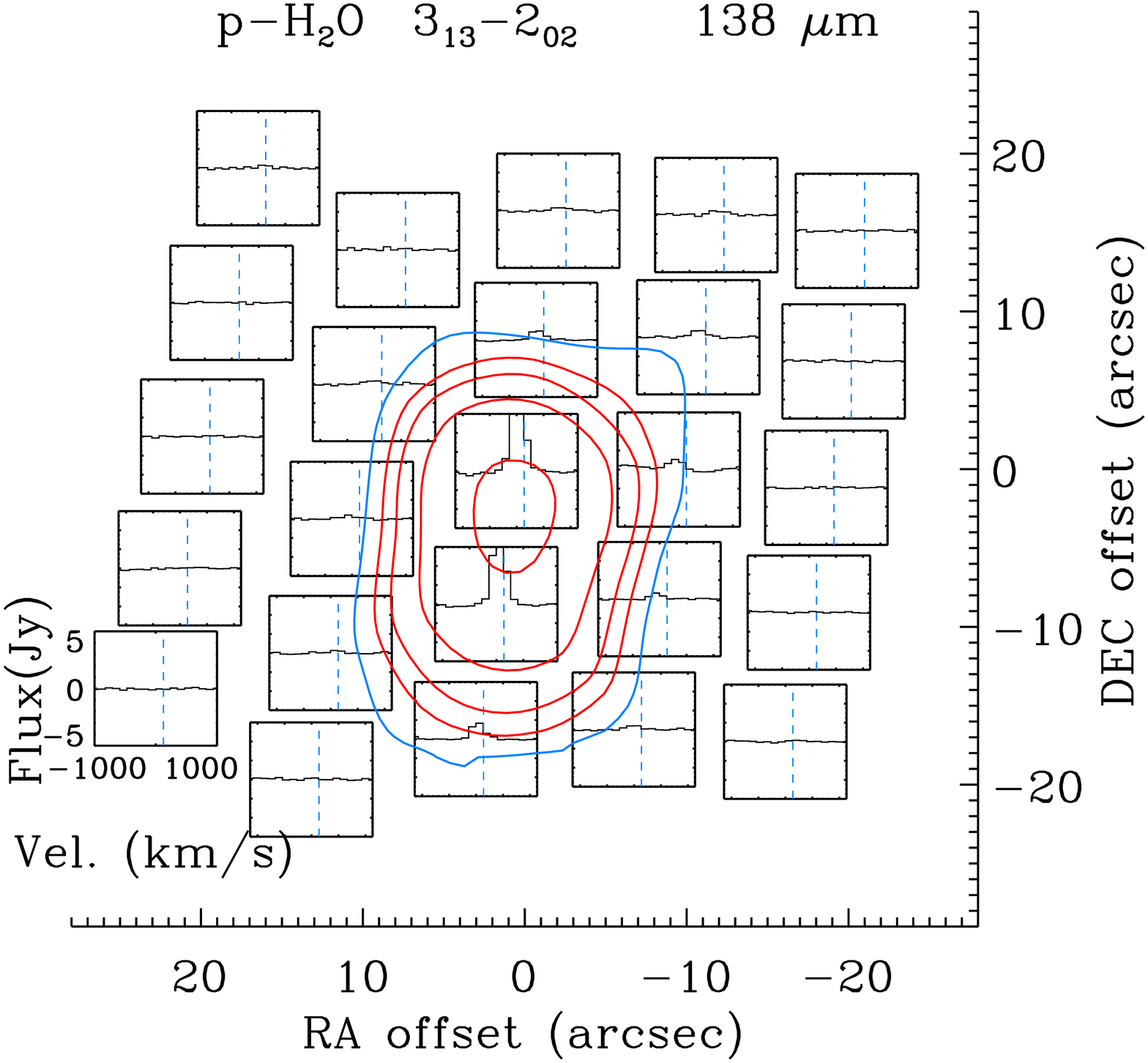}
\caption{Left: The same maps as Fig.~\ref{map_co} but for p-H$_2$O
J$_{K_{-1},K_1}$ =  6$_{06}$ -- 5$_{15}$.
The upper level energy is 643 K.
Right: The same map but for p-H$_2$O J$_{K_{-1},K_1}$ =  3$_{13}$ -- 2$_{02}$.
The upper level energy is 205 K.
}
\label{map_ph2o}
\end{figure}

\begin{figure}
\plottwo{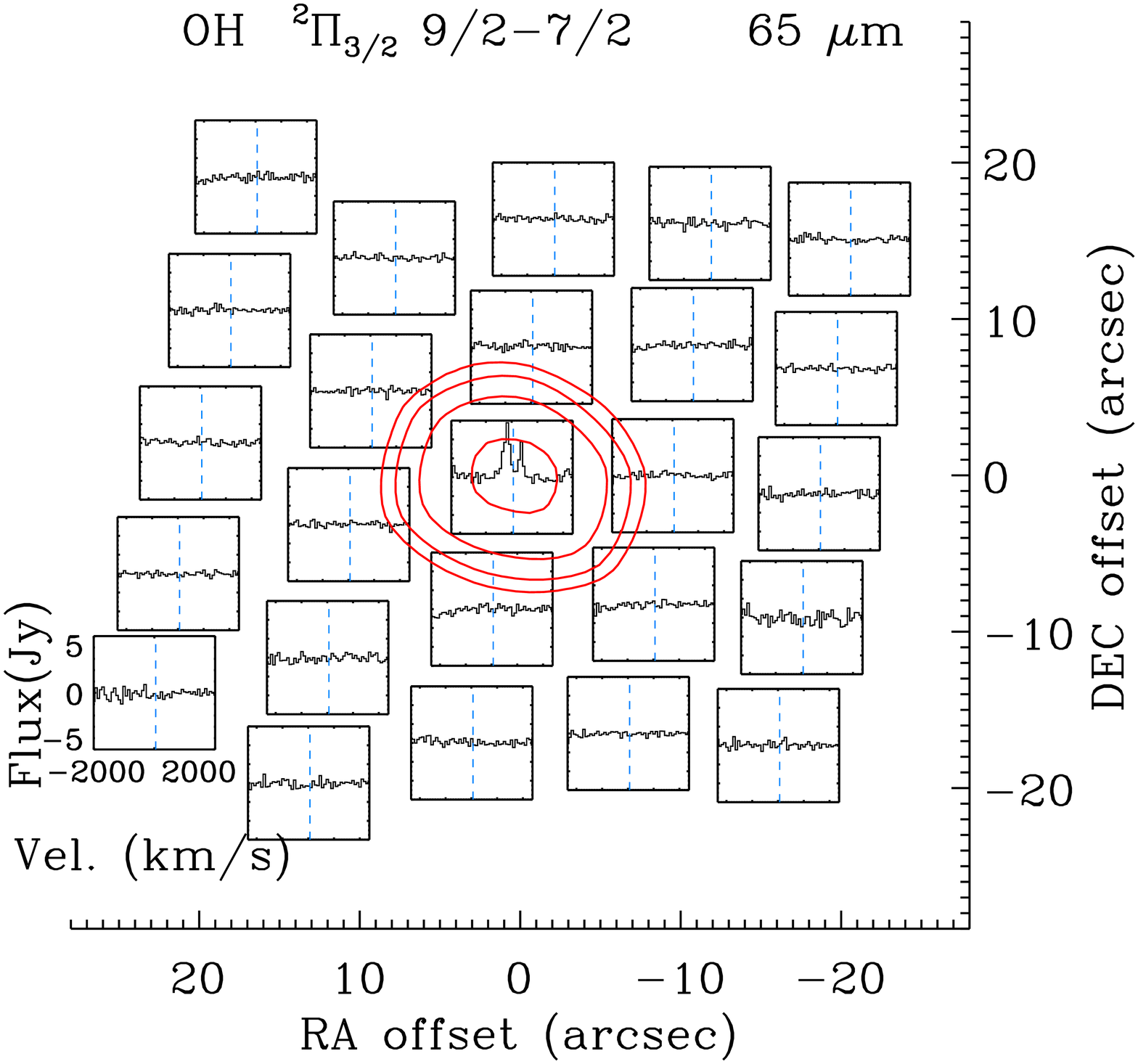}{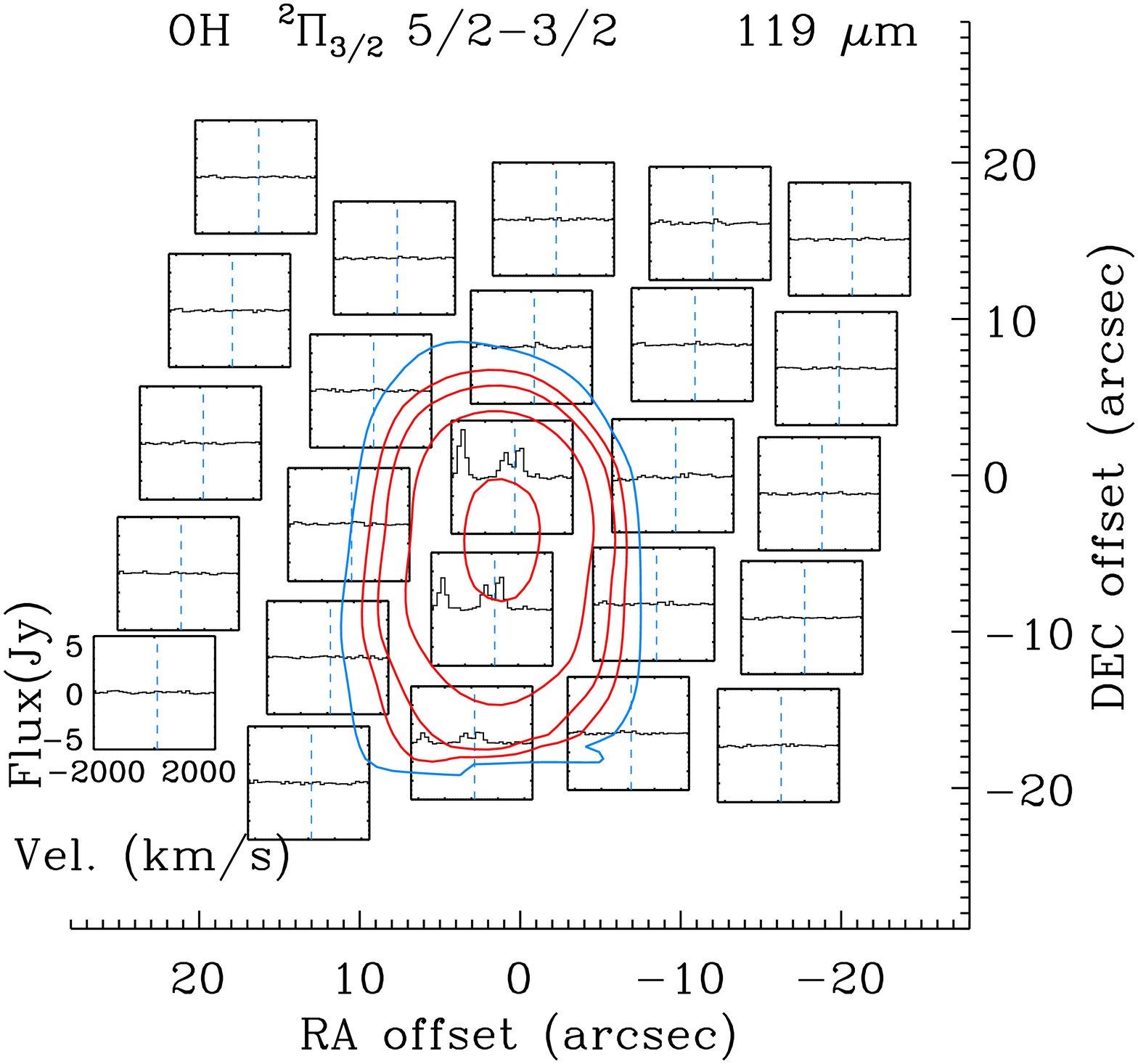}
\caption{Left: The same maps as Fig.~\ref{map_co} but for OH $^2\Pi_{3/2}$ J
$=$ 9/2 -- 7/2.
The upper level energy is 512 K for J $=$ 9/2-- -- 7/2+ and 511 K for J $=$
9/2+ -- 7/2--.
Right: The same map but for OH $^2\Pi_{3/2}$ J = 5/2 -- 3/2.
The upper level energies are 121 K for both J $=$ 5/2-- -- 3/2+ and J $=$ 5/2+
-- 3/2--.
}
\label{map_oh}
\end{figure}

\begin{figure}
\plotone{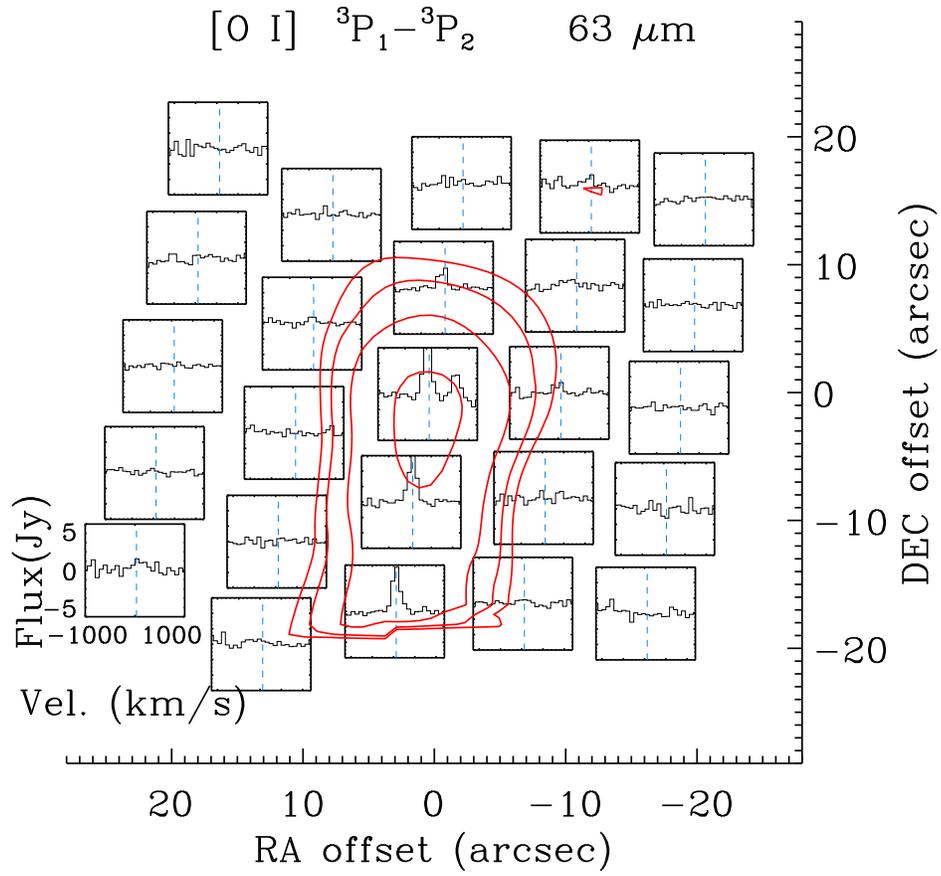}
\caption{The same maps as Fig.~\ref{map_co} but for \OI\ $^3P_1 - ^3P_2$.
The upper level energy is 228 K.
}
\label{map_oi}
\end{figure}

From Fig.~\ref{map_co} to Fig.~\ref{map_oi}, the maps of selected lines are
presented.
In each figure, a contour map is overlaid on the spectral map.
Two transitions are selected for each species to study the spatial variation in
the emission
lines depending on
energy level or wavelength; lines representing higher energy levels concentrate
on spaxel C while
lines of lower energy levels peak 4$\sim5\arcsec$ toward the south. 
In addition, the low energy lines tend to distribute more broadly compared to
lines from high energy levels.
CO (J$=$14--13, 186 $\mu$m), OH ($^2\Pi_{3/2}$ J = 5/2 -- 3/2, 119 $\mu$m),
o-H$_2$O (J$_{K_{-1},K_1}$ =  2$_{12}$--1$_{01}$, 179 $\mu$m),
p-H$_2$O (J$_{K_{-1},K_1}$ = 3$_{13}$ -- 2$_{02}$, 138 $\mu$m), and \OI\ 
($^3P_1 - ^3P_2$, 63 $\mu$m) are the most evident examples
of the spatially broadened distribution.
The broad distribution of line emission at longer wavelengths could be real
or the effect of the larger PSF at longer wavelengths.
Nevertheless, the distribution of \OI\ emission appears to be significantly different from 
the other species as the line flux
of the southernmost spaxel is as strong as that at the spaxel S, indicating a
relatively flat emission profile over several spaxels.

The line emission does not necessarily  correspond to continuum sources spatially.
However, the line emission (except for \OI\ lines) is the strongest at the spaxels C and S
as is the continuum emission. 
In addition, the peaks of the contour maps both in continuum
and line emission are shifted to the south of MM(A) similarly.
Therefore, the line emission might originate from two unresolved gas components 
located at spaxels C and S.
In order to test this idea, we decomposed the line fluxes with the same method as used
in the decomposition of the continuum (\S 3.2) by assuming that two unresolved gas components 
are located at the same positions of MM(A) and MM(B), which is not necessarily true.
The decoupled line fluxes are listed in Table~\ref{fluxtable}.
Interestingly, the sum of the decomposed molecular line fluxes is consistent with
the total fluxes over 25 spaxels within errors while the total \OI\ 63 $\mu$m flux over 25 spaxels
is greater than the sum of the decomposed fluxes by a factor of $\sim$ 2.
Therefore, for CO, OH, and H$_2$O, the extended emission beyond these two spaxels is mostly 
caused by the PSFs of the two unresolved gas components while \OI\ emission is indeed extended.
Fig.~\ref{map_oi} also shows the elongated emission along the outflow direction, which indicates that the extended \OI\ emission is connected to the outflow.

\begin{figure}
 \plotone{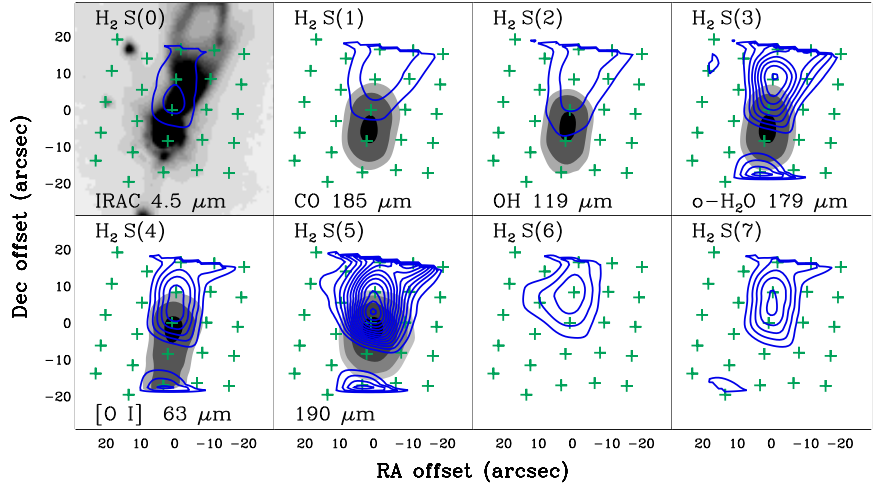}
\caption{The contour maps (blue) of the H$_2$ rotational transitions
superimposed on
 {\it spitzer} 4.5 $\mu$m image or PACS contour maps of CO, OH, o-H$_2$O, \OI, and
190 $\mu$m continuum emission (gray scale).
PACS contours have levels of 0.2, 0.3, 0.5, 0.9 times of the peak flux.
H$_2$ contours start at 10$^{-21}$ W cm$^{-2}$ and increase by 10$^{-21}$ W
cm$^{-2}$.
}
\label{map_h2}
\end{figure}

Fig.~\ref{map_h2} presents IRS line maps of the rotational transitions of H$_2$
compared with PACS emission lines and the IRAC 4.5 $\mu$m image.
The H$_2$ emission is not spatially coincident with the FIR emission.
 H$_2$ emission is detected primarily from the blue-shifted outflow to the
north,
 where $n$(H$_2$) is higher and $A\rm_V$ is lower \citep{Nis00, Dio09},
  while the extended \OI\ emission seems to originate from the position of the red-shifted outflow to
the south.
Therefore, this overall feature of the MIR and FIR emission suggests that the outflow source is 
located rather close to the surface of the associated molecular cloud.
As a result, the column density of shocked material in the blue component is not large enough to
block the MIR emission while the column density of shocked
material in the red component is large enough to obscure the MIR emission and to produce the
strong FIR emission line.

\section{Analysis}

In order to understand the physical conditions of the gas emitting the detected FIR lines, we utilize two
analysis methods: the rotational diagram and the non-LTE Large Velocity Gradient (LVG) code, RADEX. 
The rotational diagram has been conventionally adopted to provide rough idea of a temperature, 
which is called 
rotational temperature and used to explain the relative observed line intensities of a molecular species. 
However, the LVG analysis can provide more detailed physical conditions such as the kinetic temperature, 
which is not always the same as the rotation temperature, and the density of the gas.
 
\subsection{Rotational diagrams}

This simple excitation analysis assumes that the lines are optically thin.
If the populations can be fitted with a single line, the level populations 
can be characterized by a single temperature ($T_{\rm rot}$).
In the case of LTE, $T_{\rm rot}$ is the same as the kinetic temperature of the gas. A detailed 
description for this rotation diagram can be found in \citet{Gre13} as well as \citet{gol99}.
In order to produce rotational diagrams,
 we used the total flux over 25 spaxels as well as the decomposed fluxes for
the two unresolved gas components (accidentally) located at the positions of MM(A) and MM(B).
The two unresolved gas components are designated as (A) and (B) hereafter.
(Here, we note again that the two gas components, (A) and (B) are not necessarily associated with two 
continuum sources, MM(A) and MM(B).)
 
 The {\it ISO} beam and the HIFI beam were too big to resolve (A) and (B) at all.
 However, due to the much better resolution of PACS, we could decompose the fluxes of (A) and (B) to 
 produce the rotation diagrams for both gas components separately.
The errors of the total flux extracted from the whole 25 spaxels are listed in
Table~\ref{fluxtable}
 while those of the decomposed fluxes are assumed to be 20\% of the fluxes.
Since we do not know the actual emitting area of each molecule,
 we use the total number of molecules ($\mathcal N$) instead of column density.
The results of our rotational diagram analysis are summarized in
Table~\ref{rottable}.

\subsubsection{CO}

\begin{figure}
\plotone{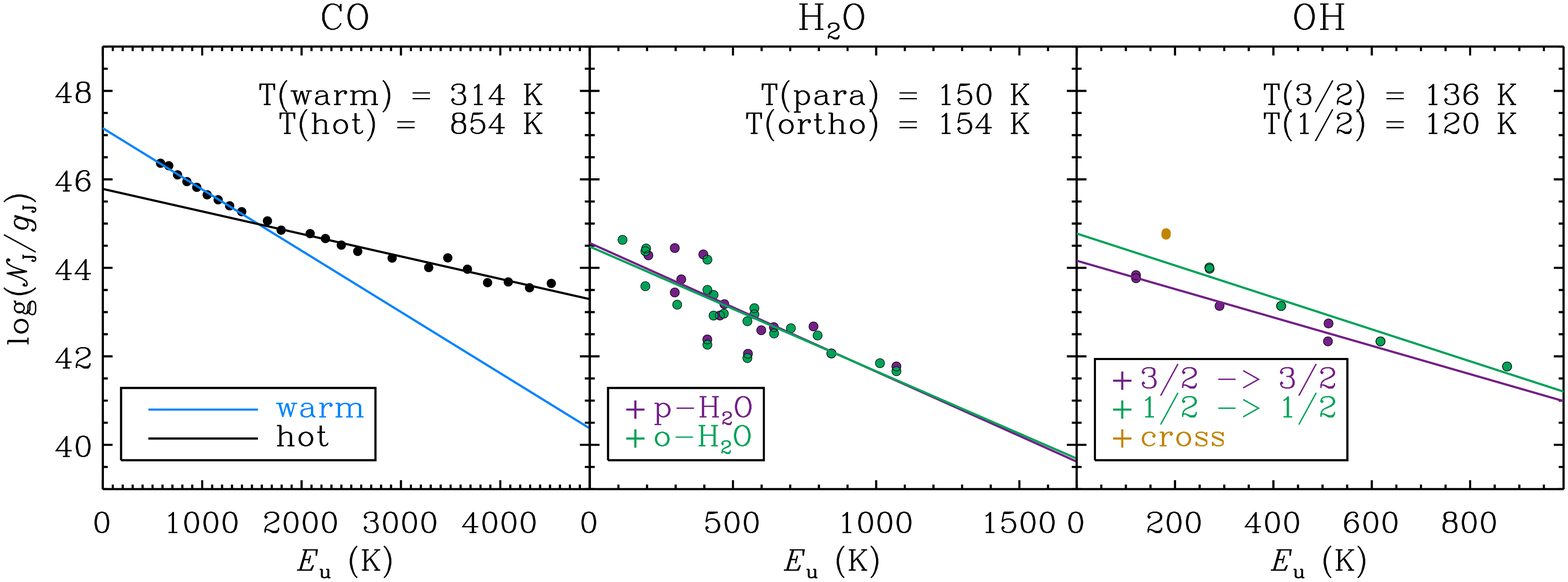}
\caption{CO,  H$_2$O, and OH rotation diagrams with the total fluxes, in units of total number of
detected molecules $\mathcal N$ divided by degeneracy g.
Left: the CO rotation diagram shows two distinct populations with a break
around 1500 K in energy.
Center: H$_2$O rotation diagram.
 p-H$_2$O and o-H$_2$O are fitted separately (purple and green are for p-H$_2$O
and o-H$_2$O, respectively.).
Right: In the OH rotation diagram, two ladders are fitted separately
(purple is for $^2\Pi_{3/2}$, and green is for $^2\Pi_{1/2}$).
}
\label{rots}
\end{figure}

The rotational CO ladder seems to contain a break at $\it E\rm_{u}$ $\sim$ 1500 K (Fig.~\ref{rots}).
Therefore, we fitted the rotational diagram with two components.
High-J CO lines ($\it E \rm_{u}$ $>$ 1500 K) are fitted by $\it T \rm_{rot}$ of 750--850 K 
and $\mathcal N$(CO) $\sim 6 \times$ 10$^{48}$ 
for the total fluxes and the fluxes of (A) while they are fitted by a lower $\it T \rm_{rot}$ of $\sim$
450 K with $\mathcal N$(CO) $\sim 3 \times$ 10$^{48}$ for the fluxes of (B).
(According to \citet{Gre13}, the break position between the two component does
not much affect the result.)
Low-J CO lines ($\it E \rm_{u}$ $<$ 1500 K) are fitted by a $\it T \rm_{rot}$ of $\sim$ 300 K
for the total fluxes measured over all 25 spaxels, as well as for the fluxes of (A).  In contrast, the fluxes
of (B) are fitted by a lower $\it T \rm_{rot}$ of $\sim$ 250 K. (The difference in the rotation temperature
between (A) and (B) is much greater than the fitting errors.)
Therefore, the gas at (A) seems hotter than the gas at (B).
The total numbers of CO molecules for this warm component are much greater than those for the hot 
component;  $\mathcal N$(CO) $\sim$ 2.8$\times$ 10$^{49}$,  1.6$\times$ 10$^{49}$, and 
1.4 $\times$ 10$^{49}$ for the total fluxes, (A), and (B), respectively (see Table~\ref{rottable}).

The warm (T$\sim$300 K) and hot (T$\sim$1000 K) CO gas components have been explained 
by a combination of PDRs and shocks \citep{Vis12}.
According to the scenario, the UV radiation from the central object can heat 
the outflow cavity walls up to 300$\sim$400 K, but shocks are necessary to heat the gas
to emit at high-J CO transitions of $\it E \rm_{u}$ $>$ 1500 K.
However, recent PACS surveys of YSOs \citep{Man13, Kar13} show 
that the UV heating along the cavity wall is a minor contributor to the 
excitation of the FIR CO fluxes. 
In addition, a more self-consistent 2-D PDR model developed by Lee et al. (in prep.) also shows 
that the UV heated outflow cavity wall cannot produce the rotational temperature of 300 K, which is universally fitted by the PACS low-J CO lines ($\it E \rm_{u}$ $<$ 1500 K), especially in Class 0 sources.
Therefore, shocks seem to be the predominant heating source in these embedded YSOs.

\subsubsection{H$_2$O}

Because of the different possible orientations of the spins of H atoms, 
H$_2$O has two kinds of states, ortho- and para-H$_2$O;
in equilibrium above a threshold ambient temperature and critical density, the
ratio of degeneracies of states is 3:1,
accounting for the state degeneracies.
Compared to CO and OH, H$_2$O shows increased scatter in its rotational
diagram  (Fig.~\ref{rots}),
probably because the water lines are usually sub-thermal due to their high
critical densities or have different opacities.

For the rotational diagram based on the full array fluxes, the o-H$_2$O
transitions are fitted by
$\it T \rm_{rot}$ = 144 $\pm$ 5 K and $\mathcal N$(H$_2$O) = (2.4 $\pm$ 0.3) $\times$ 10$^{46}$
while p-H$_2$O transitions are fitted by $\it T \rm_{rot}$ = 168 $\pm$ 6 K and
$\mathcal N$(H$_2$O) = (6.2 $\pm$ 0.9) $\times$ 10$^{45}$.
We also separated (A) and (B) in the rotational diagram, and the results are listed 
in Table~\ref{rottable}.
The rotational temperatures calculated from water lines are much lower than those calculated from
CO lines. If two species coexist in the same physical conditions, the derived rotational temperatures  
indicates the sub-thermal conditions of water lines. Therefore, in order to study the physical conditions
of the gas associated with water lines, we have to utilize a non-LTE calculation. 

In (A) and (B), $\mathcal N$(o-H$_2$O)/$\mathcal N$(p-H$_2$O) is greater than 2.
Considering the optical depth effect, it is not very different from 3.
Therefore, this may indicate that water formed at a temperature high enough for H$_2$O to
be in spin equilibrium, or the timescale is long enough to equilibrate the ortho-to-para
ratio of water after it evaporates from grain surfaces.
The rotation temperature for (A) is higher than (B), consistent
with our results from CO in the previous section.

\subsubsection{OH}

The spin-orbit interaction of OH results in two separate ladders
of rotational levels denoted as follows: 
$^{2S+1}\Lambda_J =$ $^2\Pi_{1/2}$ and $^2\Pi_{3/2}$.
Each rotational level is split by $\Lambda$ doubling and hyperfine structure
\citep{Off94}.
In the PACS spectra, typically the transitions between different
$\Lambda$ doublet levels are resolved while
the hyperfine structure is not resolved, although in some cases even the ladder
transitions are blended. 
When the total fluxes extracted from $5\times 5$ spaxels are used,
lines in the $^2\Pi_{3/2}$ ladder are fitted to $\it T \rm_{rot}$ = 115 $\pm$ 9 K with
$\mathcal N$(OH) = 7.6 $\pm$ 2.0 $\times$ 10$^{45}$,
 while the $^2\Pi_{1/2}$ transition lines are fitted to $\it T \rm_{rot}$ = 114 $\pm$ 7
K with $\mathcal N$(OH) = 4.0 $\pm$ 1.3 $\times$ 10$^{45}$  (Fig.~\ref{rots}). 
Two ladders have different y-intercepts in the rotational diagram, and
 $\mathcal N$($^2\Pi_{1/2}$)/$\mathcal N$($^2\Pi_{3/2}$) 
 is $\sim$0.6, suggesting that OH lines may not
 be optically thin, or the OH gas is not thermalized.
 (The ratio of partition functions of the two ladders is $\sim 0.15$ at $T=115$ K.)
Non-thermal effects might contribute to OH emission lines if the OH lines are optically thin.

According to \citet{Wam10}, radiative pumping via the cross ladder transitions
is more important
in the population of the $^2\Pi_{1/2}$ levels
while transitions in the $^2\Pi_{3/2}$ ladder are mostly excited by collisions.
To test the idea, we produced an LVG model including the FIR continuum
radiation from the central source as a non-thermal effect.
The result is presented in next section.
We also fitted line fluxes for (A) and (B), separately as done for CO and
H$_2$O. However, the number of OH lines for (B) is not large enough for a meaningful fitting.
The results are summarized in Table~\ref{rottable}.

\subsection{RADEX models of CO, H$_2$O, and OH emission}

According to the rotation diagrams, the gas component (A) located at the central spaxel
has higher rotational temperatures and total molecular numbers  compared to the gas component (B) 
located at the spaxel, S. 
However, the rotational diagrams cannot provide detailed information on kinetic temperatures and 
densities of the two gas components. In addition, we know that  those spatially distinct gas components 
have different kinematical components based on earlier studies 
\citep{Bac90, Dut97, Nis00, Kri11, Nis13} although our PACS spectrum does not resolve the complex kinematics.

Therefore, we used a non-LTE LVG model, RADEX \citep{Tak07}, to connect different physical conditions to the {\bf two spatially distinct} components. 
In the model, the level populations are determined by three physical
parameters:
the gas temperature $T \rm _K$, the H$_2$ density $n$(H$_2$),
 and the column density of a molecule divided by the line width, $\Delta v$.
With RADEX, we explore a wide range of physical conditions to interpret
observed line ratios.
For molecular data, the LAMDA database \citep{Sch05} was used \citep{Off94, Fau07, Yan10}.

First, we upgraded RADEX\footnote{downloaded from
http://home.strw.leidenuniv.nl/$\sim$moldata/radex.html} with the subroutine,
``newt'' \citep{Pre92, Yun09}, which is a globally used convergent Newton
method,
because the downloadable RADEX code sometimes does not easily converge to the
solution for H$_2$O and OH lines.
Second, for the OH and H$_2$O molecules, we tested the importance of
 radiative pumping by replacing a part
of the cosmic background radiation with
a blackbody radiation field emitted from the inner boundary of the envelope:
$BACK=W \times BB\rm(T)+\it (1-W)\times BB\rm{(2.7K)}$
where $BB\rm(T)$ is the Planck function of temperature $T$, and $W$ is the filling
factor of an inner source.
\citet{Kri11} adopted the inner boundary temperature of 250 K from \citet{Jor02} 
for their best-fit envelope model of L1448-MM with a power-law density structure
described with $n=n_{0}\times(\frac{r}{r_0})^{-1.5}$
($n_0$ = 1.3 $\times$ 10$^{9}$ cm$^{-3}$, $r_0$ = 20.7 AU).
(The temperature at the inner boundary is not well constrained 
because it has been derived based on the continuum at the wavelengths of 60 $\mu$m to a few millimeters, which is dominated by the outer cold envelope.)
We assumed that the radiation emitted from the inner boundary of the envelope
traveled through the outflow cavity to
reach a position in the outflow cavity wall without much attenuation.
If we assume a characteristic density at the position where the radiation
lands, the radius ($r$) from the center
to that position can be found from the envelope density profile.
Therefore, $W = (1-cos\theta) \times (1/2)$, where $\theta = sin
^{-1}(r_0/r)$,
assuming a spherical central radiation source.
Finally, the observed intensity is
$I=[BB(T\rm_{ex}) - \it BB \rm(2.7 K)] \times (1- \it exp(-\tau))$,
where $T\rm_{ex}$ and $\tau$ are the excitation temperature and the optical depth
for each transition, respectively;
we assumed that the infrared source is not located along the line-of-sight  when deriving line fluxes 
since the contribution of the flux affected by the IR source (through absorption) to the total flux within a spaxel is negligible.

\subsubsection{CO}

According to the rotation diagram, CO fluxes of both (A) and (B) seem to require multiple gas components, so we tested three combinations of physical components of gas 
(one gas component, two gas components, and the gas with a power-law temperature distribution) 
in order to check whether the non-LTE calculation also shows the same conclusion.
We fitted the observed PACS fluxes of (A) and (B), separately, assuming that all gas components 
contribute to the total flux equally. 
In this test, we adopted the line width (60 km s$^{-1}$) of the broad component, which was detected 
in the HIFI observations \citep{Kri11} and contributed dominantly to the HIFI fluxes. 
In order to find the best model, we compared flux ratios since we do not know
the actual size of the line emitting source.
We scaled total model flux to total observed flux and calculated {\bf reduced} $\chi^2$.
The best-fit models for three different combinations of gas conditions are summarized in Table~\ref{lvgmodel1}.

\begin{figure}
\plotone{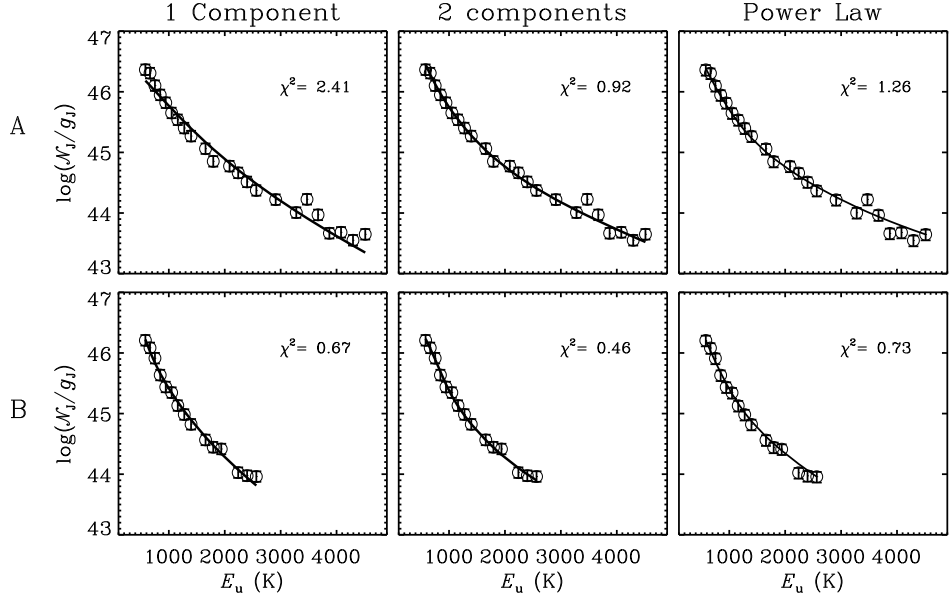}
\caption{The best-fit LVG models for CO lines of (A) (top) and (B) (bottom), respectively.
For (B), 84 $\mu$m line is excluded in the fitting.  
The first column shows one component model, the second column shows two components model, and the third column shows a model with power-law temperature distribution.
The reduced $\chi^2$ for each model is presented inside boxes.
}
\label{LVG1_CO}
\end{figure}

According to \citet{Neu12}, the PACS CO data can be fitted by one component
with a high temperature ($ T\rm_{K} \sim$ 3000 K) and low density ($n \sim$ 10$^{4-5}$
cm$^{-3}$).
However, the modeled rotation diagram of (A) with one component shows lower curvature 
compared to the observed one although the CO fluxes of (B) seems fitted well with the gas model with  $ T\rm_{K} =$ 4000 K and $n =$ 10$^4$ cm$^{-3}$ (See Fig.~\ref{LVG1_CO}).

The two gas components can fit better the CO fluxes of both (A) and (B); 
the two temperatures for (A) are both 5000 K while the two temperatures for (B) are 5000 K and 2000 K.
In the model with a power-law temperature distribution, (A) and (B) have a similar power index, b 
($\sim 3$), but the density for (A) is about 4 times higher compared to (B). 
This test shows that multi-components of gas can explain better the fluxes of both (A) and (B), and the high temperatures derived from the LVG models indicates shock origin.

\subsubsection{H$_2$O}

We also tested the three different combinations of gas components for the H$_2$O lines adopting the line width of 50 km s$^{-1}$, which is the velocity of the broad component detected by \citep{Kri11}, for all components.
As for CO lines, the two-component model fits better than the single component model for both
(A) and (B) although the single component model still fits the observed fluxes reasonably.
In the single component model, (A) requires a higher kinetic temperature (2000 K) than (B) (700 K) 
while the density for (A) is lower than that for (B). 
The best-fit model with a power-law temperature distribution for (A) has $b=0$, indicative of hot gas components are dominant emitter for the water lines.
However, the $\chi^2$ of this model is much worse than one or two components model because this
model assumed that all lines are optically thin.
The parameters for these best-fit models are summarized in Table~\ref{lvgmodel1}.

\subsubsection{OH}
For OH, we fitted fluxes only of (A)  since the majority of emission is from spaxel C, except for the
 119 $\mu$m doublet.
In addition, the collision rates for OH are available only up to $T =300$ K.
As a result, we modeled OH fluxes with a single physical component.
 
We explored models within the following parameter space:
50 $< T < 250$ K, $10^3 < \it n(\rm H_2) < 1.3\times10^{9}$ cm$^{-3}$,
and 5$\times10^{11}< N(\rm OH) < 5\times10^{17}$ cm$^{-2}$.
We assumed a line width of 50 km s$^{-1}$, which is appropriate for the broad component 
of H$_2$O gas \citep{Kri11}.  
Since it has been suggested that the FIR radiation field can play 
an important role in the excitation of 
OH, we included it in our model, as described in the first part of this section.
Although we considered only attenuated emission from the central source as the FIR radiation source 
without considering the FIR radiation from the surrounding material in situ,
the model including the effect of the FIR radiation in the OH excitation can
fit the observed fluxes better.

\begin{figure}
\plottwo{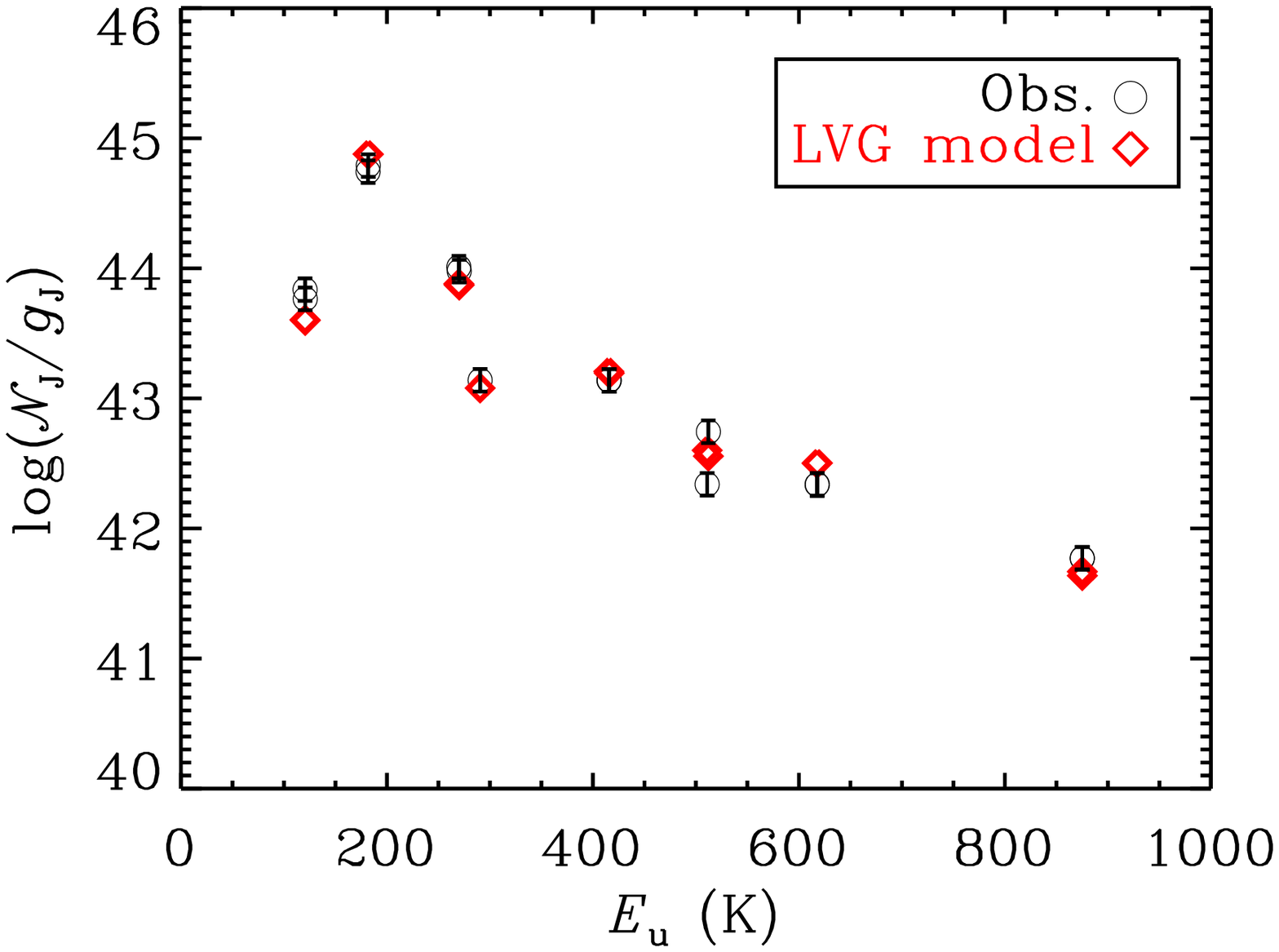}{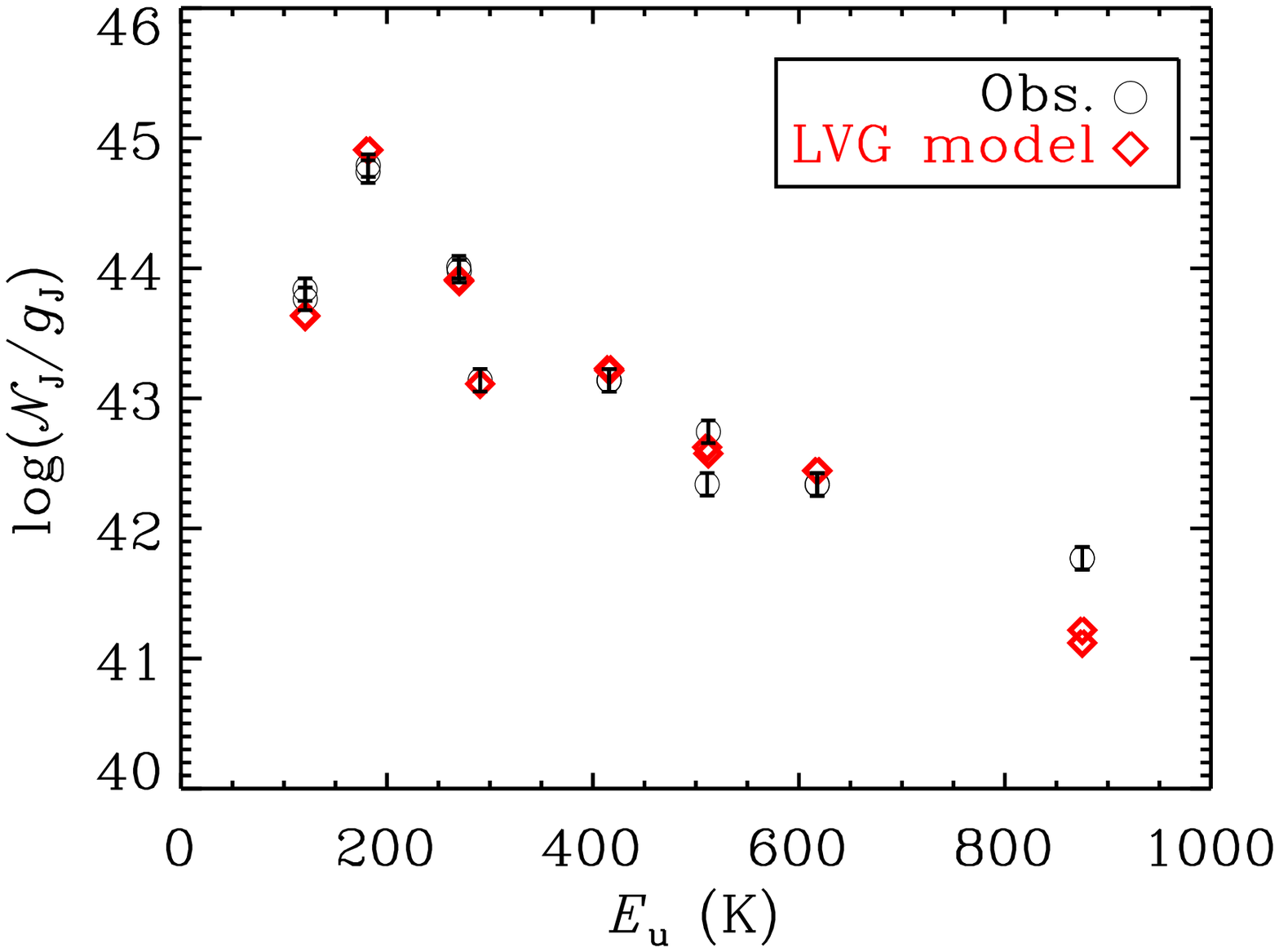}
\caption{
The best models of OH in (A).
The circles represent observed fluxes while the red diamonds display model fluxes.
Left: the best-fit LVG model with the IR-pumping effect included ($\chi^2\sim4.7$).
The observed fluxes can be well fitted when the central IR radiation is
considered in the OH level populations.
Right: the same model as the best-fit model but excluding the IR radiation
effect ($\chi^2\sim6.9$).
The line flux at the observed highest energy level is underproduced in the
model without IR-pumping.
}
\label{radexoh}
\end{figure}

Fig.~\ref{radexoh} (left) presents our best-fit OH model, where the FIR
radiation plays a role in the excitation of OH.
In the model, the temperature is 125 K, the H$_2$ density is 2$\times$10$^8$
cm$^{-3}$,
 and the OH column density is 5$\times$10$^{17}$ cm$^{-2}$.
 As described in the very first part of this section, this density would be reached 
 at a radius of 72 AU in the adopted 1-D density profile.
Therefore, the 250 K blackbody radiation field is diluted to the position of the envelope.
To examine the effect of FIR radiation on OH fluxes, we compared the same model
without the FIR radiation effect in Fig.~\ref{radexoh} (right).
At lower energy levels, the radiation effect is not prominent, but it makes a
significant difference at the highest energy level transition; the flux at the highest energy level
($ E_{\rm u} =$ 875 K) in the model with IR-pumping is greater by a factor of 5 compared to
the flux derived from the model without IR-pumping.

\begin{figure}
\plottwo{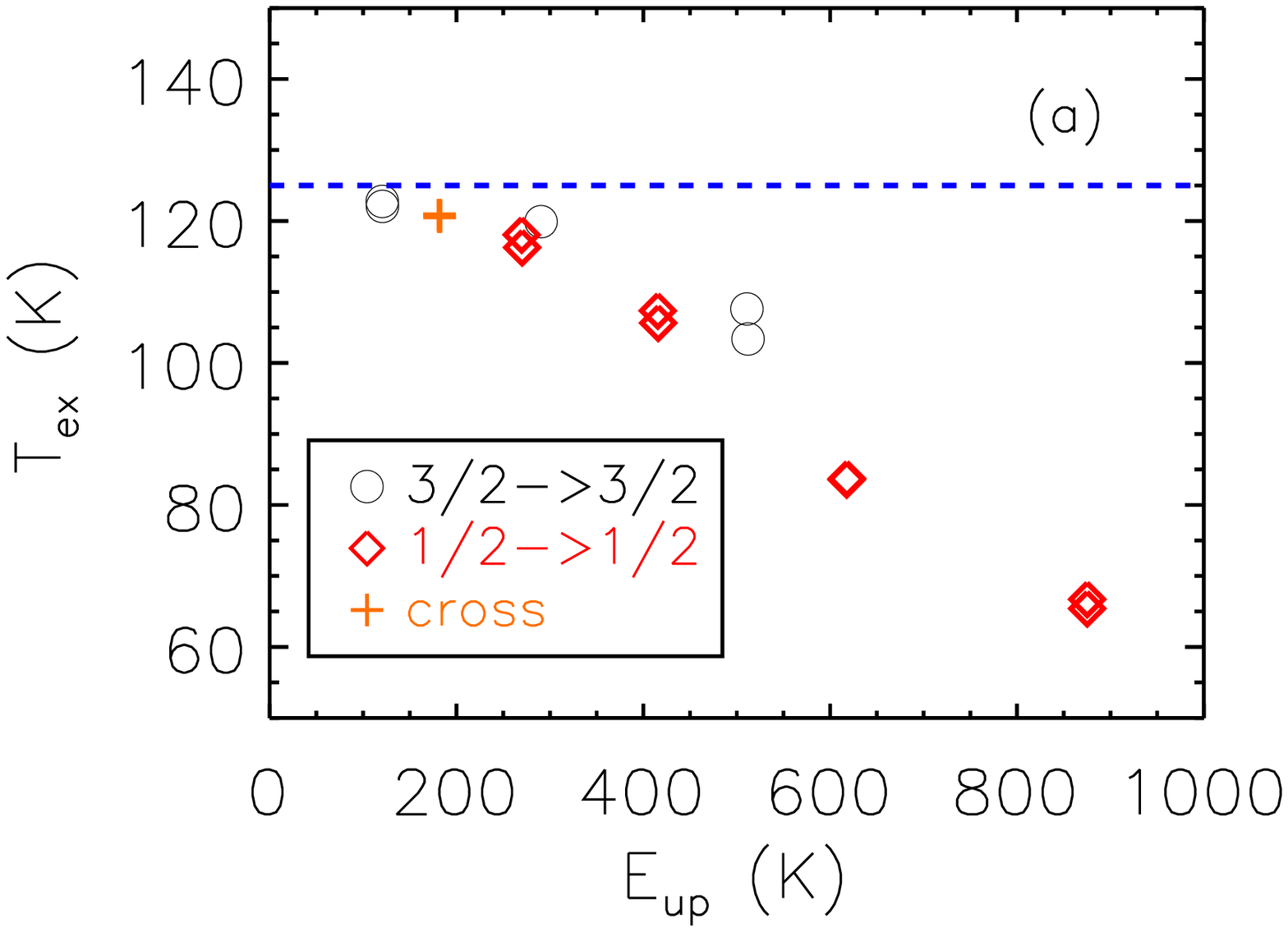}{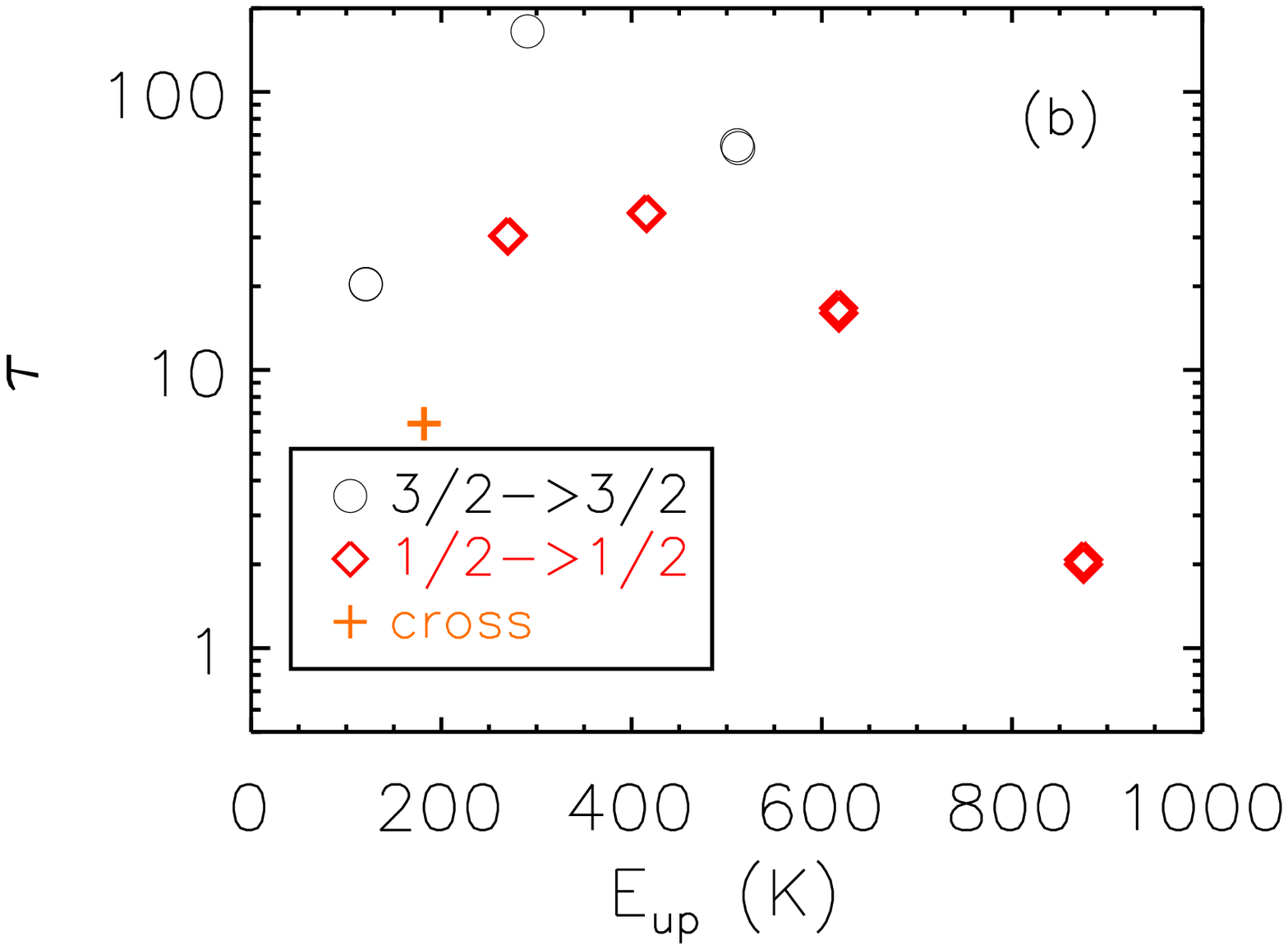}
\plotone{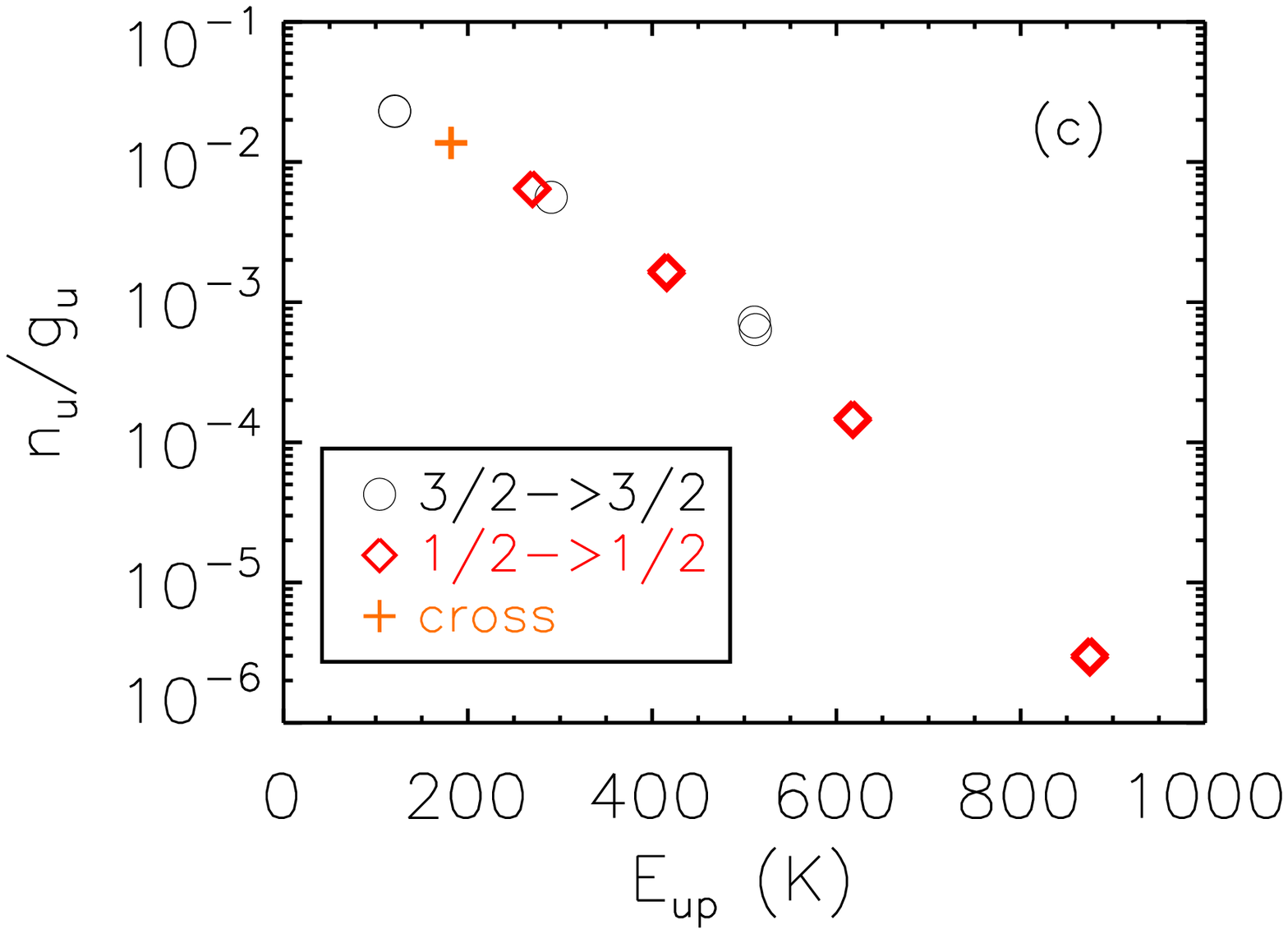}
\caption{
The excitation temperature (a), optical depth (b), and level population (c) of each line 
in the best-fit OH model with IR-pumping for (A).
The blue dashed line in (a) indicates the kinetic temperature of this model.
In (c), the upper level population is divided by the statistical weight of the upper level.
}
\label{ohtex}
\end{figure}

Although the IR effect is important for the high energy level transition, the overall fluxes are not affected much. 
The difference between y-intercepts of $^2\Pi_{3/2}$ and $^2\Pi_{1/2}$ ladders, therefore, in the rotational diagram should have other causes.
Fig.~\ref{ohtex} shows the excitation temperature and optical depth of each transition in the best-fit
model. The kinetic temperature of the model is marked as a red dotted line in the left box.
All transitions are sub-thermal, and $^2\Pi_{3/2}$ lines are extremely optically thick compared to $^2\Pi_{1/2}$ transitions. 
Therefore, the higher optical depth of $^2\Pi_{3/2}$ compared to $^2\Pi_{1/2}$ results in the separation (i.e., different y-intercepts) of two ladders in the rotational diagram, where a constant temperature ($T_{\rm rot}$) and the optically thin case are assumed.
Fig.~\ref{ohtex}(c) shows the actual level populations of the best-fit model, which results in equal populations for the two spin states.
Therefore, the separation of two ladders in the rotational diagram is caused both by optical depth effect and IR-pumping.

\subsection{Comparison with Shock Models}

Based on the {\it ISO} observations, \citet{Nis00} concluded that the CO, H$_2$O, H$_2$, and \OI\ 
IR emission in L1448-MM is caused by a non-dissociative shock with a low velocity, and the kinetic 
temperature of the shocked gas is about 1200 K.
The CO and H$_2$O line profiles are also very broad (FWHM$\sim50$ km s$^{-1}$), strongly
suggesting that the associated gas is related to shocks \citep{Kri11, Nis13}.
In addition, interferometric observations and analysis of the EHV features in L1448-MM show 
that they are likely jet-shock features \citep{Hir10}.  
Therefore, the derived excitation conditions and resolved kinematical components infer that shocks play 
an important role in L1448-MM.

The shocks produced by the interaction between the outflow and the envelope
heat the gas,
resulting in emission.
Since the shock could dissociate the gas in the dense envelope, which consists
mainly of molecular gas,
 the shock could change relative abundances among the species H$_2$, H, O, CO,
and H$_2$O.
Therefore, the relative intensities of these emission lines are indicative of
the type and speed
of the shock waves and of the physical conditions in the gas.
We compared our line fluxes to the calculation by \citet{Flo10} for both C- and J-shocks.
In the comparisons, we consider {\it a single physical component} for
all the  emission.
The calculations by \citet{Flo10} cover shock velocities from  10 to 40 km s$^{-1}$
 (the shock with 40 km  s$^{-1}$ is only for the C-shock) and
hydrogen densities $n(\rm H)$ of 2 $\times$ 10$^4$ $\sim$ 2 $\times$ 10$^5$
cm$^{-3}$.
The magnetic field strength in the pre-shock gas is given as $b[n(\rm H)
(cm^{-3})]^{1/2} \mu$G,
 with $b$ = 1 in the C-shock and $b$ = 0.1 in the J-shock.

\begin{figure}
\plottwo{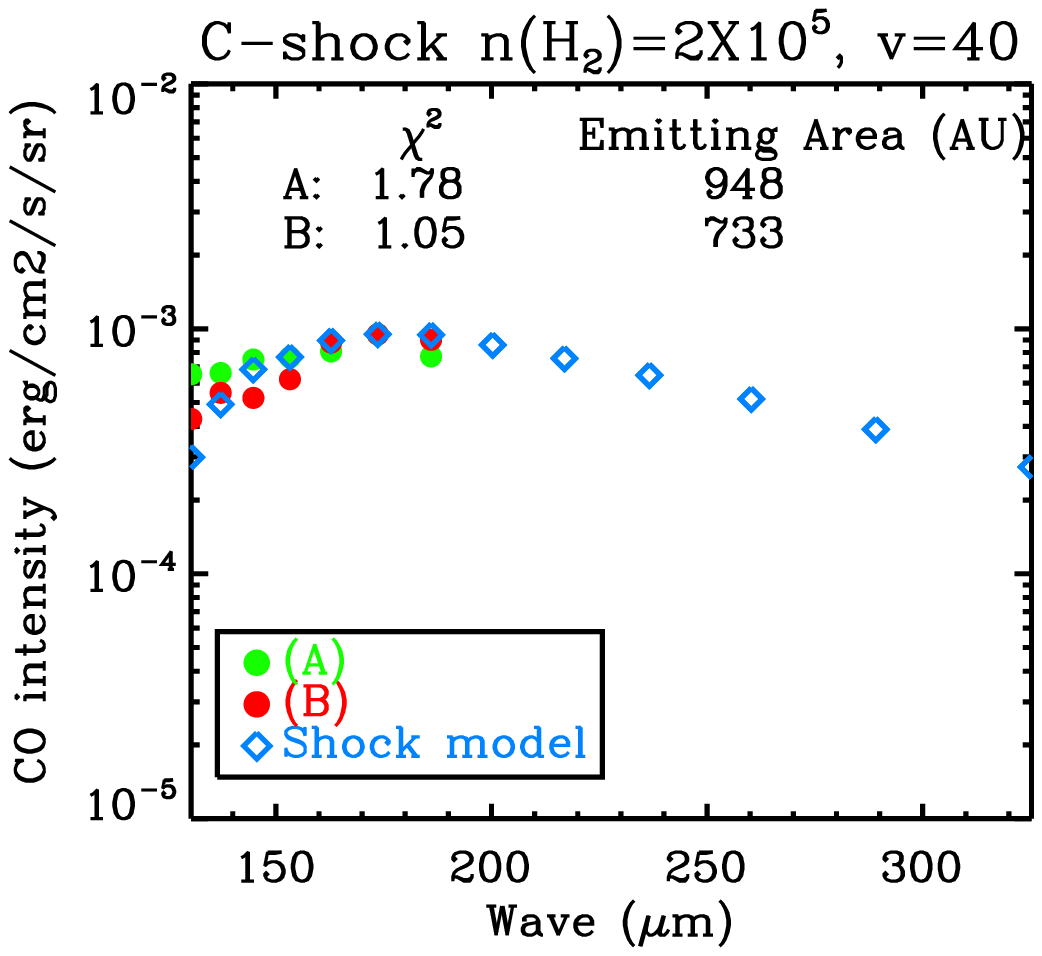}{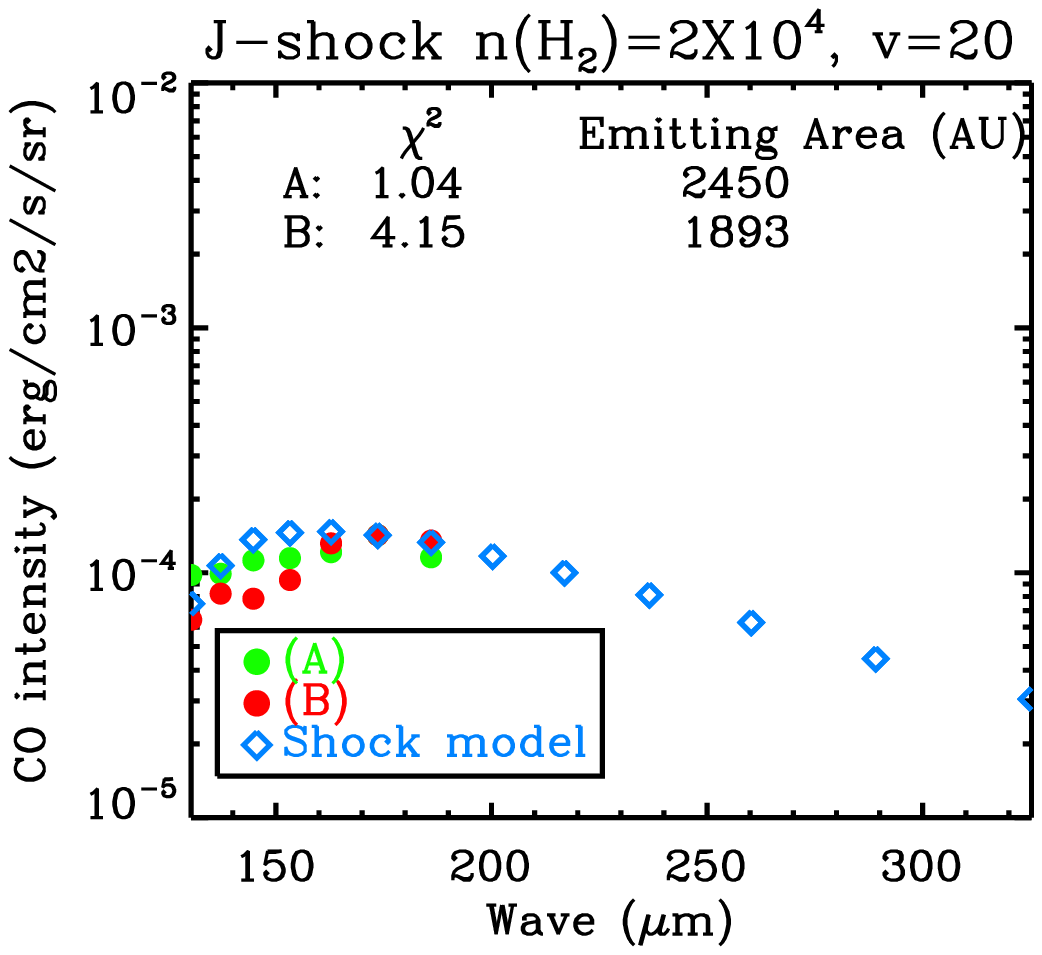}
\caption{
Comparisons of observed CO line fluxes with shock model fluxes 
(left for the best-matched C-shock model and right for the best-matched J-shock model.) T
he model parameters of
density and velocity are presented at the top of boxes. $\chi^2$ and the emitting areas to fit the observed
fluxes are presented inside boxes.
}
\label{co_shock}
\end{figure}

The relative CO line fluxes of (A) are well matched by the J-shock model with $n{\rm (H_2)}$ = 2$\times10^4$ cm$^{-3}$ and $v$ = 20 km s$^{-1}$.  The C-shock model with $n{\rm(H_2)}$ = 2$\times10^5$ cm$^{-3}$ and $v$ = 40 km s$^{-1}$ also reproduces the observed line flux ratios reasonably well.  In the case of (B), the relative line fluxes are well fitted by the C-shock model with $n{\rm (H_2)}$ = 2$\times10^5$ and $v$ = 40 km s$^{-1}$ (Fig.~\ref{co_shock}).  The diameter of emitting areas derived from these models are about 1000$\sim$2000 AU.  

\begin{figure}
\plottwo{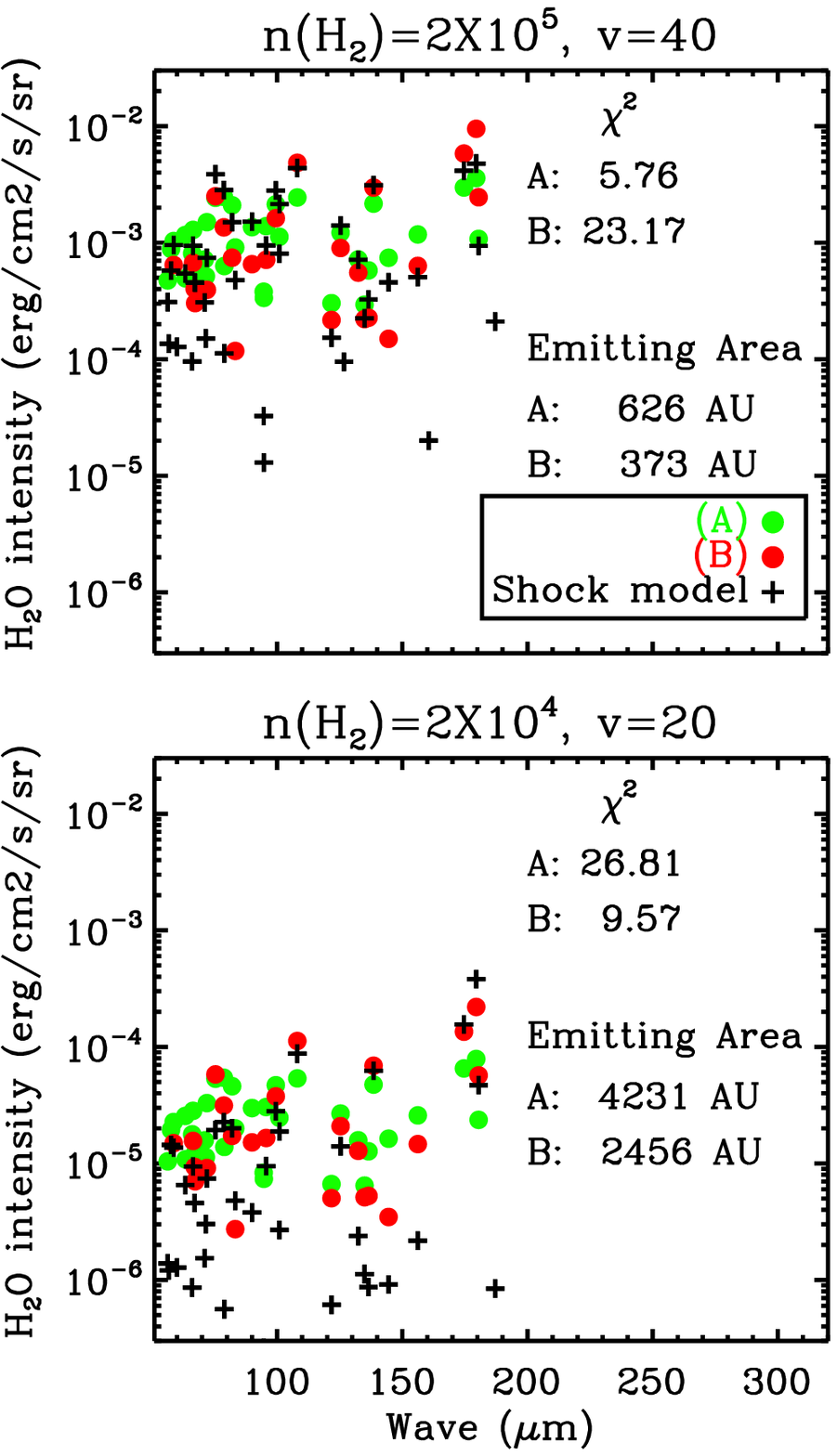}{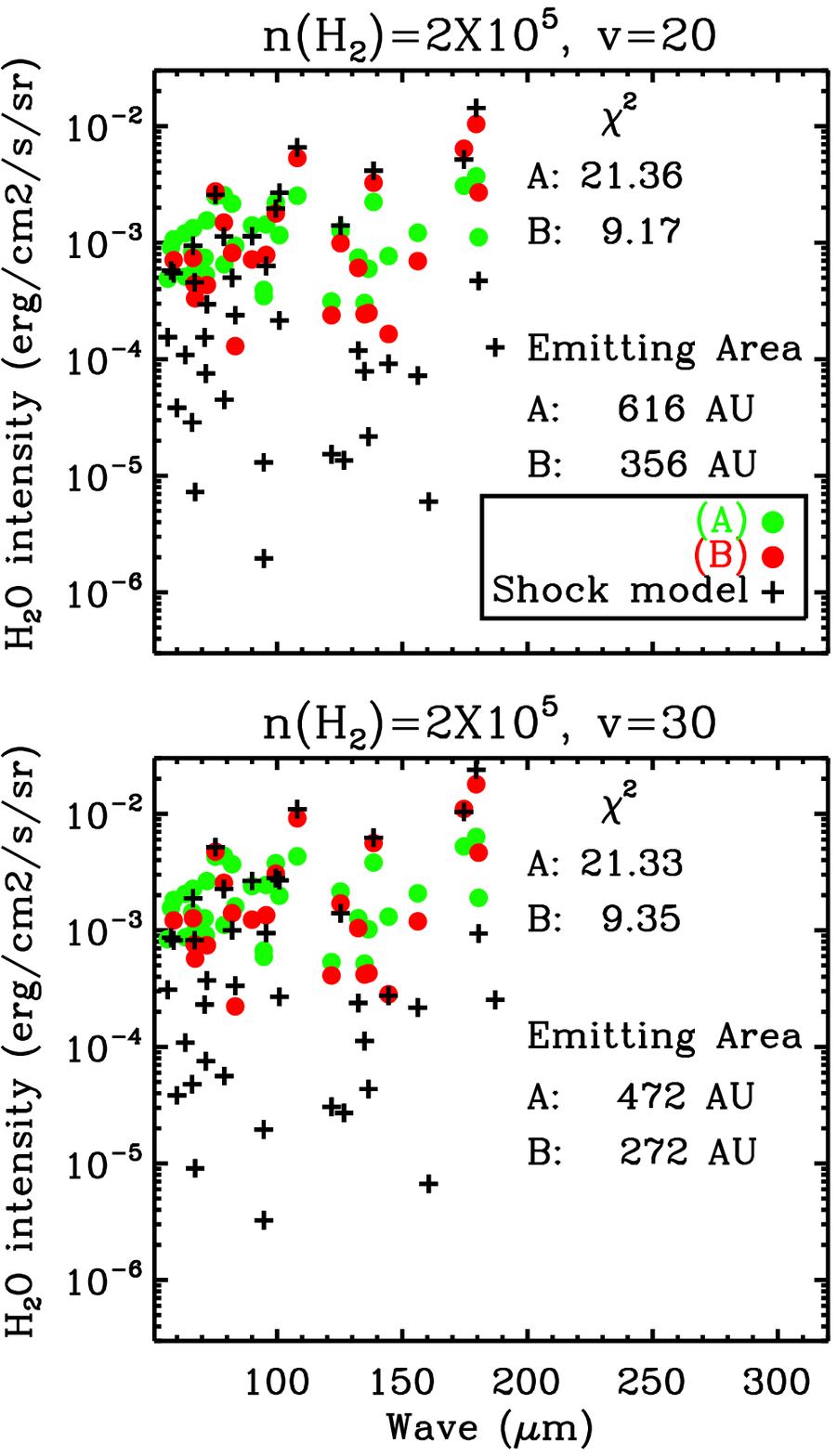}
\caption{
The same plot as Fig.~\ref{co_shock} but for H$_2$O 
(left for the best-matched C-shock models and right for the best-matched J-shock models.)}
\label{h2o_shock}
\end{figure}

The relative H$_2$O line fluxes of (A) are well matched by the C-shock model with $n{\rm(H_2)}$ = 2$\times10^5$ cm$^{-3}$ and $v$ = 40 km s$^{-1}$.  For (B), the C-shock model with $n{\rm(H_2)}$ = 2$\times 10^4$ cm$^{-3}$ and $v$ = 20 km s$^{-1}$ and the J-shock model with $n{\rm(H_2)}$ = 2$\times10^5$ cm$^{-3}$ in $v$= 20 km s$^{-1}$ and 30 km s$^{-1}$ fit well the observed flux ratios (Fig.~\ref{h2o_shock}).  
When we consider that H$_2$O line fluxes are highly scattered, (B) can be also explained by the shock model for (A).
In that case, the relative fluxes of CO and H$_2$O at both (A) and (B) are reproduced by the C-shock model with $n{\rm (H_2)}$ = 2$\times10^5$ and 40 km s$^{-1}$, 
supporting the idea that H$_2$O and CO 
are excited at the same physical conditions as suggested in \citet{Kar13}.

\begin{figure}
\plottwo{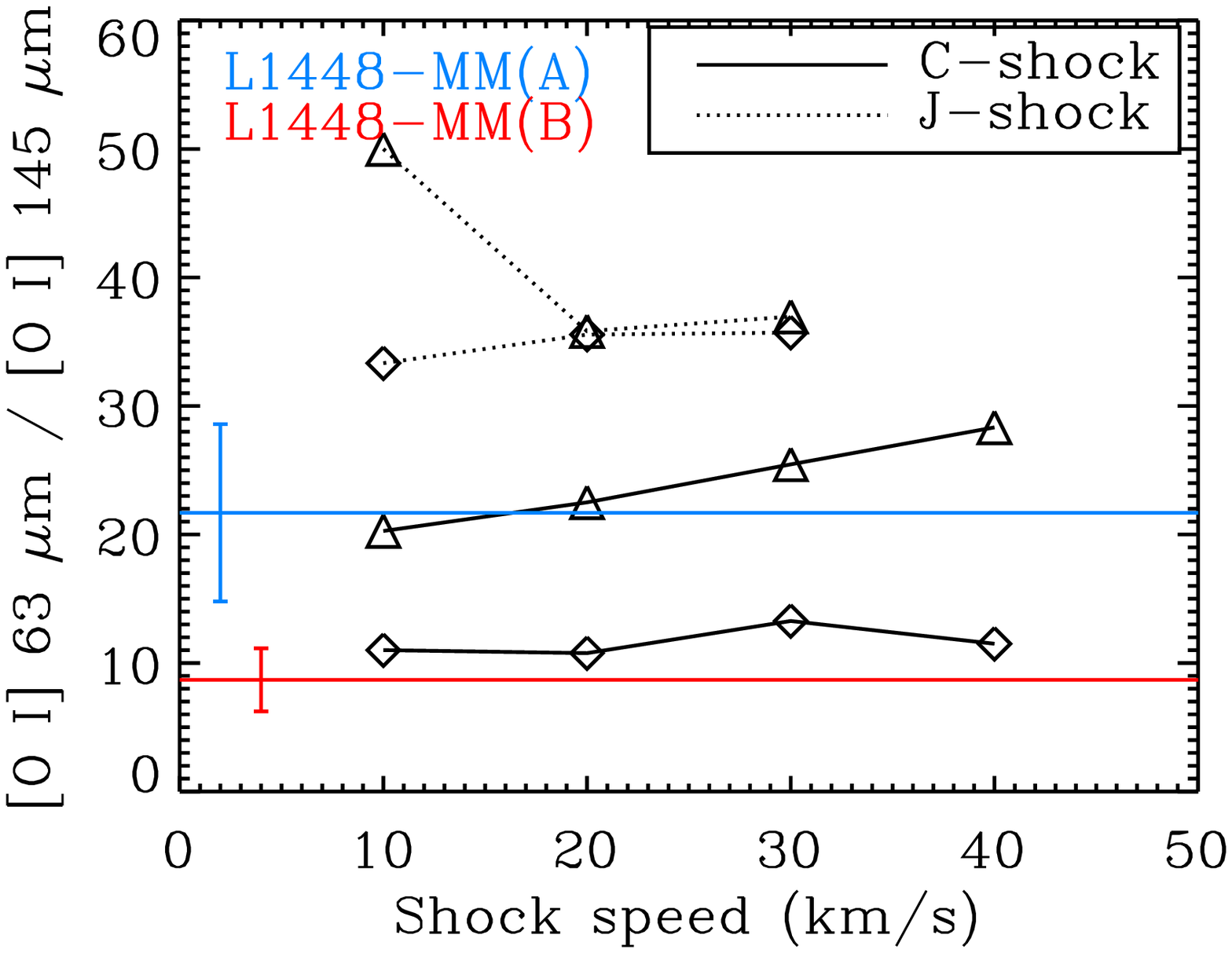}{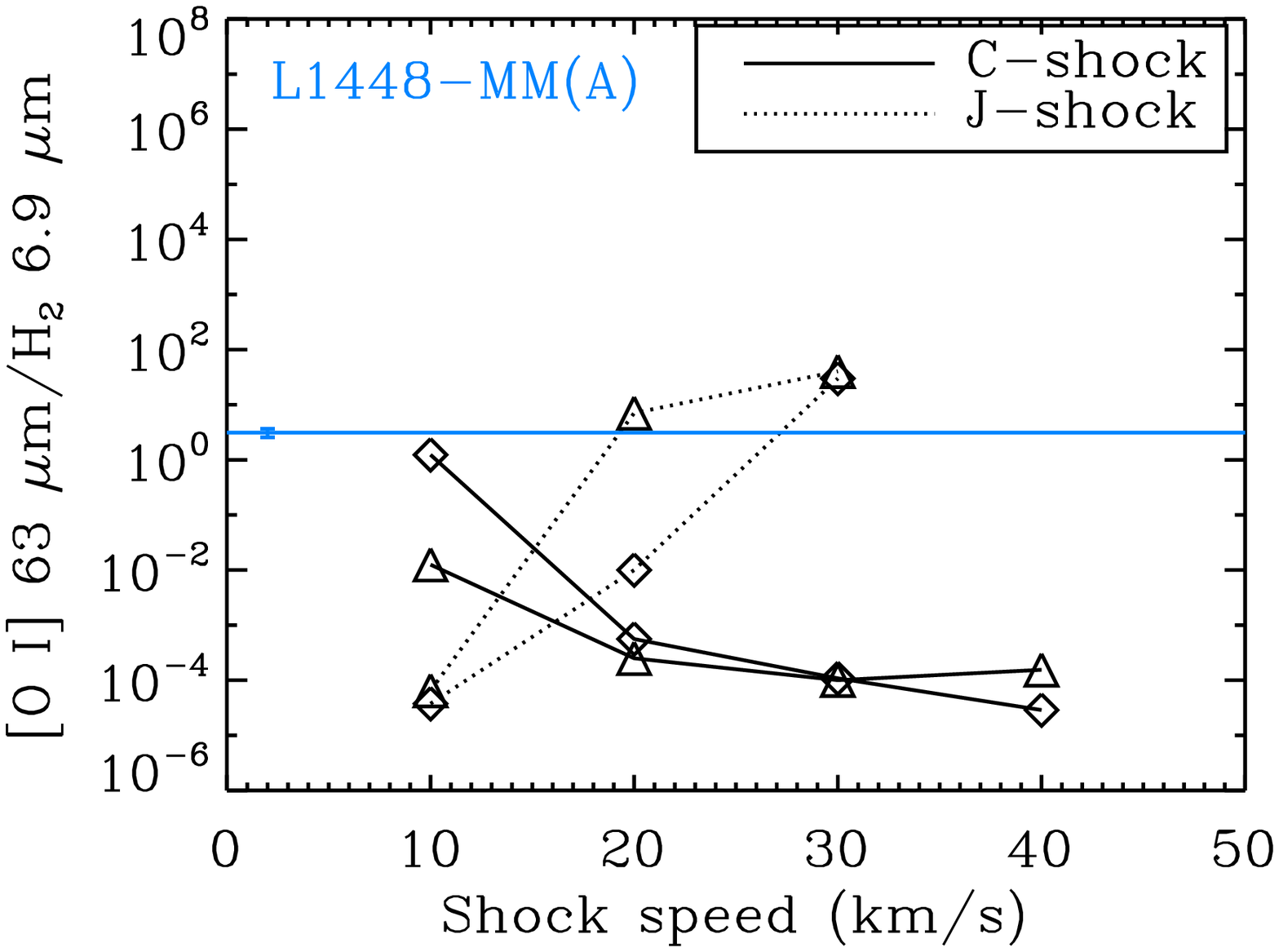}
\caption{
The line flux ratios as a function of shock conditions \citep{Flo10}.
The shock parameters are the shock speed (km s$^{-1}$) and preshock H$_2$
density (cm$^{-3}$).
Solid lines and dashed lines represent C- and J-shock models, respectively.
Diamonds refer to 2 $\times$ 10$^4$ cm$^{-3}$ and triangles to 2 $\times$
10$^5$ cm$^{-3}$ of hydrogen density.
Horizontal lines indicate the values of L1448-MM and
 vertical lines represent the uncertainties.
 Blue and red colors represent the flux ratios for (A) and (B), respectively.
 Left: The line flux ratio of \OI\ 63 $\mu$m to 145 $\mu$m.
 Right: The line intensity ratios of \OI\ 63 $\mu$m with respect to H$_2$ S(5)
6.9 $\mu$m pure rotation transition.
H$_2$ fluxes are convolved with PACS spaxels.
}
\label{shock_oi}
\end{figure}

Due to its low upper level energy of 230 K,
the \OI\ 63 $\mu$m line is easily excited, and thus, it is an important cooling
channel in the postshock
region.  As a result, it is one of the best tracers of shocks in dense
environments \citep{Gia01}.
We calculated the \OI\ 63/145 $\mu$m ratio for (A) and (B), which 
are $\sim$ 22 and 9, respectively.
In (A), flux ratio corresponds to low density C-shock model
 while flux ratio of (B) fits to a high density C-shock model as seen in
Fig.~\ref{shock_oi} (left).
Therefore, the flux ratios of \OI\ lines are consistent with the C-shock model.

However, the [O I] flux is severely under-estimated by the C-type shock models; an emitting region of  $> 10^4$ $\times$ 10$^4$ AU$^2$, corresponding to $> 50$ spaxels, would be required. Alternatively, more than 50 individual shocks would be required to generate this amount of emission. If a J-shock model is adopted (with $v$ = 30 km s$^{-1}$ and $n{\rm (H_2)}$ = 2$\times10^4$ cm$^{-3}$ for (A) and 2$\times10^5$ cm$^{-3}$ for (B)), the emitting area is 300 $\times$ 300 AU$^2$ (1/40 spaxel) for (A) and 400 $\times$ 400 AU$^2$ (1/25 spaxel) for (B). Considering the spatial distribution of the [O I] emission covering more than four spaxels (Fig.~\ref{map_oi}), neither a single C-shock nor a single J-shock can reproduce the [O I] absolute fluxes and extent, indicative of multiple shock components. 

\citet{Flo10} also calculated several H$_2$ line strengths under the same shock
conditions
and compared the intensities with those of \OI\ emission.
In (A), the ratio is fitted to a J-shock model with shock speed of 20--30
km s$^{-1}$ and a low density C-shock model with $v\rm_{shock}$ $<$ 20 km s$^{-1}$ 
simultaneously (Fig.~\ref{shock_oi}). 
In (B), however, the observed fluxes cannot be compared with the model flux ratios
since the H$_2$ fluxes avoids the (B) position.
Therefore, the ratio of \OI\ to H$_2$ emission could not constrain the shock
characteristics in (B).

These comparisons suggest that a C-shock model can explain most of emission in
L1448-MM, but the \OI\ absolute fluxes and the \SiII\ emission detected in the IRS spectra 
indicate that a dissociative J-shock should also exist in this region.
According to \citet{Neu89}, \SiII\ emission cannot be produced by a
non-dissociative shock, but could be produced by either a dissociative shock or PDR.
The dissociative shock tracers such as \SiII\ and \FeII\ often go close together
 with non-dissociative
shock tracers (e.g., \cite{Neu09}), and the dissociative apex of a bow shock that is flanked by
non-dissociative shocks is suggested to explain this feature.
Therefore, we cannot designate a single type of shock in the simple planar models for L1448-MM;
we require more sophisticated multi-dimensional shock models with different initial conditions 
to explain the relative emission of each species in this complicated region.

\subsection{Luminosities}

Table~\ref{lumtable} presents luminosities for the lines detected in L1448-MM
as well as continuum.
The total line luminosity accounts for only $\sim$ 0.7\% of the total FIR luminosity 
in the PACS range. Therefore, the dominant cooling occurs by the continuum radiation.
According to \citet{Nis99} and \citet{Gia01}, the FIR line luminosities of L1448-MM calculated from the 
{\it ISO} observations are greater than what we derive from our PACS observations 
by factors of 2 to 8 depending on species although the relative luminosities among different species are
similar. However,  {\it ISO} and PACS both show that  the line cooling 
mainly occurs through H$_2$O emission. 
In contrast, the FIR continuum luminosity 
obtained by {\it ISO} is much smaller than what derived by PACS by a factor of 20, resulting in
a higher fraction of line luminosity to the continuum luminosity in FIR. 
These differences in luminosities between {\it ISO} and PACS are probably caused by 
the low sensitivity and large beam of {\it ISO} compared to PACS. 
In L1448-MM, molecular emission extends beyond the FOV of PACS.
Therefore, the {\it ISO} beam, larger than the PACS FOV, possibly picked up a
significant amount of the extended molecular emission.

In the PACS line luminosity, CO, H$_2$O, and OH emission arises mostly from (A);
$\sim 70\%$ of the total CO and H$_2$O fluxes are emitted from (A), 
while (B) supplies $\sim 30\%$ of the CO and H$_2$O line cooling, and
most of OH line emission ($\sim 90\%$) is concentrated on (A).
However, the sum of \OI\ line luminosity of (A) and (B) has just $\sim
65\%$ of the total luminosity calculated over the whole 5$\times$5 spaxels, 
indicative of broadly extended emission in \OI.
If we assume that the gas components of (A) and (B) are associated with MM(A)
and MM(B), respectively, the ratio of the FIR line luminosity in the PACS range 
to the bolometric luminosity ($L_{\rm mol}/L_{\rm bol}$) both for MM(A) and MM(B) is  
$\sim 4\times 10^{-3}$, indicating that both sources are in the Class 0 stage.
According to \citet{Gia01}, $L_{\rm mol}/L_{\rm bol} > 1\times 10^{-3}$ for  Class 0 objects while the ratio is smaller than $5\times 10^{-4}$ for Class I and II sources.
According to our best-fit LVG models, 
H$_2$O and CO emit the majority ($> 50-80\%$)
of their luminosity  in the PACS wavelength range.

Table~\ref{relalumtable} shows the fractional contribution of each species to the
FIR line luminosity as a total and in (A) and (B).
\citet{van11} suggested that H$_2$O might not be the dominant coolant in YSOs.
For L1448-MM, however, water is the primary outflow coolant
 and this is consistent with the results for the Class 0 protostar NGC 1333 IRAS 4B
\citep{Her12},
where H$_2$O is responsible for 72\%\ of line luminosity.
H$_2$O occupies $\sim 50\%$ and CO fills $\sim 30\%$ of the line luminosity in
the PACS wavelength range in L1448-MM.
For CO, the warm ($E\rm_{u}<$ 1500 K) and hot ($E\rm_{u} >$ 1500 K) components
are responsible for $\sim$65\%\ and $\sim$35\%\ of the CO line luminosity
in the PACS range, respectively, based on the results of rotation diagrams.
OH and \OI\ contribute $\sim$ 10\%\ and $\sim$ 5\%\ of the line
luminosity, respectively.
In L1448-MM, the cooling through molecular emission is significantly larger
than the cooling via atomic emission,
 which is consistent with the characteristic of the Class 0 YSOs \citep{Nis02, Her12}.
Bright H$_2$O emission and dim \OI\ emission in L1448-MM is indicative of low
dissociation rate of
H$_2$O, or the fast formation process of H$_2$O in the postshock gas.
Alternatively, \OI\ and H$_2$O emission possibly arises from unassociated gas
components, i.e., the \OI\ emission is attributed to a more extended gas while the H$_2$O
emission is localized around the YSOs.

\section{Discussion}

\subsection{Multiple Sources in L1448-MM}

Previously, L1448-MM was known as a single YSO with a prominent outflow.
However, \citet{Jor07}, \citet{Tob07}, and \citet{Hir11} suggested the
existence of a secondary YSO, named L1448-MM(B)
while \citet{Mau10} suggested that the secondary point-like source at {\it Spizer} bands and 3 mm
might be a result of a shock on the outflow cavity wall.
However, we conclude that L1448-MM(B) is a YSO based on its millimeter SED
(Fig.~\ref{SED}) and the detection of the double peak structure of CO$_2$ ice absorption feature 
at 15.2 $\mu$m (Fig.~\ref{co2ice}) \citep{Pon08}.

The SEDs (Fig.~\ref{SED}) of MM(A) and MM(B) rising into the FIR indicate that both sources are very embedded.
\citet{Hir10} suggested that MM(B) was likely less obscured in the MIR compared to MM(A).
This is consistent with a much higher FIR flux level in MM(A) than in
MM(B), indicating less material toward MM(B).
From the separated SEDs and the photometric data points in \citet{Gre13}, we calculated the bolometric luminosities 
($L_{\rm bol}$) of MM(A) and MM(B) as 5.5 and 1.7 $L_\sun$, respectively, which do not 
add up to $L_{\rm bol}$ of 8.4 $L_\sun$ calculated over the whole region \citep{Gre13}.
For the calculation, single-dish (sub)millimeter fluxes for two sources were separated based on the flux 
ratios derived from the interferometric observations. (The interferometric continuum fluxes were not 
included in the calculation of $L_{\rm bol}$ and $T_{\rm bol}$ 
although they do not affect the results at all.)
The derived $L_{\rm bol}$  for each source is rather sensitive to this flux separation at (sub)millimeter. 
Therefore, it should not be considered very accurate. 
The calculated bolometric temperatures ($T_{\rm bol}$) of MM(A) and MM(B) are 49 and 80 K,
respectively, indicating that MM(A) is more embedded and in the earlier evolutionary stage.
$L_{\rm bol}/L_{\rm smm}$ are 18.7 and 48.7 for MM(A) and MM(B), respectively,  which also
suggests that MM(A) is more embedded. Note that MM(B) is classified as Class I 
by ($T_{\rm bol}$) but Class 0 by $L_{\rm bol}/L_{\rm smm}$ while MM(A) is classified as 
Class 0 by both criteria.

MM(A) is separated from MM(B) by 8\farcs17 that is equivalent to
$\sim$2000 AU.
\citet{Mun01} divided embedded multiple systems into three groups:
independent envelope, common envelope, and common disk systems.
As seen in the submm continuum maps \citep{Shi00},
 two sources seem associated with only one dense core.
Therefore, L1448-MM may be a common envelope system which has one primary core
in gravitational contraction,
and where objects are separated by 250 -- 3000 AU.
Although the possible detection of the CO$_2$ gas line at 14.98 $\mu$m \citep{Dar98} toward MM(A)
 is indicative of a hotter region, 
 the column density of CO$_2$ ice, calculated from the IRS 15.2 $\mu$m
CO$_2$ ice feature, is $\sim$2 $\times$ 10$^{18}$ cm$^{-2}$ in both YSOs.
This might also support the idea that they are in a common envelope.

\subsection{Shocked Gas in L1448-MM}
The outflow activity in the two sources appears significantly different.
\citet{Hir10} has reported that the weak CO outflow possibly associated with
MM(B) is nearly perpendicular to that of MM(A),
and its small momentum flux is comparable to those of outflows by Class I
objects.
However, the MM(A) outflow elongates in the SE-NW direction.
Since MM(B) is about $8\arcsec$ south of MM(A),
fluxes in spaxel S are likely contaminated by the outflow emission from the
MM(A).
The continuum emission of MM(A) is greater than that of MM(B) at $\lambda \ge 20$ $\mu$m.
Lines of high energy levels also peak in the position of MM(A) while the emission peaks of
lines of lower energy levels shift to the south.
These features suggest that MM(A) may be the primary outflow source in
the direction of SE-NW.
Then the gas traced in our PACS observations may be heated predominantly by the jet/outflow 
driven by MM(A).
If this is true, the molecular gas in (A) and (B) rather correspond to 
the blue and red wings of the L1448-MM(A) jet, which were detected by the CO and 
SiO transitions \citep{Nis00, Hir10}.

\begin{figure}
\plottwo{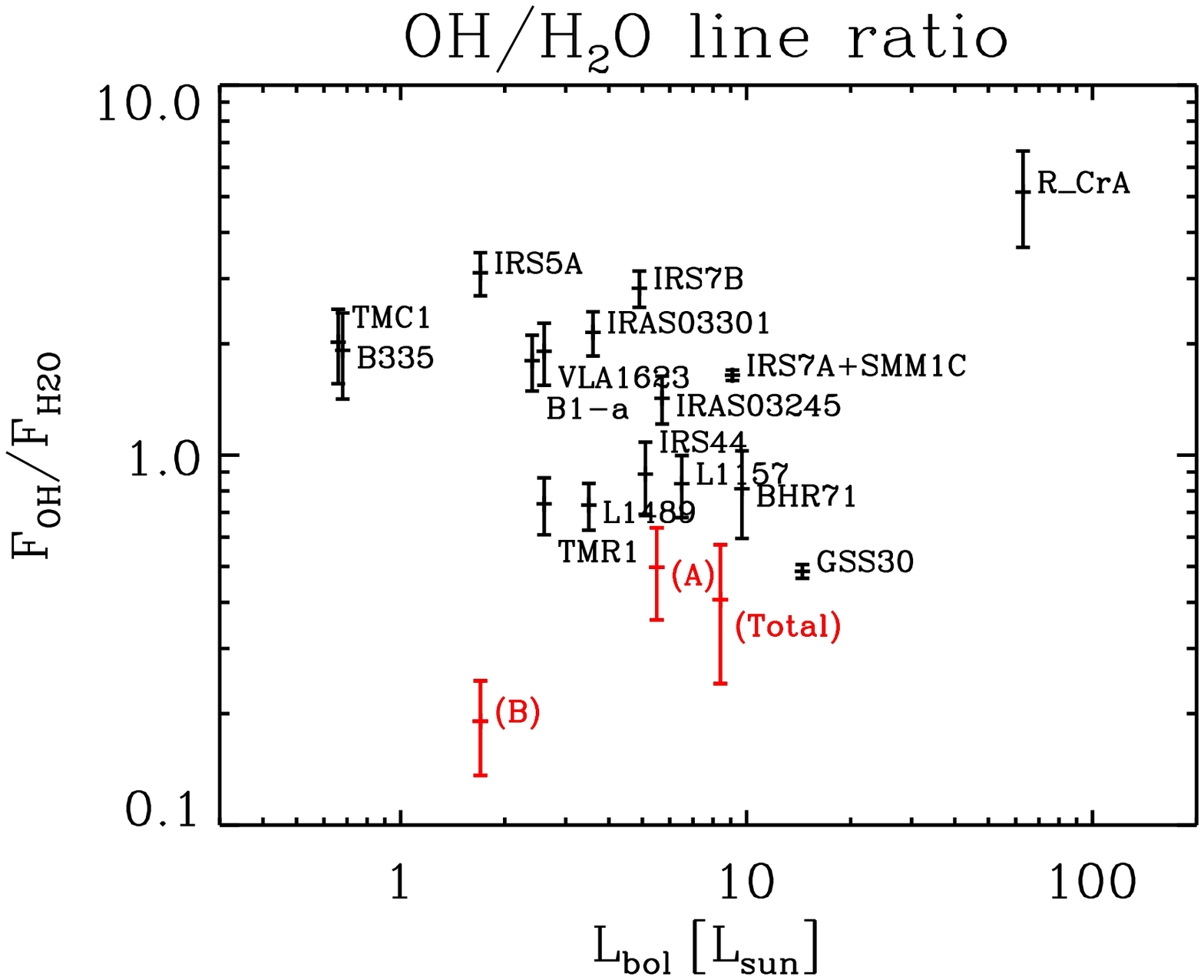}{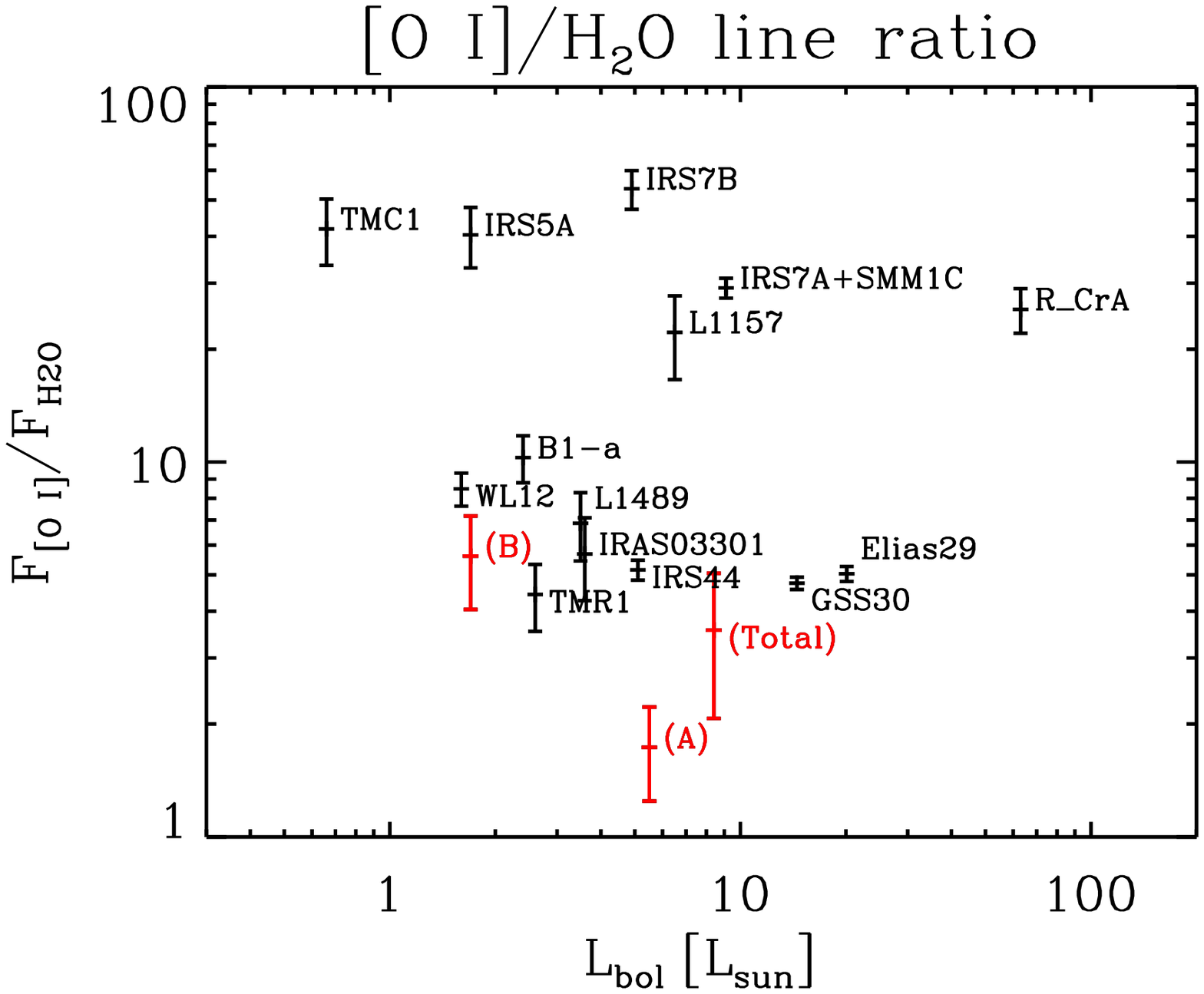}
\caption{
OH and \OI\ line fluxes relative to H$_2$O line flux toward the DIGIT embedded sources \citep{Gre13}.
Left: the flux ratio of the OH (84 $\mu$m, $E_{\rm u}$ = 291 K) and the o-H$_2$O (75 $\mu$m, $E_{\rm u}$ = 305 K)
Right: The flux ratio of the \OI\ (63 $\mu$m, $E_{\rm u}$ = 227 K) and o-H$_2$O (66 $\mu$m, $E_{\rm u}$ = 410 K).
The minimal ratios for L1448-MM indicate that the photodissociation is not important in the region.
}
\label{oh_oi_h2o_ratio}
\end{figure}

According to our excitation analyses of (A) and (B), 
the PACS FIR line fluxes are possibly produced mainly by 
shocked hot gas components, rather than by the UV-heated gas along the outflow cavity.
According to \citep{Yil10}, up to J=10--9, the contribution of the broad wing component increases with J, hinting that shocks may play a more important role in higher J transitions traced by the PACS.
However, if the UV photons play an important role in this region, we should expect an enhancement of the \OI\ and OH line 
fluxes compared to H$_2$O. In order to check the relative emission among \OI, OH, and H$_2$O, 
as done in Lindberg et al. (subm.), we calculated the flux ratios between 
the OH (84 $\mu$m, $E_{\rm u}$ = 291 K) and o-H$_2$O (75 $\mu$m, $E_{\rm u}$ = 305 K) lines,
which have small PSFs and similar upper-level energies, 
as well as the flux ratios between \OI\ (63 $\mu$m, $E_{\rm u}$ = 227 K) and o-H$_2$O (66 $\mu$m, $E_{\rm u}$ = 410 K) lines, toward the DIGIT embedded sources (Fig.~\ref{oh_oi_h2o_ratio}).
According to the comparisons, L1448-MM has the minimum flux ratios among the DIGIT embedded sources, supporting that the observed line emission in L1448-MM is mainly 
from shocks. 

Comparing to the recently observed Class 0 objects, NGC 1333 IRAS4B
 (hereafter, IRAS4B) and Serpens SMM1(hereafter, SMM1), L1448-MM is similar to IRAS4B in molecular emission; water dominates in the line cooling and \OI\ is dim. In addition, the best model for IRAS4B with the non-LTE analysis 
suggests the physical conditions of $T_{\rm K}\sim$1500 K and $n(\rm H_2$) $\sim$3$\times$10$^6$ cm$^{-3}$.
In SMM1, however, CO is the main coolant and \OI\ is relatively brighter than the other two sources, 
and the derived physical conditions are $T_{\rm K}\sim$800 K 
and $n(\rm H_2)\ge$ 5$\times$10$^6$ cm$^{-3}$.
Therefore, the derived physical conditions for L1448-MM are more similar to those of IRAS4B. 
In addition, \CII\ (158 $\mu$m) is detected in SMM1 but not detected in IRAS4B and L1448-MM.  
$L_{\rm OH}$/$L_{\rm H_2O}$ of L1448-MM and IRS4B are $\sim$0.2 while that  of SMM1 is 0.4.
$L_{\rm OI}$/$L_{\rm H_2O}$ are 0.06$\sim$0.1, 0.01, and 0.6 for L1448-MM, IRS4B, and SMM1, respectively. (\citet{Goi12} mentioned that the strong \OI\ and OH emission towards SMM1 is similar to that of HH46, a Class I source \citep{Kem10a}.)
Therefore, the FIR line emission in L1448-MM and NGC 1333 IRAS4B might have a similar origin, i.e., non-dissociative shocks shielded from UV radiation play a larger role in the excitation than dissociative shocks \citep{Her12}.

\section{Summary $\&$ Conclusion}

Contour maps show that FIR line emission from
 low energy levels is toward the south
 while H$_2$ emission in the MIR peaks toward the north.
Most of FIR molecular line emission is from the unresolved two gas components, 
(A) and (B), both of which might be heated by the jet shock of MM(A).
 For CO and H$_2$O ($L_{\rm CO}+L_{\rm H_2O}$), $\sim70\%$ of cooling occurs in (A)
 while most OH emission ($\sim 90\%$ of $L_{\rm OH}$) concentrates on (A).
 Differently from other species, the \OI\ emission extends more broadly beyond
the two positions to the south.

 According to the simple rotational diagram model,
CO seems to have two temperature components (warm and hot),
which have been attributed by previous studies
to the PDR and shock, respectively. 
The rotational temperatures and total number of molecules of the two CO
components in (A) are higher than those measured in (B).
 This tendency is true for OH and H$_2$O as well.
 Therefore, the gas in (A) is hotter and has more of the excited 
molecules than does (B),
 according to the rotation diagram analysis.
In the case of H$_2$O, the derived ortho-to-para ratio is close to 3,
 indicating that H$_2$O might have formed in the hot postshock gas,
or the timescale is long enough to equilibrate the ortho-to-para
ratio of water after its evaporation from grain surfaces.
For OH, $\mathcal N$($^2\Pi_{1/2}$)/$\mathcal N$($^2\Pi_{3/2})\sim$0.6, which is greater 
than the ratio of partition functions of the two ladders by a factor of 4 at 115 K. This is possibly due to 
 the IR-pumping in the $^2\Pi_{1/2}$ transitions and/or higher optical depths of 
 the $^2\Pi_{3/2}$ transitions.

According to our non-LTE LVG analyses, the PACS CO and H$_2$O emission arises from shocked gas 
(rather than photo-heated gas) and requires multiple gas components with different physical conditions.

The non-LTE LVG model shows the sub-thermal condition of OH.
All OH lines except the highest energy level ($ E_{\rm u} =$ 875 K) transition are optically thick, and 
the optical depths of the $^2\Pi_{3/2}$ transitions are higher than those of the $^2\Pi_{1/2}$ transitions.
Therefore, the displacement between two ladders in the rotation diagram is caused by the higher 
optical depths of the $^2\Pi_{3/2}$ transitions.
In addition, the LVG model supports the IR-pumping processes for OH transitions
because the OH line flux of $ E_{\rm u} =$ 875 K is 
much better fitted when the FIR radiation from the central source is included.
In contrast, the IR-pumping is not very important for the H$_2$O lines.

Our best-fit LVG models predict that  (50--80)\%\ of the molecular line emission is produced in the PACS wavelength range depending on models.
The continuum luminosity observed in the PACS range is $\sim$ 50\%\ of
$L\rm_{bol}$. 
The modeled cooling luminosities are $L_{\rm CO} \sim (1.1-2.4) \times 10^{-2}$
$L_{\sun}$ and $L_{\rm H_2O}\sim (2.0-4.5) \times 10^{-2}$ $L_{\sun}$
 while HIFI observations predict $L_{\rm CO}\sim$ 0.5 $\times 10^{-2}$ $L_{\sun}$ and 
 $L_{\rm H_2O} \sim 2-4  \times 10^{-2}$ $L_{\sun}$ for CO and H$_2$O, respectively
\citep{Kri11}.
Both models show that the major line cooling occurs at the wavelengths $>$ 60
$\mu$m, which is consistent with the {\it ISO} result \citep{Nis00}.

In comparisons with shock models, the PACS molecular emission can be explained by a C-shock,
but the atomic emission such as PACS \OI\ and {\it Spitzer}/IRS \SiII\ requires a J-shock, indicative of
multiple shocks in L1448-MM.
 
In conclusion, our study of L1448-MM with the PACS spectra shows that the atomic and
molecular line observations at the FIR wavelengths are very important to understand
in detail the energy budget and excitation conditions in the embedded YSOs.

\acknowledgments

Support for this work, part of the {\it Herschel} Open Time Key Project
Program, was provided by NASA through an award
issued by the Jet Propulsion Laboratory, California Institute of
Technology.
Jeong-Eun Lee was supported by the Basic Science Research Program through the National Research Foundation of Korea (NRF) funded by the Ministry of Education of the Korean government (grant number NRF-2010-0008704 and NRF-2012R1A1A2044689). 
This work was also supported by the Korea Astronomy
and Space Science Institute (KASI) grant funded by the Korea government(MEST).


\clearpage

\begin{figure}
\plotone{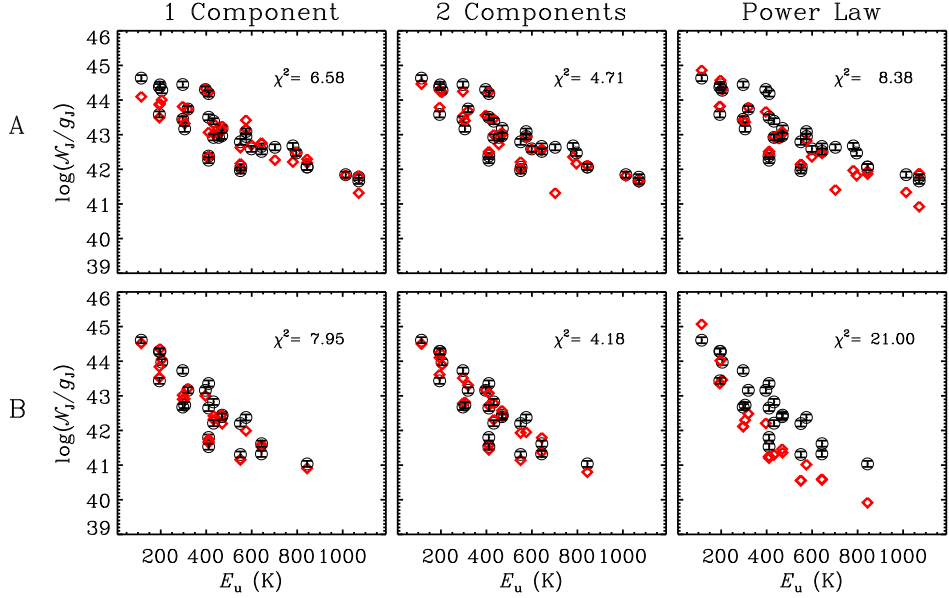}
\caption{The same as Fig.~\ref{LVG1_CO} but for H$_2$O.
}
\label{LVG1_H2O}
\end{figure}

\begin{deluxetable}{llcccccc}
\tablewidth{0pt}
\tablecaption{Detected lines in {\it Herschel}/PACS spectrum of L1448-MM}
\scriptsize
\tablehead{
\colhead{Species} 		& \colhead{Transition}	 	& \colhead{$E\rm_u$ (K)}	 &
\colhead{$A\rm_{ul}$ (cm$^{-1}$)} & \colhead{$\lambda$ ($\mu$m)} &
\multicolumn{3}{c}{Flux $(10^{-18}$ W m$^{-2}$) } \\
\cline{6-8}
\colhead{} &\colhead{} & \colhead{} & \colhead{} & \colhead{} &
 \colhead{$5\times$5\tablenotemark{a}} & \colhead{(A)\tablenotemark{b}} &
\colhead{(B)\tablenotemark{c}}
 }
\startdata
CO&40-39&      4512.67&   0.00461300& 65.69&          67$\pm$          26&
    77&-\\
 &39-38&      4293.64&   0.00436500& 67.34&          60$\pm$          19&
   55&-\\
 &38-37&      4079.98&   0.00412000& 69.07&          74$\pm$          22&
   67&-\\
 &37-36&      3871.69&   0.00387800& 70.91&          61$\pm$          25&
   58&-\\
 &36-35&      3668.78&   0.00363800& 72.84&         104$\pm$          39&
  104&-\\
 &35-34&      3471.27&   0.00340400& 74.89&          91$\pm$          27&
  166\tablenotemark{d}&-\\
 &34-33&      3279.15&   0.00317500& 77.06&         123$\pm$          36&
   88&-\\
 &33-32&      3092.45&   0.00295200& 79.36&          83$\pm$          27&-&-\\
 &32-31&      2911.15&   0.00273500& 81.81&         121$\pm$          36&
  110&          12\\
 &30-29&      2564.83&   0.00232100& 87.19&         172$\pm$          54&
  117&          45\\
 &29-28&      2399.82&   0.00212600& 90.16&         214$\pm$          64&
  139&          40\\
 &28-27&      2240.24&   0.00194000& 93.35&         214$\pm$          65&
  167&          38\\
 &27-26&      2086.12&   0.00176100& 96.77&         229$\pm$          75&
  182&-\\
 &25-24&      1794.23&   0.00143200&104.44&         195$\pm$          56&
  152&          59\\
 &24-23&      1656.47&   0.00128100&108.76&         234$\pm$          68&
  203&          65\\
 &22-21&      1397.38&   0.00100600&118.58&         288$\pm$          81&
  217&          77\\
 &21-20&      1276.05&  0.000883300&124.19&         328$\pm$          92&
  236&          90\\
 &20-19&      1160.20&  0.000769500&130.37&         350$\pm$          99&
  256&         100\\
 &19-18&      1049.84&  0.000665000&137.20&         381$\pm$         108&
  259&         128\\
 &18-17&      944.970&  0.000569500&144.78&         422$\pm$         120&
  294&         122\\
 &17-16&      845.590&  0.000482900&153.27&         433$\pm$         123&
  300&         146\\
 &16-15&      751.720&  0.000405000&162.81&         503$\pm$         142&
  318&         206\\
 &15-14&      663.350&  0.000335400&173.63&         547$\pm$         155&
  373&         222\\
 &14-13&      580.490&  0.000273900&186.00&         507$\pm$         144&
  303&         211\\

\\
\\
\\
\\
\\
\\
\\
\\
OH& $\frac{1}{2}$,$\frac{9}{2}$-$\frac{1}{2}$,$\frac{7}{2}$ &      875.100&
 2.18200& 55.89&    87$\pm$ 32&          70\tablenotemark{e}&-\\
 & &      875.100&      2.17500& 55.95&        88$\pm$          33&
70\tablenotemark{e}&-\\
 & $\frac{3}{2}$,$\frac{9}{2}$-$\frac{3}{2}$,$\frac{7}{2}$ &      512.100&
1.27600& 65.13&       312$\pm$          96&         332&-\\
 & &      510.900&      1.26700& 65.28&         141$\pm$          41&
129&-\\
 &  $\frac{1}{2}$,$\frac{7}{2}$-$\frac{1}{2}$,$\frac{5}{2}$ &      617.600&
 1.01400& 71.17&       100$\pm$          29&          76&          13\\
 & &      617.900&      1.01200& 71.22&        101$\pm$          29&
76&          13\\
 & $\frac{1}{2}$,$\frac{1}{2}$-$\frac{3}{2}$,$\frac{3}{2}$ &      181.900&
0.0360600& 79.12&       190$\pm$          59&         172&          26\\
 & &      181.700&    0.0359800& 79.18&         167$\pm$          49&
154&          39\\
 &$\frac{3}{2}$,$\frac{7}{2}$-$\frac{3}{2}$,$\frac{5}{2}$ &      290.500&
0.520200& 84.60&        251$\pm$          73&         207&          29\\
 &$\frac{1}{2}$,$\frac{3}{2}$-$\frac{3}{2}$,$\frac{5}{2}$  &      270.200&
0.00927000& 96.31&       37$\pm$          21&-&-\\
 & &      269.800&   0.00925000& 96.37&         15$\pm$          17&-&-\\
 &$\frac{3}{2}$,$\frac{5}{2}$-$\frac{3}{2}$,$\frac{3}{2}$ &      120.700&
0.138800&119.23&     202$\pm$          57&         124&          77\\
 & &      120.500&     0.138000&119.44&        249$\pm$          73&
145&         102\\
 &$\frac{1}{2}$,$\frac{3}{2}$-$\frac{1}{2}$,$\frac{1}{2}$&      270.200&
0.0648300&163.12&    53$\pm$          16&          46&           7\\
 & &      269.800&    0.0645000&163.40&         58$\pm$          17&
49&           2\\

\\

p-H$_2$O &4$_{31}$--3$_{22}$&      552.300&      1.45200& 56.33&
89$\pm$          42&          81&-\\
 &9$_{19}$--8$_{08}$&      1324.00&      2.48600& 56.77&         125$\pm$
   39&-\tablenotemark{f}&-\\
 &4$_{22}$--3$_{13}$&      454.300&     0.378500& 57.64&         120$\pm$
   37&         151&-\\
 &7$_{26}$--6$_{15}$&      1021.00&      1.33800& 59.99&          69$\pm$
   24&-\tablenotemark{f}&-\\
 &8$_{08}$--7$_{17}$&      1070.60&      1.74200& 63.46&          61$\pm$
   23&          84&-\\
 &3$_{31}$--2$_{20}$&      410.400&      1.22200& 67.09&         137$\pm$
   41&          94&          24\\
 &5$_{24}$--4$_{13}$&      598.800&     0.667900& 71.07&         131$\pm$
   40&         123&-\\
 &7$_{17}$--6$_{06}$&      843.800&      1.17800& 71.54&          96$\pm$
   28&          88&-\\
 &6$_{15}$--5$_{24}$&      781.100&     0.452600& 78.93&          96$\pm$
   32&         108&-\\

\\
\\
\\
\\
\\
\\
\\
 &6$_{06}$--5$_{15}$&      642.700&     0.713200& 83.28&         165$\pm$
   48&         156&           7\\
 &3$_{22}$--2$_{11}$&      296.800&     0.352400& 89.99&         298$\pm$
   95&         233&          39\\
 &5$_{15}$--4$_{04}$&      469.900&     0.446000& 95.63&         276$\pm$
   78&         238&          43\\
 &2$_{20}$--1$_{11}$&      195.900&     0.260700&100.98&         342$\pm$
  117&-\tablenotemark{f}&-\\
 &4$_{04}$--3$_{13}$&      319.500&     0.172700&125.35&         285$\pm$
   81&         209&          54\\
 &3$_{31}$--3$_{22}$&      410.400&    0.0784800&126.71&          24$\pm$
    8&-&-\\
 &3$_{13}$--2$_{02}$&      204.700&     0.125100&138.53&         520$\pm$
  147&         371&         181\\
 &4$_{13}$--3$_{22}$&      396.400&    0.0331600&144.52&         124$\pm$
   38&         127&           9\\
 &3$_{22}$--3$_{13}$&      296.800&    0.0524600&156.19&         239$\pm$
   70&         202&          38\\
 &4$_{13}$--4$_{04}$&      396.400&    0.0372600&187.11&          32$\pm$
   13&-&-\\

\\
o-H$_2$O &4$_{32}$--3$_{21}$&      550.400&      1.37400& 58.70&
184$\pm$          53&         177&          39\\
 &8$_{18}$--7$_{07}$&      1070.70&      1.75100& 63.32&         210$\pm$
   67&         199&-\\
 &7$_{16}$--6$_{25}$&      1013.20&     0.950800& 66.09&         125$\pm$
   43&         139&-\\
 &3$_{30}$--2$_{21}$&      410.700&      1.24300& 66.44&         259$\pm$
   76&         221&          41\\
 &3$_{30}$--3$_{03}$&      410.700&   0.00850500& 67.27&         127$\pm$
   37&         124&          18\\
 &7$_{07}$--6$_{16}$&      843.500&      1.15700& 71.95&         276$\pm$
   80&         257&          24\\
 &3$_{21}$--2$_{12}$&      305.300&     0.331800& 75.38&         617$\pm$
  176&         416&         152\\
 &4$_{23}$--3$_{12}$&      432.200&     0.483800& 78.74&         522$\pm$
  147&         424&          82\\
 &6$_{16}$--5$_{05}$&      643.500&     0.749100& 82.03&         417$\pm$
  118&         359&          45\\
 &6$_{25}$--6$_{16}$&      795.500&     0.174000& 94.64&         113$\pm$
   32&          65&-\\
 &4$_{41}$--4$_{32}$&      702.300&     0.152800& 94.71&          68$\pm$
   19&          57&-\\
 &5$_{05}$--4$_{14}$&      468.100&     0.390200& 99.49&         547$\pm$
  170&         366&          99\\
 &5$_{14}$--4$_{23}$&      574.700&     0.156600&100.91&         247$\pm$
   90&         192&-\\
 &2$_{21}$--1$_{10}$&      194.100&     0.256400&108.07&         735$\pm$
  210&         419&         296\\

\\
\\
\\
\\
\\
\\
\\
 &4$_{32}$--4$_{23}$&      550.400&     0.122900&121.72&          52$\pm$
   15&          52&          13\\
 &4$_{23}$--4$_{14}$&      432.200&    0.0808400&132.41&         173$\pm$
   50&         123&          33\\
 &5$_{14}$--5$_{05}$&      574.700&    0.0757900&134.93&          70$\pm$
   22&          50&          13\\
 &3$_{30}$--3$_{21}$&      410.700&    0.0661900&136.50&         117$\pm$
   34&          99&          13\\
 &5$_{32}$--5$_{23}$&      732.100&    0.0817400&160.51&          48$\pm$
   14&-&-\\
 &3$_{03}$--2$_{12}$&      196.800&    0.0504800&174.63&         824$\pm$
  233&         510&         355\\
 &2$_{12}$--1$_{01}$&      114.400&    0.0559300&179.53&        1173$\pm$
  334&         615&         579\\
 &2$_{21}$--2$_{12}$&      194.100&    0.0305800&180.49&         337$\pm$
   96&         184&         149\\

\\
$\rm \OI$&$^3$P$_1$--$^3$P$_2$&      227.712&      0.0000891& 63.18&
922\tablenotemark{g}$\pm$275 & 383	& 230	\\
$\rm \OI$&$^3$P$_0$--$^3$P$_1$&      326.579&      0.0000175& 145.53&
84\tablenotemark{h}$\pm$30  & 18	& 26	\\

\enddata
\tablecomments{Emissions of CO $\&$ OH (84 $\mu$m) and CO $\&$ o-H$_2$O (113
$\mu$m) are detected,  but not measured since they are blended.
CO J=13-12 is detected but not included in our analysis  because of low
responsivity of PACS in this wavelength range.}
\tablenotetext{a} {From whole 5$\times$5 spaxel}
\tablenotetext{b} {From the unresolved gas component located at the position of L1448-MM(A)}
\tablenotetext{c} {From the unresolved gas component located at the position of L1448-MM(B)}
\tablenotetext{d}{The line in the spectrum extracted from the central spaxel seems contaminated by an unknown feature, which possibly increases the EW of the line.}
\tablenotetext{e}{This rotation transition was not split in the spectrum extracted from the central spaxel while it was split in the spectrum extracted over 25 spaxels.}
\tablenotetext{f}{We excluded the lines because the line shapes are too strange to be fitted by a Gaussian profile.}
\tablenotetext{g}{Equivalent width measured from 4 spaxels}
\tablenotetext{h}{Equivalent width measured from 3 spaxels}
\label{fluxtable}
\end{deluxetable}


\begin{deluxetable}{llcccccccc}
\tablecolumns{10}
\tablewidth{0pc}
\tabletypesize{\scriptsize}
\tablecaption{Summary of the rotation diagram analysis}
\tablehead{
\colhead{} & \colhead{} &
\multicolumn{2}{c}{$5\times$5} & \colhead{}&
\multicolumn{2}{c}{(A)} & \colhead{}&
\multicolumn{2}{c}{(B)} \\
\cline{3-4} \cline{6-7} \cline{9-10}
\colhead{ Species } &
\colhead{Component} &
\colhead{$T\rm_{rot}$} &
\colhead{$\mathcal N$(molecule)\tablenotemark{a} } & \colhead{} &
\colhead{$T\rm_{rot}$} &
\colhead{$\mathcal N$(molecule)\tablenotemark{a} } & \colhead{} &
\colhead{$T\rm_{rot}$} &
\colhead{$\mathcal N$(molecule)\tablenotemark{a} } \\
\colhead{} & \colhead{} &
\colhead{(K)} & \colhead{(10$\rm^{47}$)} & \colhead{} &
\colhead{(K)} & \colhead{(10$\rm^{47}$)} & \colhead{} &
\colhead{(K)} & \colhead{(10$\rm^{47}$)}
}
\startdata

CO & Hot & 758$\pm$54 & 65$\pm$19 & & 854$\pm$42 & 55$\pm$10
& & 455$\pm$37 & 31$\pm$13 \\
& Warm & 293$\pm$30 & 275$\pm$98 & & 314$\pm$24 & 164$\pm$41 &
& 250$\pm$15 & 136$\pm$34\\

H$_2$O & Para & 168$\pm$6 & 0.06$\pm$0.01 & & 150$\pm$5 & 0.06$\pm$0.01
& & 74$\pm$3 & 0.06$\pm$0.01 \\
& Ortho & 144$\pm$5 & 0.24$\pm$0.03 & & 154$\pm$4 &
0.15$\pm$0.02 & & 88$\pm$2 & 0.13$\pm$0.02 \\

OH &$^2\Pi_{3/2}$-$^2\Pi_{3/2}$ & 115$\pm$9 & 0.076$\pm$0.020 & & 136$\pm$9
& 0.044$\pm$0.008 & & -\tablenotemark{b} & -\tablenotemark{b}\\ 
&$^2\Pi_{1/2}$-$^2\Pi_{1/2}$ & 114$\pm$7 & 0.040$\pm$0.013 & & 120$\pm$4 &
0.027$\pm$0.005 & & -\tablenotemark{b} & -\tablenotemark{b}\\ 

\enddata
\tablenotetext{a}{The total number of molecules}
\tablenotetext{b}{The number of data points for this fitting is not large enough for a meaningful result.}
\label{rottable}
\end{deluxetable}

\clearpage

\begin{deluxetable}{lccccccccccccc}
\tablecolumns{12}
\tablewidth{0pc}
\tabletypesize{\scriptsize}
\tablecaption{The best-fit LVG model parameters for three different combinations of gas components}
\tablehead{
\colhead{} & \colhead{} &
\multicolumn{3}{c}{1 Component}  & \colhead{}&
\multicolumn{3}{c}{2 Components}  & \colhead{}&
\multicolumn{3}{c}{Power Law} \\
\cline{3-5} \cline{7-9} \cline{11-13}
\colhead{ Species } &
\colhead{Spatial} &
\colhead{$T\rm_{kin}$} &
\colhead{$n(\rm H_2$)} &
\colhead{$N$(mole)} & \colhead{$\chi^2$} &
\colhead{$T\rm_{kin}$} &
\colhead{$n(\rm H_2$)} &
\colhead{$N$(mole)} & \colhead{$\chi^2$} &
\colhead{$n(\rm H_2$)} &
\colhead{b}  &
\colhead{$N$(mole)} & \colhead{$\chi^2$} \\
\colhead{} & \colhead{Component} &
\colhead{(K)} & \colhead{(cm$^{-3}$)} & \colhead{ (cm$^{-2}$)} & \colhead{} &
\colhead{(K)} & \colhead{(cm$^{-3}$)} & \colhead{ (cm$^{-2}$)} & \colhead{} &
\colhead{(cm$^{-3}$)} & & \colhead{ (cm$^{-2}$)} \\
}
\startdata

CO	&	A	&	5000	&	10$^5$	&	1.9$\times 10^{14}$	& 2.41 &5000	& $10^{6}$	&	1.9$\times 10^{11}$	& 0.92 &	2.51$\times 10^{7}$	& 2.95 & & 1.26\\
	&		&	   		&			&						&      &5000	& $10^{4}$	&	6.0$\times 10^{13}$ &		\\

		&	B	&	4000	&	10$^4$	&	1.9$\times 10^{18}$	& 0.67 &5000	& $10^{4}$	&   6.0$\times 10^{18}$ & 0.46 &6.31$\times 10^{6}$ & 3.30 & & 0.73\\
	&		&	   		&			&							& 		&2000	& $10^{7}$	&	1.9$\times 10^{16}$ &		\\
	\hline

H$_2$O&	A	&	2000	&	10$^6$	&	6.3$\times 10^{17}$	& 6.58 &100		& $10^{5}$	&	6.3$\times 10^{18}$	& 4.71 &	2.51$\times 10^{7}$	& 0.0 &$10^{13}$&  8.38\\
	&		&	   		&			&						&      &5000	& $10^{8}$	&	6.3$\times 10^{14}$ &		\\
	&	B	&	700		&	10$^7$	&	6.3$\times 10^{15}$	& 7.95 &1000	& $10^{5}$	&   6.3$\times 10^{17}$ & 4.18 &	6.31$\times 10^{6}$ & 3.2 &$10^{13}$&  21.00\\
	&		&	   		&			&						& 		&300	& $10^{4}$	&	6.3$\times 10^{16}$ &		\\
	\hline
	
OH &	A	&	125	&2$\times$10$^8$&	5.0$\times 10^{17}$ & 4.67	\\

\enddata
\label{lvgmodel1}
\end{deluxetable}

\begin{deluxetable}{cccc}
\tablewidth{0pt}
\tablecaption{Luminosities in FIR\tablenotemark{a} for the full map and each position}
\tablehead{
\colhead{Species} 		& \colhead{$5\times5$}	 	&
\colhead{(A)} &
\colhead{(B)}
}
\startdata
$L_{\rm CO}$       &	0.96 	&	0.70  &	0.27 		\\
$L_{\rm H_2O}$& 	1.73 	&	1.21  &	0.39 		\\
$L_{\rm OH}$       &	0.34 	&	0.30	&	0.05 		\\
$L_{\rm OI} $	   &	0.17 	&	0.07 	&	0.04 		\\
$L_{\rm continuum}$ &	430 & 349\tablenotemark{b} &	90\tablenotemark{b} 	\\
\enddata
\tablenotetext{a}{In units of  10$^{-2}$ $L_{\sun}$}
\tablenotetext{b}{The FIR luminosities of MM(A) and MM(B)}

\label{lumtable}
\end{deluxetable}

\begin{deluxetable}{cccc}
\tablewidth{0pt}
\tablecaption{Fractional contribution to total line cooling\tablenotemark{a} for full map and each position}
\tablehead{
\colhead{Species} 		& \colhead{$5\times5$}	 	&
\colhead{(A)} &
\colhead{(B)}
}
\startdata
$L_{\rm CO}$ 	       &	30$\%$	&	31$\%$ 	&	36$\%$		\\
$L_{\rm H_2O}$ 	& 	54$\%$	&	53$\%$	&	52$\%$   	\\
$L_{\rm OH}$  	       &	11$\%$	&	13$\%$	&	7$\%$		\\
$L_{\rm OI}$  	       &	5$\%$	&	3$\%$	&	5$\%$		\\
\enddata
\tablenotetext{a}{$L_{\rm total, line}=L_{\rm H_2O}+L_{\rm CO}+L_{\rm OH}+L_{\rm OI}$}
\label{relalumtable}
\end{deluxetable}

\clearpage

\appendix

\section{Line flux and error calculation}

Reduced data with HIPE v6.1 are flux calibrated more accurately while HIPE v8.1
provide higher S/N spectra.
Therefore, we calculated equivalent width  from HIPE v8.1 reduction $(EW_8$)
and multiplied the equivalent width by the local continuum of HIPE v6.1
reduction $(F\rm_{conti,6}$) to obtain line flux $(LF)$ over 5$\times$5 spaxels.

$LF = EW\rm_8 \times \it F\rm_{conti,6} = \rm\frac{\it LF\rm_8}{\it F\rm_{conti,8}} \times
\it F\rm_{conti,6}$,

where $\it F\rm_{conti,6}$ were measured from a sum over the whole 5$\times$5 spaxels while
$EW\rm_8$ were measured from the spectrum extracted over only two spaxels (C and S) 
except for [OI] lines.
For the [OI] lines, $EW\rm_8$ were measured from the sum over 4 spaxels with clear detection.

According to the error propagation, the line flux errors in
Table~\ref{fluxtable} are calculated with the equation,

$\delta LF =   \sqrt{(\frac{F\rm_{conti,6}}{\it F\rm_{conti,8}}\delta \it LF\rm_8)^2 +
(\frac{\it LF\rm_8 \times \it F\rm_{conti,6}}{\it F\rm_{conti,8}^2}\delta \it F\rm_{conti,8})^2 +
(\frac{\it LF\rm_8}{\it F\rm_{conti,8}}\delta \it F\rm_{conti,6})^2}$.

We assumed that continuum uncertainty is 20$\%$ of the flux.

\end{document}